\documentclass[aps,prd,superscriptaddress,nofootinbib]{revtex4-1}

\usepackage{graphicx} 
\usepackage{bm} 
\usepackage{hyperref} 
\usepackage[usenames]{color} 
\usepackage{multirow}

\newcommand{\numu}{\ensuremath{\nu_{\mu}}}                   
\newcommand{\numubar}{\ensuremath{\overline{\nu}_{\mu}}}                   
\newcommand{\nue}{\ensuremath{\nu_{e}}}                   
\newcommand{\nuebar}{\ensuremath{\overline{\nu}_{e}}}                   
     
\newcommand{\deltacp}{\ensuremath{\delta_{CP}}}

\begin{document}



\title{
A Long Baseline Neutrino Oscillation Experiment \\ 
Using J-PARC Neutrino Beam  
and Hyper-Kamiokande 
}




\date{\today}

\begin{abstract}
Hyper-Kamiokande will be a next generation underground water Cherenkov detector with 
a total (fiducial) mass of 0.99 (0.56) million metric tons, 
approximately 20 (25) times larger than that of Super-Kamiokande.
One of the main goals of Hyper-Kamiokande is the study of $CP$ asymmetry in the lepton sector using accelerator neutrino and anti-neutrino beams.

In this document, the physics potential of a long baseline neutrino experiment using the Hyper-Kamiokande detector and a neutrino beam from the J-PARC proton synchrotron is presented.
The analysis has been updated from the previous Letter of Intent [K. Abe et al., arXiv:1109.3262 [hep-ex]], based on the experience gained from the ongoing T2K experiment.
With a total exposure of 7.5~MW $\times$ 10$^7$ sec integrated proton beam power (corresponding to $1.56\times10^{22}$ protons on target with a 30~GeV proton beam)
to a $2.5$-degree off-axis neutrino beam
produced by the J-PARC proton synchrotron,
it is expected that the $CP$ phase $\deltacp$ can be determined to better than 19 degrees 
for all possible values of $\deltacp$,
and $CP$ violation can be established with a statistical significance 
of more than $3\,\sigma$ ($5\,\sigma$) for $76\%$ ($58\%$) of the $\,\deltacp$ parameter space.
\end{abstract}


\newcommand{\BERN}{\affiliation{University of Bern, Albert Einstein Center for Fundamental Physics, Laboratory for High Energy Physics (LHEP), Bern, Switzerland}}
\newcommand{\BOSTON}{\affiliation{Boston University, Department of Physics, Boston, Massachusetts, U.S.A.}}
\newcommand{\UBC}{\affiliation{University of British Columbia, Department of Physics and Astronomy, Vancouver, British Columbia, Canada }}
\newcommand{\UCDAVIS}{\affiliation{University of California, Davis, Department of Physics, Davis, California, U.S.A.}}
\newcommand{\UCI}{\affiliation{University of California, Irvine, Department of Physics and Astronomy, Irvine, California, U.S.A.}}
\newcommand{\CSU}{\affiliation{California State University, Department of Physics, Carson, California, U.S.A.}}
\newcommand{\SACLAY}{\affiliation{IRFU, CEA Saclay, Gif-sur-Yvette, France}}
\newcommand{\CHONNAM}{\affiliation{Chonnam National University, Department of Physics, Gwangju, Korea}}
\newcommand{\DONGSHIN}{\affiliation{Dongshin University, Department of Physics, Naju, Korea}}
\newcommand{\DUKE}{\affiliation{Duke University, Department of Physics, Durham, North Carolina, U.S.A.}}
\newcommand{\DURHAM}{\affiliation{University of Durham, Science Laboratories, Durham, United Kingdom}}
\newcommand{\LLR}{\affiliation{Ecole Polytechnique, IN2P3-CNRS, Laboratoire Leprince-Ringuet, Palaiseau, France}}
\newcommand{\EDINBURGH}{\affiliation{University of Edinburgh, School of Physics and Astronomy, Edinburgh, United Kingdom}}
\newcommand{\ETHZ}{\affiliation{ETH Zurich, Institute for Particle Physics, Zurich, Switzerland}}
\newcommand{\GENEVA}{\affiliation{University of Geneva, Section de Physique, DPNC, Geneva, Switzerland}}
\newcommand{\HAWAII}{\affiliation{University of Hawaii, Department of Physics and Astronomy, Honolulu, Hawaii, U.S.A.}}
\newcommand{\IMPERIAL}{\affiliation{Imperial College London, Department of Physics, London, United Kingdom}}
\newcommand{\BARI}{\affiliation{INFN Sezione di Bari and Universit\`a e Politecnico di Bari, Dipartimento Interuniversitario di Fisica, Bari, Italy}}
\newcommand{\NAPOLI}{\affiliation{INFN Sezione di Napoli and Universit\`a di Napoli, Dipartimento di Fisica, Napoli, Italy}}
\newcommand{\PADOVA}{\affiliation{INFN Sezione di Padova and Universit\`a di Padova, Dipartimento di Fisica, Padova, Italy}}
\newcommand{\ROME}{\affiliation{INFN Sezione di Roma, Roma, Italy}}
\newcommand{\INR}{\affiliation{Institute for Nuclear Research of the Russian Academy of Sciences, Moscow, Russia}}
\newcommand{\ISU}{\affiliation{Iowa State University, Department of Physics and Astronomy, Ames, Iowa, U.S.A.}}
\newcommand{\IPMU}{\affiliation{University of Tokyo, Kavli Institute for the Physics and Mathematics of the Universe (WPI), Todai Institutes for Advanced Study, Kashiwa, Chiba, Japan}}
\newcommand{\KEK}{\affiliation{High Energy Accelerator Research Organization (KEK), Tsukuba, Ibaraki, Japan}}
\newcommand{\KOBE}{\affiliation{Kobe University, Department of Physics, Kobe, Japan}}
\newcommand{\KYOTO}{\affiliation{Kyoto University, Department of Physics, Kyoto, Japan}}
\newcommand{\LNF}{\affiliation{Laboratori Nazionali di Frascati, Frascati, Italy}}
\newcommand{\LANCASTER}{\affiliation{Lancaster University, Physics Department, Lancaster, United Kingdom}}
\newcommand{\LIVERPOOL}{\affiliation{University of Liverpool, Department of Physics, Liverpool, United Kingdom}}
\newcommand{\LANL}{\affiliation{Los Alamos National Laboratory, New Mexico, U.S.A.}}
\newcommand{\LSU}{\affiliation{Louisiana State University, Department of Physics and Astronomy, Baton Rouge, Louisiana, U.S.A. }}
\newcommand{\MADRID}{\affiliation{University Autonoma Madrid, Department of Theoretical Physics, Madrid, Spain}}
\newcommand{\MIYAGI}{\affiliation{Miyagi University of Education, Department of Physics, Sendai, Japan}}
\newcommand{\NAGOYA}{\affiliation{Nagoya University, Nagoya, Japan}}
\newcommand{\NCNR}{\affiliation{National Centre for Nuclear Research, Warsaw, Poland}}
\newcommand{\OKAYAMA}{\affiliation{Okayama University, Department of Physics, Okayama, Japan}}
\newcommand{\OCU}{\affiliation{Osaka City University, Department of Physics, Osaka, Japan}}
\newcommand{\OXFORD}{\affiliation{Oxford University, Department of Physics, Oxford, United Kingdom}}
\newcommand{\PITTSBURGH}{\affiliation{University of Pittsburgh, Department of Physics and Astronomy, Pittsburgh, Pennsylvania, U.S.A.}}
\newcommand{\REGINA}{\affiliation{University of Regina, Department of Physics, Regina, Saskatchewan, Canada}}
\newcommand{\RIO}{\affiliation{Pontif{\'\i}cia Universidade Cat{\'o}lica do Rio de Janeiro, Departamento de F\'{\i}sica, Rio de Janeiro, Brazil}}
\newcommand{\ROCHESTER}{\affiliation{University of Rochester, Department of Physics and Astronomy, Rochester, New York, U.S.A.}}
\newcommand{\QMUL}{\affiliation{Queen Mary University of London, School of Physics and Astronomy, London, United Kingdom}}
\newcommand{\RHUL}{\affiliation{Royal Holloway University of London, Department of Physics, Egham, Surrey, United Kingdom}}
\newcommand{\SAOPAULO}{\affiliation{Universidade de S\~ao Paulo, Instituto de F\'{\i}sica, S\~ao Paulo, Brazil}}
\newcommand{\SHEFFIELD}{\affiliation{University of Sheffield, Department of Physics and Astronomy, Sheffield, United Kingdom}}
\newcommand{\SNU}{\affiliation{Seoul National University, Department of Physics, Seoul, Korea}}
\newcommand{\SEOYEONG}{\affiliation{Seoyeong University, Department of Fire Safety, Gwangju, Korea }}
\newcommand{\STONYBROOK}{\affiliation{State University of New York at Stony Brook, Department of Physics and Astronomy, Stony Brook, New York, U.S.A.}}
\newcommand{\RAL}{\affiliation{STFC, Rutherford Appleton Laboratory, Harwell Oxford, and Daresbury Laboratory, Warrington, United Kingdom}}
\newcommand{\SKKU}{\affiliation{Sungkyunkwan University, Department of Physics, Suwon, Korea}}
\newcommand{\TOHOKU}{\affiliation{Research Center for Neutrino Science, Tohoku University, Sendai, Japan}}
\newcommand{\ERI}{\affiliation{University of Tokyo, Earthquake Research Institute, Tokyo, Japan}}
\newcommand{\KAMIOKA}{\affiliation{University of Tokyo, Institute for Cosmic Ray Research, Kamioka Observatory, Kamioka, Japan}}
\newcommand{\RCCN}{\affiliation{University of Tokyo, Institute for Cosmic Ray Research, Research Center for Cosmic Neutrinos, Kashiwa, Japan}}
\newcommand{\TOKYO}{\affiliation{University of Tokyo, Department of Physics, Tokyo, Japan}}
\newcommand{\TITECH}{\affiliation{Tokyo Institute of Technology, Department of Physics, Tokyo, Japan}}
\newcommand{\TRIUMF }{\affiliation{TRIUMF, Vancouver, British Columbia, Canada}}
\newcommand{\TORONTO}{\affiliation{University of Toronto, Department of Physics, Toronto, Ontario, Canada}}
\newcommand{\WARSAW}{\affiliation{University of Warsaw, Faculty of Physics, Warsaw, Poland}}
\newcommand{\WARWICK}{\affiliation{University of Warwick, Department of Physics, Coventry, United Kingdom}}
\newcommand{\WASHINGTON}{\affiliation{University of Washington, Department of Physics, Seattle, Washington, U.S.A.}}
\newcommand{\WINNIPEG}{\affiliation{University of Winnipeg, Department of Physics, Winnipeg, Manitoba, Canada}}
\newcommand{\VT}{\affiliation{Virginia Tech, Center for Neutrino Physics, Blacksburg, Virginia, U.S.A.}}
\newcommand{\WROCLAW}{\affiliation{Wroclaw University, Faculty of Physics and Astronomy, Wroclaw, Poland}}
\newcommand{\YORK}{\affiliation{York University, Department of Physics and Astronomy, Toronto, Ontario, Canada}}

\BERN
\BOSTON
\UBC
\UCDAVIS
\UCI
\CSU
\SACLAY
\CHONNAM
\DONGSHIN
\DUKE
\DURHAM
\LLR
\EDINBURGH
\ETHZ
\GENEVA
\HAWAII
\IMPERIAL
\BARI
\NAPOLI
\PADOVA
\ROME
\INR
\ISU
\KEK
\KOBE
\KYOTO
\LNF
\LANCASTER
\LIVERPOOL
\LANL
\LSU
\MADRID
\MIYAGI
\NAGOYA
\NCNR
\OKAYAMA
\OCU
\OXFORD
\PITTSBURGH
\REGINA
\RIO
\ROCHESTER
\QMUL
\RHUL
\SAOPAULO
\SHEFFIELD
\SNU
\SEOYEONG
\STONYBROOK
\RAL
\SKKU
\TOHOKU
\ERI
\KAMIOKA
\RCCN
\IPMU
\TOKYO
\TITECH
\TRIUMF 
\TORONTO
\WARSAW
\WARWICK
\WASHINGTON
\WINNIPEG
\VT
\WROCLAW
\YORK

\author{K.~Abe}\KAMIOKA\IPMU
\author{H.~Aihara}\TOKYO\IPMU
\author{C.~Andreopoulos}\LIVERPOOL
\author{I.~Anghel}\ISU
\author{A.~Ariga}\BERN
\author{T.~Ariga}\BERN
\author{R.~Asfandiyarov}\GENEVA
\author{M.~Askins}\UCDAVIS
\author{J.J.~Back}\WARWICK
\author{P.~Ballett}\DURHAM
% ...

\collaboration{The Hyper-Kamiokande Working Group}\noaffiliation

\maketitle

\tableofcontents
\newpage


\vspace*{4cm}
\begin {figure}[htbp]
  \begin{center}
    \includegraphics[width=\textwidth]{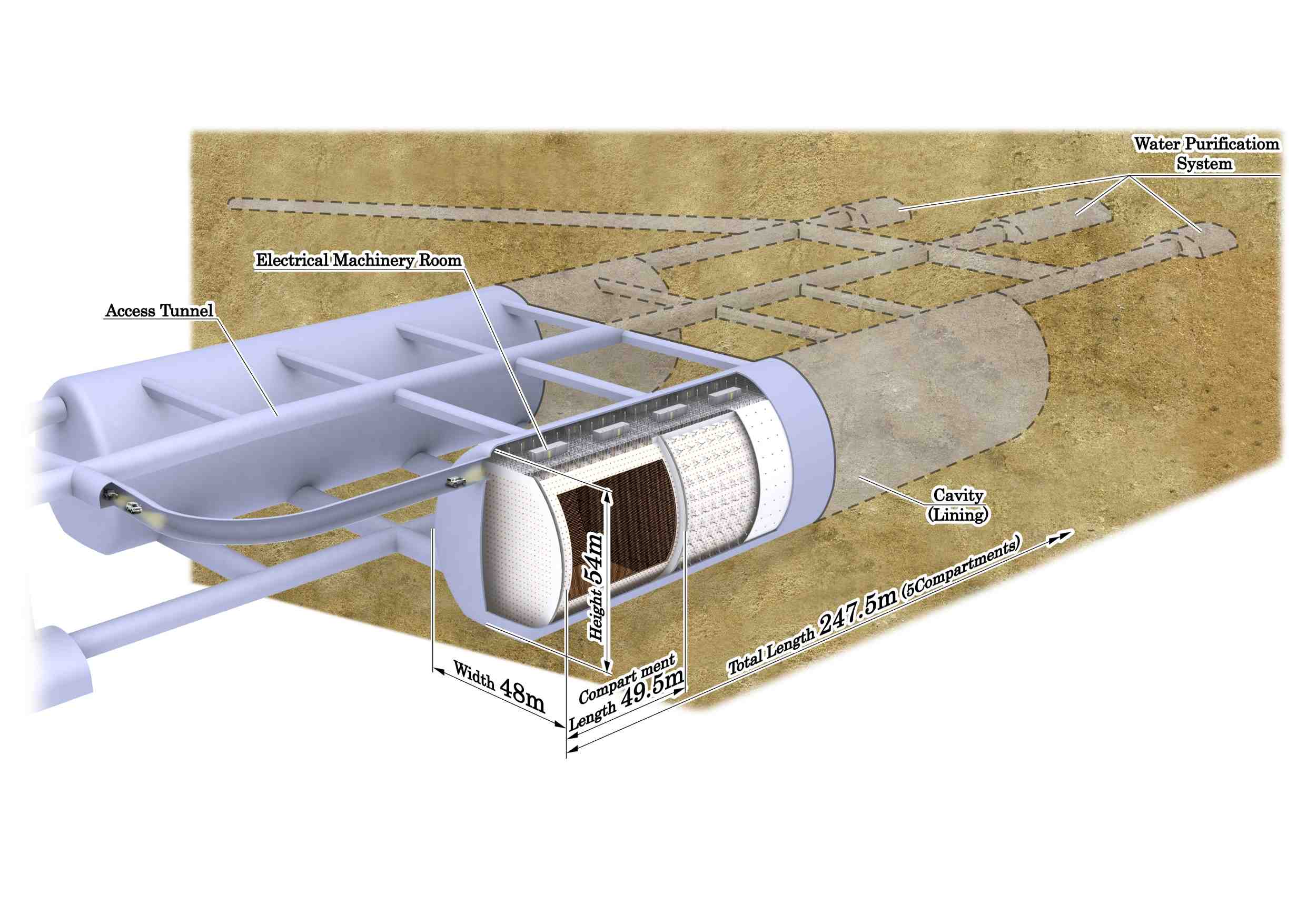}
    \caption{Schematic view of the Hyper-Kamiokande detector~\cite{Abe:2011ts}.}
    \label{fig:hk_schematic}
  \end{center}
\end {figure}
\clearpage

\section{Introduction}
The goal of particle physics is to discover and understand the fundamental laws of nature.
The Standard Model (SM), which is the current paradigm of elementary particles and their interactions, gives a successful account of the
experimental data to date \cite{Agashe:2014kda}.
Yet, deeper insights are still needed to answer more fundamental questions.
For instance, why does a gauge  structure of 
$SU(3)_C \otimes SU(2)_L \otimes U(1)_Y$ among  the strong, weak and
electromagnetic interactions exist?
Why is there a three generation structure of fundamental fermions
and what are the origins of the masses and generation mixings of quarks and leptons?
To address these questions physics beyond the SM (BSM)  is required.

The discovery of neutrino oscillations by the Super-Kamiokande (Super-K) 
experiment in 1998~\cite{Fukuda:1998mi}  opened a new window to explore BSM physics.
Evidence of neutrino oscillations is one of the most convincing experimental proofs known today that shows the existence of BSM physics at work.
The mixing parameters of neutrinos 
were found
to be remarkably different from those of quarks, which suggests the presence of an unknown flavor symmetry waiting to be explored.
The extremely small masses of neutrinos compared with those of their charged partners
lead to the preferred scenario of a seesaw mechanism~\cite{Minkowski:1977sc, GellMann:1980vs, Yanagida:1979as, Mohapatra:1979ia}, in which small neutrino masses are 
a reflection of the ultra-high energy scale of BSM physics.

Furthermore, a theoretical framework called Leptogenesis points to the intriguing possibility that 
$CP$ asymmetries related to flavor mixing among the three generations of neutrinos may have played an important role 
in creating the observed matter-antimatter asymmetry in the universe~\cite{Fukugita:1986hr}.
This makes a study of the full picture of neutrino masses and mixings and the
measurement of the $CP$ asymmetry
in the neutrino sector among the most important and urgent subjects in today's 
elementary particle physics world.

$CP$ asymmetry in the neutrino sector can only be seen if all the three mixing angles governing neutrino oscillations differ from zero. 
The Super-K detector has successfully measured all three angles. The angle $\theta_{23}$ was first measured in atmospheric neutrino observations~\cite{Fukuda:1998mi}, $\theta_{12}$ was constrained in solar neutrino observations~\cite{Fukuda:2001nk}, and the evidence of non-zero $\theta_{13}$ was found in T2K~\cite{Abe:2011sj} which used Super-K as the far detector.
In 2013, T2K established $\nu_\mu \to \nu_e$ oscillation with 7.3\,$\sigma$ 
significance, leading the way towards $CP$ violation measurements in neutrinos~\cite{Abe:2013hdq}. 
The highly successful Super-K program indicates that Hyper-Kamiokande (Hyper-K)
is well placed to discover $CP$ violation.


\subsection{Introduction to Neutrino Oscillations and $CP$ violation}
Throughout this document, unless stated otherwise, we consider 
the standard three flavor neutrino framework. 
The 3$\times3$ unitary matrix $U$ which describes the mixing of 
neutrinos~\cite{Maki:1962mu} (that is often referred to as the Maki-Nakagawa-Sakata-Pontecorvo (MNSP) or
Maki-Nakagawa-Sakata (MNS)~\cite{Pontecorvo:1967fh,Maki:1962mu} matrix) 
relates the flavor and mass eigenstates of neutrinos: 
\begin{eqnarray} 
\nu_\alpha = \sum_{i=1}^3 U_{\alpha i} \nu_i, 
\ \ (\alpha = e, \mu, \tau),
\end{eqnarray} 
where $\nu_\alpha  (\alpha = e, \mu, \tau)$ and 
$\nu_i  (i = 1,2,3)$ denote, respectively, flavor and 
mass eigenstates of neutrinos. 
Using the standard parametrization, which can be found, e.g. in 
Ref.~\cite{Agashe:2014kda}, $U$ can be expressed as,  
\begin{eqnarray}
U 
& = &
\left(
\begin{array}{ccc}
1 & 0   & 0 \\
0 & c_{23} & s_{23} \\
0 & -s_{23} & c_{23} \\
\end{array}
\right)
\left(
\begin{array}{ccc}
c_{13} & 0   & s_{13}e^{-i\deltacp} \\
0 & 1 & 0 \\
-s_{13}e^{i\deltacp} & 0 & c_{13} \\
\end{array}
\right)
\left(
\begin{array}{ccc}
c_{12} & s_{12}  & 0 \\
-s_{12} & c_{12} & 0 \\
0 & 0 & 1 \\
\end{array}
\right)
\times
\left(
\begin{array}{ccc}
1 & 0 & 0 \\
0 & e^{i\frac{\alpha_{21}}{2}} & 0 \\
0 & 0 & e^{i\frac{\alpha_{31}}{2}} \\
\end{array}
\right)
\label{eq:mixing}
\end{eqnarray}
where $c_{ij} \equiv \cos\theta_{ij}$, $s_{ij} \equiv \sin\theta_{ij}$, 
and $\deltacp$ --- often called the Dirac $CP$ phase ---,  
is the Kobayashi-Maskawa type $CP$ phase~\cite{Kobayashi:1973fv} 
in the lepton sector. 
On the other hand, the two phases, $\alpha_{21}$ and $\alpha_{31}$, 
--- often called Majorana $CP$ phases --- 
have physical meaning only if neutrinos are of Majorana 
type~\cite{Schechter:1980gr}.
While the Majorana $CP$ phases can not be observed in neutrino 
oscillation~\cite{Bilenky:1980cx,Doi:1980yb},
they can be probed by lepton number violating processes such as 
neutrinoless double beta decay.

In vacuum,
the oscillation probability of $\nu_\alpha \to  \nu_\beta$ ($\alpha,\beta = e,\mu, \tau$)
for ultrarelativistic neutrinos is given by, 
\begin{eqnarray}
&&P(\nu_\alpha \to  \nu_\beta)   =  
\left|~\sum_{i=1}^3 U^*_{\alpha i} ~U_{\beta i} 
\text{e}^{-i\frac{m^2_i}{2E}L}~\right|^2 \nonumber \\
&= &
\delta_{\alpha \beta} 
-4 \sum_{i>j} {\Re}(U^*_{\alpha i} U_{\alpha j} U_{\beta i} U^*_{\beta j} )
\sin^2\left(  \frac{\Delta m^2_{ij}}{4E}L \right) 
+2 \sum_{i>j} 
{\Im}(U^*_{\alpha i} U_{\alpha j} U_{\beta i} U^*_{\beta j} )
\sin \left( \frac{\Delta m^2_{ij}}{2E}L \right),
\label{eq:prob_vac}
\end{eqnarray}
where $E$ is the neutrino energy, $L$ is the baseline, 
$\Delta m^2_{ij} \equiv m_i^2-m_j^2$ $(i,j=1,2,3)$ is the mass squared
differences with $m_i$ and $m_j$ being the neutrino masses. 
For the $CP$ conjugate channel, $\bar{\nu}_\alpha \to  \bar{\nu}_\beta$, 
the same expression in Eq.~(\ref{eq:prob_vac}) holds, but the matrix $U$ 
is replaced by its complex conjugate (or equivalently $\deltacp \to -\deltacp$ in Eq.~(\ref{eq:mixing})), 
resulting in the third term in Equation~\ref{eq:prob_vac} switching sign.
For neutrinos traveling inside matter, coherent forward scattering induces
an asymmetry between the oscillation probabilities of neutrinos and antineutrinos unrelated to the intrinsic $CP$ violation.

Since there are only three neutrinos, only two mass squared 
differences, $\Delta m^2_{21}$ and $\Delta m^2_{31}$, for example, 
are independent. 
Therefore, for a given energy and baseline, 
there are six independent parameters, namely, 
three mixing angles, one $CP$ phase, and two mass squared differences, 
in order to describe neutrino oscillations.
Among these six parameters, 
$\theta_{12}$ and $\Delta m^2_{21}$ have been measured
by solar~\cite{Ahmad:2002jz,Ahmad:2001an,Abe:2010hy} 
and reactor~\cite{Eguchi:2002dm,Araki:2004mb,Abe:2008aa}
neutrino experiments.
The parameters $\theta_{23}$ and $|\Delta m^2_{32}|$ 
(only its absolute value)
have been measured by 
atmospheric neutrino \cite{Ashie:2005ik,Ashie:2004mr}
and accelerator neutrino experiments~\cite{Ahn:2006zza,Adamson:2011ig,Abe:2012gx,Abe:2014ugx}.
Recently, \(\theta_{13}\) has also been measured by 
accelerator~\cite{Abe:2011sj,Adamson:2011qu,Abe:2013xua,Abe:2013hdq} and reactor 
experiments~\cite{Abe:2011fz,Ahn:2012nd,An:2012eh, An:2013zwz,Abe:2014lus}.
The relatively large value of $\theta_{13}$ opens the window
to explore the $CP$ phase, $\deltacp$, in neutrino oscillation.

The oscillation probability from $\nu_\mu$ to $\nu_e$ in accelerator experiments
is expressed, 
to the first order of the matter effect, as follows~\cite{Arafune:1997hd}:
\begin{eqnarray}
P(\numu \to \nue) & = & 4 c_{13}^2s_{13}^2s_{23}^2 \cdot \sin^2\Delta_{31}  \nonumber \\
& & +8 c_{13}^2s_{12}s_{13}s_{23} (c_{12}c_{23}\cos\deltacp - s_{12}s_{13}s_{23})\cdot \cos\Delta_{32} \cdot \sin\Delta_{31}\cdot \sin\Delta_{21} \nonumber \\
& & -8 c_{13}^2c_{12}c_{23}s_{12}s_{13}s_{23}\sin\deltacp \cdot \sin\Delta_{32} \cdot \sin\Delta_{31}\cdot \sin\Delta_{21} \nonumber \\
& & +4s_{12}^2c_{13}^2(c_{12}^2c_{23}^2 + s_{12}^2s_{23}^2s_{13}^2-2c_{12}c_{23}s_{12}s_{23}s_{13}\cos\deltacp)\cdot \sin^2\Delta_{21} \nonumber \\
& & -8c_{13}^2s_{13}^2s_{23}^2\cdot \frac{aL}{4E_\nu} (1-2s_{13}^2)\cdot \cos\Delta_{32}\cdot \sin\Delta_{31} \nonumber \\
& & +8 c_{13}^2s_{13}^2s_{23}^2 \frac{a}{\Delta m^2_{31}}(1-2s_{13}^2)\cdot\sin^2\Delta_{31}, \label{Eq:cpv-oscprob}
\end{eqnarray}
\noindent where $\Delta_{ij}$ is $\Delta m^2_{ij}\, L/4E_\nu$, 
and $a \mathrm{[eV^2]}= 7.56\times 10^{-5} \times \rho \mathrm{[g/cm^3]} \times E_\nu[\mathrm{GeV}] $.
The corresponding probability for a $\numubar \to \nuebar$ transition is obtained by replacing $\deltacp \rightarrow -\deltacp$
and $a \rightarrow -a$.
The third term, containing $\sin\deltacp$, is the $CP$ violating term which flips sign between $\nu$ and $\bar{\nu}$ and thus introduces $CP$ asymmetry if $\sin\deltacp$ is non-zero.
The last two terms are due to the matter effect. 
As seen from the definition of $a$, the amount of asymmetry due to the matter effect is proportional to the neutrino energy at a fixed value of $L/E_\nu$.

The magnitude of the $CP$ violation in neutrino oscillation 
can be characterized by the probabilities between neutrino and anti-neutrino channels, 
which, in vacuum, is given by~\cite{Barger:1980jm,Pakvasa:1980bz}, 
\begin{eqnarray}
\Delta P_{\alpha\beta} \equiv P(\nu_\alpha \to  \nu_\beta)
- P(\bar\nu_\alpha \to  \bar\nu_\beta)
= 16 J_{\alpha \beta} \sin \Delta_{12}
\sin \Delta_{23} \sin \Delta_{31},
\label{eq:DeltaP}
\end{eqnarray}
and 
\begin{eqnarray}
J_{\alpha \beta} \equiv 
{\Im}(U_{\alpha 1}U^*_{\alpha 2} U^*_{\beta 1} U_{\beta 2} )
= \pm J_{CP}, \ \ 
J_{CP} \equiv s_{12}c_{12}s_{23}c_{23}s_{13}c_{13}^2\sin\deltacp
\label{eq:Jarlskog} 
\end{eqnarray}
with positive (negative) sign for (anti-)cyclic permutation of the
flavor indices $e$, $\mu$ and $\tau$.  
The parameter $J_{CP}$ is the lepton analogue of 
the $CP$-invariant factor for quarks, 
the unique and phase-convention-independent measure 
for $CP$ violation~\cite{Jarlskog:1985ht}.
In matter with constant density, the same expressions in 
Eqs.~(\ref{eq:prob_vac})--(\ref{eq:Jarlskog}) hold, but the
mixing angles $\theta_{ij}$ and $\Delta m^2_{ij}$ must be replaced
by the effective ones in matter. 
Using the current best fitted values of mixing parameters~\cite{Capozzi:2013csa}, we get 
$J_{CP} \simeq 0.034 \sin \deltacp$, or   
\begin{equation}
\Delta P_{\alpha\beta} 
\simeq  \pm 0.55  \sin \deltacp \sin \Delta_{12}
\sin \Delta_{23} \sin \Delta_{31}. 
\label{eq:DeltaP2}
\end{equation}
Thus, a large $CP$ violation effects are possible in the neutrino oscillation.

In general, it is considered that $CP$ violation in the neutrino sector which can be observed in the low energy regime, namely, in neutrino oscillation, does not directly imply the $CP$ violation required at high energy for the successful leptogenesis in the early universe. 
It has been discussed, however, that they could be related to each other and the $CP$ violating phase in the MNS matrix could be responsible also for the generation of the observed baryon asymmetry through leptogenesis in some scenarios. 
For example, in~\cite{Pascoli:2006ie,Pascoli:2006ci}, in the context of the seesaw mechanism, it has been pointed out that assuming the hierarchical mass spectrum for right handed Majorana neutrinos with the lightest mass to be $\lesssim 5\times 10^{12}$~GeV, observed baryon asymmetry could be generated through the leptogenesis if $|\sin\theta_{13}\sin\delta_{CP}|\gtrsim 0.1$, which is compatible with the current neutrino data. 
Hence, measurement of $CP$ asymmetry in neutrino oscillations may provide a clue for understanding the origin of matter-antimatter asymmetry of the Universe.

When we measure $\theta_{23}$ with the survival probability $P(\numu \to \numu)$ which is proportional to $\sin^22\theta_{23}$ to first order, 
\begin{eqnarray*}
P(\nu_\mu \rightarrow \nu_\mu) &\simeq& 1-4c^2_{13}s^2_{23} [1-c^2_{13}s^2_{23}]\sin^2(\Delta m^2\, L/4E_\nu) \\
&\simeq & 1-\sin^22\theta_{23}\sin^2(\Delta m^2\, L/4E_\nu), \hspace{2cm} \textrm{(for $c_{13}\simeq1$)}
\end{eqnarray*}
there is an octant ambiguity:  either $\theta_{23} \le 45^\circ $  (in the first octant) or $\theta_{23} > 45^\circ $  (in the second octant).
By combining the measurement of $P(\numu \to \nue)$, the $\theta_{23}$ octant can be determined.

\subsection{Expected results from T2K, NO$\nu$A and reactor experiments}
The T2K experiment \cite{Abe:2011sj} first found evidence for the 
parameter $\theta_{13}$ by studying $\nu_\mu \to \nu_e$ oscillations. The 
reactor experiments subsequently verified the observation and precisely 
measured the parameter~\cite{Abe:2011fz,An:2012eh,Ahn:2012nd}. 
Then, in 2013, T2K established $\nu_\mu \to \nu_e$ oscillation with 7.3\,$\sigma$ significance~\cite{Abe:2013hdq}, by which a measurement of $CP$ violation in neutrinos becomes realistic.

 The T2K experiment is based on a neutrino beam (mainly $\nu_\mu$)  generated
 at J-PARC from a 30\,GeV proton beam incident on a 90~cm long carbon target.
 The neutrino beam is observed in a  $2.5^\circ$ off-axis direction so that the
 average neutrino energy $E_\nu$ is peaked at the first oscillation maximum,
 with a multi purpose detector (ND280) consisting of a fully active tracker for
 charged particles and lead-scintillator calorimeters for photons,  immersed in
 a $0.2$-T magnetic field; it is used for characterizing the initial beam
 composition and flux and for determining the relevant cross sections. The far detector, Super-K located 295~km away in the Kamioka mine also sits $2.5^\circ$ off-axis.
An on-axis detector made  of iron-scintillator tracker modules is used to monitor the beam direction and profile on a daily basis.
T2K has been approved for $7.8 \times 10^{21}$ protons-on-target (POT).  It has
been running at 240\,kW beam power and the J-PARC upgrade plan for the Main Ring accelerator (MR) calls for operation at 750kW by FY 2017. Hence by 2020, one expects to have accumulated the approved POT.

The NO$\nu$A experiment at Fermilab, is also exploiting an off-axis beam from the existing 120~GeV Main Injector, initially 
starting at 350\,kW. With the upgrades to the Booster, the beam power will reach 700~kW.
Both near and far detectors are identical liquid scintillator tracking
calorimeters with wavelength shifter read out, respectively 330 tons and 14,000
tons in weight. The near detector is at 1.01\,km while the far detector is at 810~km. A 14~mrad off-axis angle is chosen so that the $\langle E_\nu \rangle$ is 2~GeV, centered on the first oscillation maximum for $\nu_\mu$ to $\nu_e$ oscillation (400~$\rm km/GeV$).

T2K has been fully operational since 2010, while NOvA started data taking with a fully operational far detector in the summer 2014.

The updated physics goals for T2K~\cite{T2KPACreport,Abe:2014tzr} are focused on the search for evidence of
$CP$ violation in the MNS mixing matrix. Combining the value of $\theta_{13}$
obtained from the reactor experiments, such as Daya Bay, RENO and Double Chooz, which are not sensitive to the $CP$ violation phases, with those obtained from $\nu_e$ appearance, which are highly correlated to the $CP$ phases, T2K  will search for:
\begin{itemize}
\item Signal of a $CP$ violation phase
\item Precision measurement of the MNSP mixing matrix elements $\Delta m^2_{32}$ to $10^{-4}$~$\rm eV^2$, $\sin^2 2\theta_{23}$ to 0.01, determination of the $\theta_{23}$ octant
\item Provide experimental data useful to improve the mass hierarchy (the sign of $\Delta m^2_{32}$) sensitivity of other experiments.
\end{itemize}
The goals of NO$\nu$A are similar, but because of the longer baseline, NO$\nu$A has more sensitivity to the mass hierarchy through the matter oscillation terms.

The two experiments, T2K and NO$\nu$A, are complementary and a combined analysis will produce the best chances of observing  $\deltacp$, 
the sign of $\Delta m^2_{32}$ and $\theta_{23}$ octant.
Referring to the expression for the appearance probability, one notes the strong correlations between the three quantities $\deltacp$, 
$\sin^2 \theta_{23}$ and $\Delta m^2_{32}$. 
Recent studies of T2K and NO$\nu$A combined analyses together with the precise $\theta_{13}$ values by reactor experiments
indicate that by 2020 one could establish the presence of a $CP$ phase at the 1.5 to 2.5 $\sigma$ level, the sign of $\Delta m^2_{32}$ at the 1 to 3 $\sigma$ level and the $\theta_{23}$ octant at the 1.5 to 2 $\sigma$ level if $|\theta_{23} -45^\circ|> 4^\circ$~\cite{T2KPACreport,Abe:2014tzr}. 
In Table~\ref{tab:intro:osci}, we summarize the expected sensitivity of T2K and
NO$\nu$A for the $CP$ phase, the sign of $\Delta m^2_{32}$ and the $\theta_{23}$
octant by 2020 together with the current knowledge of neutrino oscillation parameters in the 
Particle Data Book (PDG) 2014~\cite{Agashe:2014kda} and the $1\,\sigma$ range calculated by a global fit~\cite{Capozzi:2013csa}.

\begin{table}[tb]
\begin{center}
\caption{The expected sensitivity of T2K and NO$\nu$A for the $CP$ phase, the sign of $\Delta m^2_{32}$ and the $\theta_{23}$ octant by 2020. As a reference, the current knowledge of neutrino oscillation parameters in PDG 2014~\cite{Agashe:2014kda} and the $1\,\sigma$ range calculated by a global fit~\cite{Capozzi:2013csa} in the case of $\Delta m_{32}^2>0$ are listed.
}
\label{tab:intro:osci}
\begin{tabular}{lllll}
\hline\hline
Parameter&  T2K \& NO$\nu$A in 2020 & PDG 2014 & $1\,\sigma$ range by a global fit \\ 
\hline\hline
$\deltacp$ & $ \neq 0$ at $1.5 \sim 2.5$\,$\sigma$ if $\deltacp = 1.5 \pi$ & unknown &  $1.12\pi \sim 1.77\pi$ \\
$sign (\Delta m_{32}^2)$ &  determination at $1.5 \sim 3$\,$\sigma$ & unknown & unknown \\
$|\Delta m_{32}^2|$ $\rm (eV^2)$& $\pm 0.04 \times 10^{-3}$ &  $(2.44 \pm 0.06) \times 10^{-3}$ & $(2.37 \sim 2.49) \times 10^{-3}$ \\
$\Delta m_{21}^2$ $\rm (eV^2)$& not sensitive  &  $(7.53 \pm 0.18) \times 10^{-5}$& $(7.32 \sim 7.80) \times 10^{-5}$ \\
$\theta_{23}$ octant & determination at  $1.5 \sim 2$\,$\sigma$ if $|\theta_{23} -45^\circ|> 4^\circ$ & unknown & $<$45$^\circ$ \\ 
$\sin^2 2 \theta_{23}$ &   $\pm 0.01 $  &  $0.999^{+0.001}_{-0.018}$ &  $0.97 \sim 1.00$  \\
$\sin^2 2 \theta_{12}$ &   not sensitive  &  $0.846 \pm 0.021$ &  $0.83 \sim 0.88$  \\
$\sin^2 2 \theta_{13}$ &   not precise  &  $0.093 \pm 0.008$ &  $0.084 \sim 0.099$\\
\hline\hline
\end{tabular}
\end{center}
\end{table}


The reactor neutrino oscillation experiments are an alternative and complementary way to measure the
$\theta_{13}$ angle.  
Currently three experiments,
Daya Bay~\cite{An:2012eh} in China,
Double Chooz~\cite{Abe:2011fz} in France and
RENO~\cite{Ahn:2012nd} in Korea are running.
All three experiments use liquid scintillator detectors, and place detectors at the optimum (far) distance
for oscillation, as well as at near distances to measure the un-oscillated flux thus canceling the systematics
due to the source flux uncertainty.  
The reaction used is inverse beta-decay, $\bar \nu_e p \to e^+ n$,
in which the delayed neutron capture signal (typically by the Gd nuclei doped in the scintillator) follows the prompt positron signal.


The strength and complementary of the reactor experiments lie in the fact that they are pure $\theta_{13}$
measurements, since the effects of the $\Delta m^2_{21}$ term, matter effect and those sensitive to $CP$ phase
are negligible at the distance of their measurements.  The survival probability is directly $1 - \sin^2 2\theta_{13}
\sin^2 (\Delta m^2_{31} L /4E)$, where $L$ is the distance and $E$ the neutrino energy.
By combining this $\theta_{13}$ measurement and the accelerator $\nu_e$ appearance probability,
one can have a handle on the effect of the $CP$ violation phase, as already hinted in the most recent T2K publication~\cite{Abe:2013hdq}.
In the next few years, the three experiments will improve the statistical and systematic uncertainties
and ultimately aim for $\sin^2 2\theta_{13}$ measurement at the level of 5\% precision.

\subsection{Anticipated Neutrino Physics Landscape in the 2020s}
Before Hyper-K commences data taking in $\sim 2025$, we expect Super-K, T2K,
NOvA, KamLAND, Double Chooz, Daya Bay, RENO experiments and cosmological
observations will advance our understanding of neutrino physics. 
In addition to accelerator and reactor experiments, Super-K will provide
precise measurements of neutrino oscillation parameters from atmospheric
neutrino observations, and will look for the mass hierarchy and the octant of $\theta_{23}$.
Cosmological observations will provide the information on neutrino masses. 
An observation by KamLAND-Zen or other experiments of neutrino-less double $\beta$ decay in the next 10 years would be evidence that the neutrino is a Majorana particle with the inverted mass hierarchy.
Following the progress, we definitely need a new experiment to discover $CP$ violation in neutrinos, and to unambiguously establish the mass hierarchy and $\theta_{23}$ octant. For the purposes, we propose the Hyper-K experiment with the J-PARC neutrino beam.

\subsubsection{Uniqueness of Hyper-Kamiokande with the J-PARC neutrino beam}
Hyper-K is a successor of Super-K and has various physics objectives listed in Table~\ref{tab:intro:phys}: search for $CP$ violation in neutrinos, precise study of neutrino oscillations including determination of mass hierarchy and $\theta_{23}$ octant, search for nucleon decay and observation of cosmic origin neutrinos. In this document, we focus on neutrino $CP$ violation. 
The uniqueness of Hyper-K is listed as follows.
\begin{itemize}
\item The experiment will operate in the same beam line as T2K with the same off-axis configuration. The feature of the neutrino beam and the operation of the high power beam are well understood.
\item The experiment will have high statistics of neutrino events thanks to the large fiducial mass and the high power J-PARC neutrino beam. 
\item The systematc errors are already well understood based on Super-K and T2K which makes reliable extrapolations.
\end{itemize}
With this uniqueness, Hyper-K is one of the most sensitive experiment to probe neutrino $CP$ violation, which will be reported in this paper.
A direct test of $CP$ violation is to measure both neutrino and antineutrino appearance probabilities in a model independent way.
Although the sensitivity of $CP$ violation is relating to a determination of the mass hierarchy, the mass hierarchy could be determined by the atmospheric neutrino measurement in Hyper-K and several measurements by other experiments mentioned in the next subsection.

\begin{table}[tb]
\begin{center}
\caption{Summary of the proposed experiments in the 2020s. The ``atm.'' means atmospheric neutrinos, and MH means ``Mass-Hierarchy''.}
\label{tab:intro:2020}
\begin{tabular}{lccccc}
\hline\hline
Experiment (Place) &  $\nu$ source  & Fiducial mass (kt) & Energy (MeV) & baseline (km) & physics targets \\ 
\hline\hline
Hyper-K (Japan)  &  beam     &560                 & 600 & 295 & $CP$, MH, $\theta_{23}$, $\theta_{13}$, $\Delta m_{32}^2$ \\
                            &  atm.       &560                 &   $100 \sim 10^{6}$      & $10 \sim 10,000$ & MH, $CP$, $\theta_{23}$, $\Delta m_{32}^2$  \\  
\hline
LBNE (US)           & beam      & 34   &  $1,000 \sim 5,000$     & 1300 & MH, $CP$, $\theta_{23}$, $\theta_{13}$, $\Delta m_{32}^2$  \\  
LBNO (EU)           & beam     & $ 20 \to 100$   &  $1,000 \sim 10,000$         & 2300 & MH, $CP$, $\theta_{23}$, $\theta_{13}$ , $\Delta m_{32}^2$ \\  
JUNO (China)      & reactor    & 20                  &  $1 \sim 10$    & $\sim 50$ & MH, $\theta_{12}$,  $\Delta m_{21}^2$, $\Delta m_{31}^2$  \\  
RENO50 (Korea) & reactor    & 10                  &  $1 \sim 10$    & 47    & MH, $\theta_{12}$, $\Delta m_{21}^2$, $\Delta m_{31}^2$   \\  
PINGU (South pole)         & atm.        & $\sim 6,000$  &  $1,000 \sim 10^{6}$    & $10 \sim 10,000$ & MH, $\theta_{23}$ and , $\Delta m_{32}^2$    \\  
ORCA (EU)          & atm.        & $\sim 2,000$  &  $1,000 \sim 10^{6}$    & $10 \sim 10,000$ & MH, $\theta_{23}$, $\Delta m_{32}^2$    \\  
INO (India)           & atm.        & 50                  &  $1,000 \sim 10^{6}$    & $10 \sim 10,000$ & MH, $\theta_{23}$, $\Delta m_{32}^2$    \\  

\hline\hline
\end{tabular}
\end{center}
\end{table}

\subsubsection{Other planned experiments: LBNE, LBNO and others}
Several new experiments throughout the world are proposed to start taking data in the 2020s.
The LBNE experiment in US and the LBNO experiment in Europe are accelerator based experiments to study $CP$ violation, the mass hierarchy and neutrino oscillations precisely. The projected neutrino beam powers are $\sim 1$\,MW, similar to J-PARC.
They adopt a longer baseline than that of Hyper-K which results in the better sensitivity for the mass hierarchy thanks to the larger matter effect. 
Their far detectors are Liquid Ar TPCs, which require intense R\&D to realize large scale detectors of $O(10)$~kton, while the technology for water Cherenkov detectors of $O(100)$~kton is more established for Hyper-K.
In addition to the technology, the understanding of detector systematics is more advanced for water Cherenkov detectors.
Much smaller far detectors of LBNO and LBNE result in less statistics of neutrino events. 
Due to the larger statistics, the better understanding of systematics and smaller matter effects relative to $CP$ violating effects, 
Hyper-K has better sensitivity for $CP$ violation.

The next generation reactor neutrino experiments, JUNO in China and RENO50 in Korea, are proposed.
The main purpose of these experiments are to determine the mass hierarchy. The $CP$ violation sensitivity in Hyper-K is greatly improved with knowledge of  the mass hierarchy. The atmospheric neutrino experiments, PINGU, ORCA and INO, also focus on the mass hierarchy, and their measurements would
represent a positive synergy for Hyper-K.

In Table~\ref{tab:intro:2020}, the summary of the proposed experiments in the 2020s is listed with Hyper-K.

\subsection{Overall Science goals of the Hyper-Kamiokande project} 

\label{sec:int:pays}
In addition to the long baseline neutrino oscillation experiment that is the main focus of this document,
Hyper-K will provide rich programs in a wide range of science~\cite{Abe:2011ts}.
The scope of the project includes observation of atmospheric and solar neutrinos, proton decays, and neutrinos from other astrophysical origins.
The physics potential of Hyper-K is summarized in Table~\ref{tab:intro:phys}.

\begin{table}[btp]
\caption{Physics targets and expected sensitivities of the Hyper-Kamiokande experiment.} 	
\label{tab:intro:phys}
\begin{center}
\begin{tabular}{lll} \hline \hline
Physics Target & Sensitivity & Conditions \\
\hline \hline
Neutrino study w/ J-PARC $\nu$~~ && 7.5\,MW $\times$ $10^7$ sec\\
$-$ $CP$ phase precision & $<19^\circ$ & @ $\sin^22\theta_{13}=0.1$, mass hierarchy known \\
$-$ $CPV$ discovery coverage & 76\% (3\,$\sigma$), 58\% ($5\,\sigma$) & @ $\sin^22\theta_{13}=0.1$, mass hierarchy known \\
$-$ $\sin^2\theta_{23}$ & $\pm 0.015$ & 1$\sigma$ @ $\sin^2\theta_{23}=0.5$ \\
\hline
Atmospheric neutrino study && 10 years observation\\
$-$ MH determination & $> 3\,\sigma$ CL & @ $\sin^2\theta_{23}>0.4$ \\
$-$ $\theta_{23}$ octant determination & $> 3\,\sigma$ CL & @ $\sin^2\theta_{23}<0.46$ or $\sin^2\theta_{23}>0.56$ \\\hline
Nucleon Decay Searches && 10 years data \\
$-$ $p\rightarrow e^+ + \pi^0$ & $1.3 \times 10^{35}$ yrs (90\% CL UL) &\\
 & $5.7 \times 10^{34}$ yrs ($3\,\sigma$ discovery) &\\
$-$ $p\rightarrow \bar{\nu} + K^+$ & $3.2 \times 10^{34}$ yrs (90\% CL UL) &\\
 & $1.2 \times 10^{34}$ yrs ($3\,\sigma$ discovery) &\\ 
\hline
Astrophysical neutrino sources && \\
$-$ $^8$B $\nu$ from Sun & 200 $\nu$'s / day & 7.0\,MeV threshold (total
	 energy) w/ osc.\\
$-$ Supernova burst $\nu$ & 170,000$\sim$260,000 $\nu$'s & @ Galactic center (10 kpc)\\ 
 & 30$\sim$50 $\nu$'s & @ M31 (Andromeda galaxy) \\ 
$-$ Supernova relic $\nu$ & 830 $\nu$'s / 10 years & \\
$-$ WIMP annihilation at Sun & & 5 years observation\\
 ~~($\sigma_{SD}$: WIMP-proton spin & $\sigma_{SD}=10^{-39}$cm$^2$ & @ $M_{\rm WIMP}=10$\,GeV, $\chi\chi\rightarrow b\bar b$ dominant\\
 ~~~~dependent cross section)& $\sigma_{SD}=10^{-40}$cm$^2$ & @ $M_{\rm WIMP}=100$\,GeV, $\chi\chi\rightarrow W^+ W^-$ dominant\\
 \hline \hline
\end{tabular}
\end{center}
\end{table}

\section{The Hyper-Kamiokande Detector} 
Hyper-Kamiokande
is to be the third generation water Cherenkov detector in Kamioka, 
designed for a wide variety of neutrino studies and nucleon decay searches.
Its total (fiducial) water mass of one (0.56) million tons would be approximately 20 (25) times larger than
that of Super-Kamiokande.
Table~\ref{tab:detector-parameters} 
summarizes the baseline design parameters of the Hyper-K detector.
The design of the detector is briefly summarized in this section.

%

\begin{table}[htdp]
\caption{Parameters of the Hyper-Kamiokande baseline design.}
\begin{center}
\begin{tabular}{lll} \hline \hline
Detector type & & Ring-imaging water Cherenkov detector \\ \hline 
Candidate site & Address & Tochibora mine \\
& & Kamioka town, Gifu, JAPAN \\
& Lat. & $36^\circ21'20.105''$N $^{\dagger}$ \\
& Long. & $137^\circ18'49.137''$E $^{\dagger}$\\
& Alt. & 508 m \\
& Overburden & 648 m rock (1,750 m water equivalent)  \\
& Cosmic Ray Muon flux & $\sim$ 8 $\times$ 10$^{-7}$ sec$^{-1}$cm$^{-2}$  \\
& Off-axis angle for the J-PARC $\nu$ & $2.5^\circ$ (same as Super-Kamiokande)  \\
& Distance from the J-PARC & 295 km (same as Super-Kamiokande)  \\ \hline 
Detector geometry & Total Water Mass & 0.99 Megaton  \\
 & Inner Detector (Fiducial) Mass & 0.74 (0.56) Megaton  \\
 & Outer Detector Mass & 0.2 Megaton  \\ \hline 
Photo-multiplier Tubes & Inner detector & 99,000 20-inch $\phi$ PMTs \\
& & 20\% photo-coverage \\ 
& Outer detector & 25,000 8-inch $\phi$ PMTs \\ \hline 
Water quality & light attenuation length & $>100$ m @ 400 nm  \\
 & Rn concentration & $<1$ mBq/m$^3$ \\ \hline 
\hline 
\multicolumn{3}{r}{$^{\dagger}$ World geographical coordination system}\\
\end{tabular}
\end{center}
\label{tab:detector-parameters} 
\end{table}

\subsection{Site, caverns, and tanks} \label{sec:site}



\begin{figure}[tb]
  \includegraphics[scale=0.25]{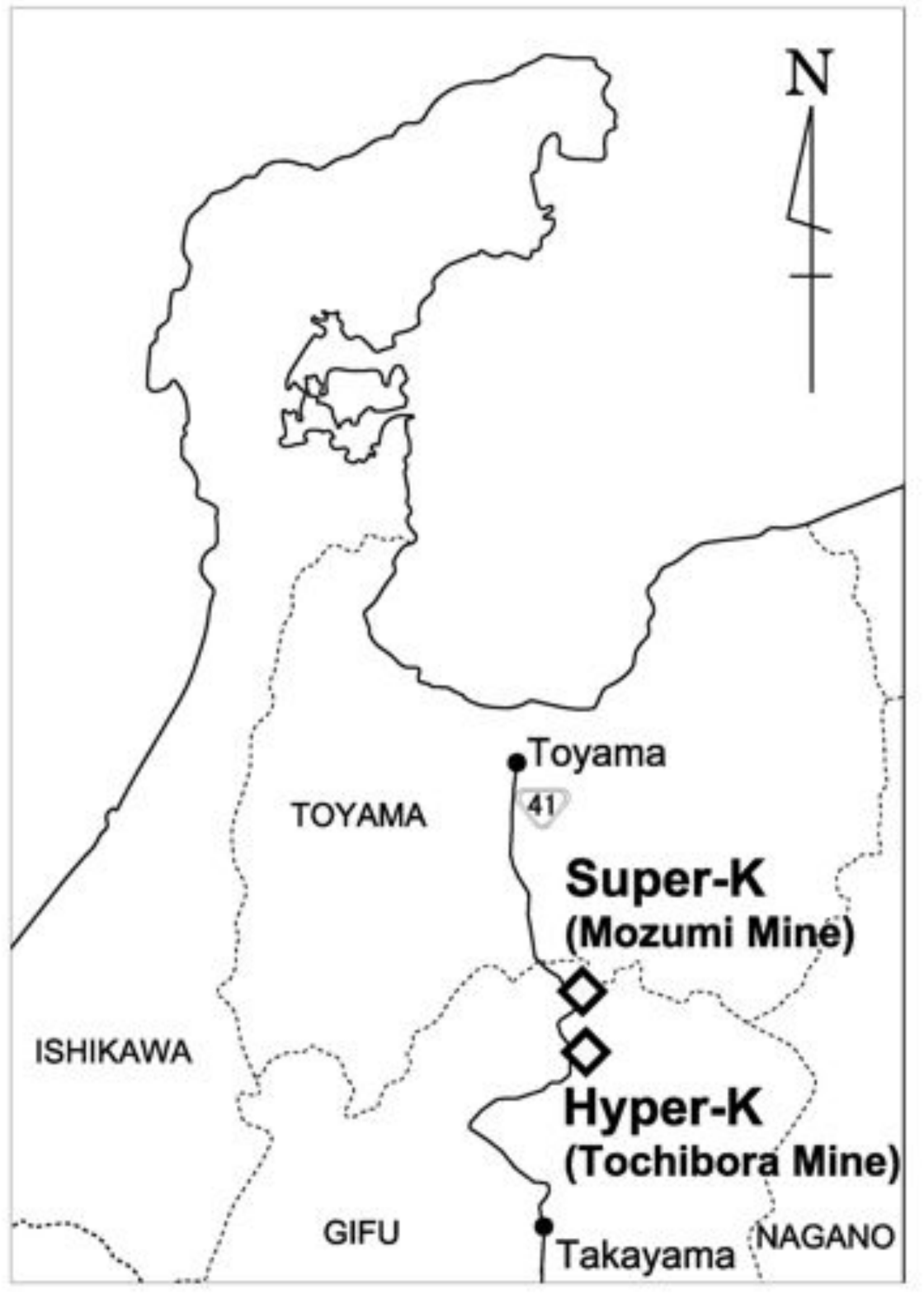}
  \includegraphics[scale=0.25]{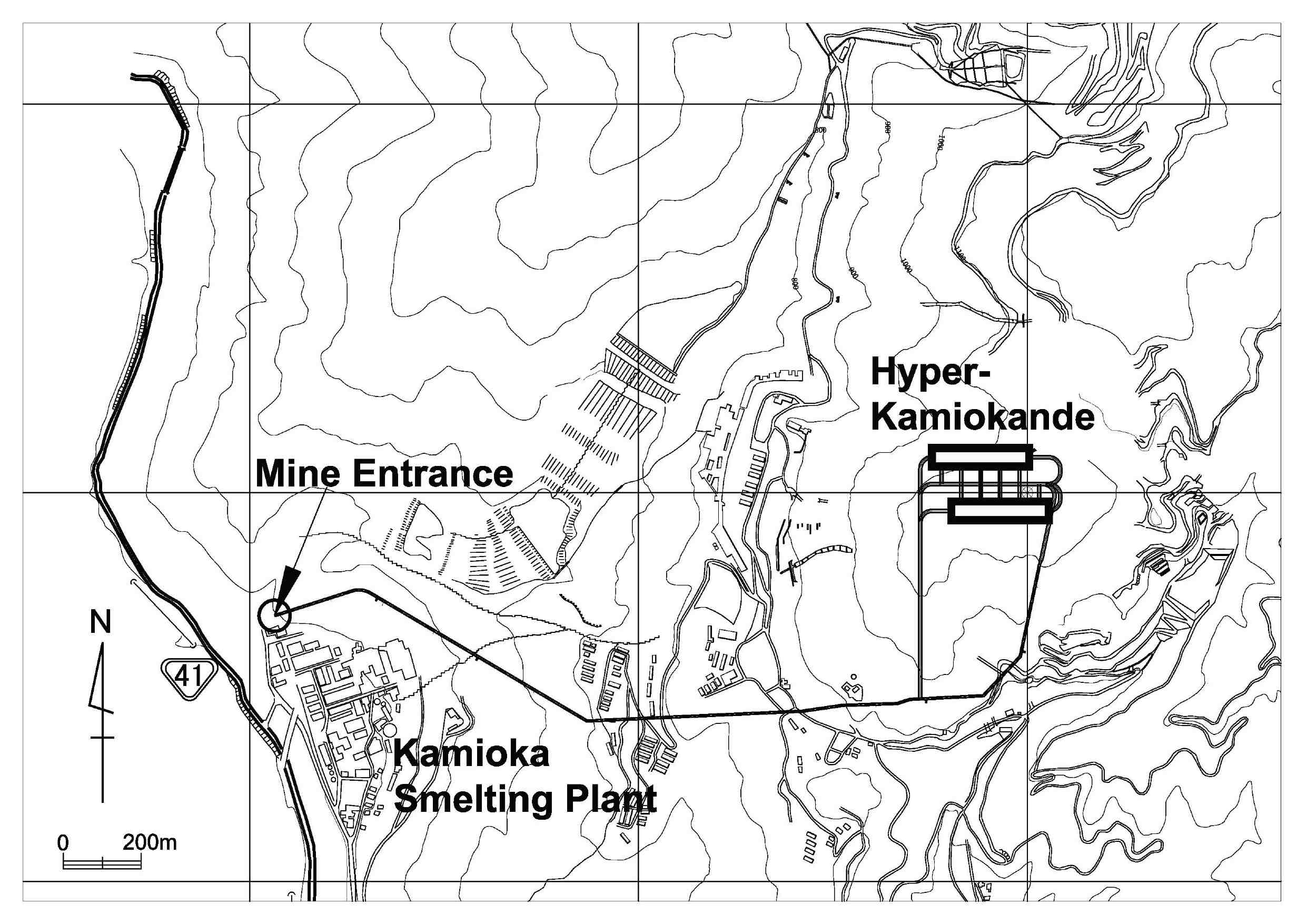}
  \caption{The candidate site map.  The site is located 8\,km south
    of the Super-K site as shown in the left panel.
    The map of the Tochibora mine is shown in the right panel.}
  \label{fig:map}
\end{figure}

The Hyper-K detector candidate site,
located 8\,km south of Super-K,
is in the Tochibora mine
of the Kamioka Mining and Smelting Company,
near Kamioka town in the Gifu Prefecture, Japan,
as shown in Fig.~\ref{fig:map}.
The J-PARC neutrino beamline is designed so that
the existing Super-Kamiokande detector and the Hyper-K candidate site in Tochibora mine
have the same off-axis angle.
The experiment site is accessible via a drivable,
$\sim$2.6\,km long, (nominally) horizontal mine tunnel.
The detector will lie under the peak of Mt. Nijuugo-yama,
with an overburden of 648\,meters of rock
or 1,750\,meters-water-equivalent (m.w.e.).
%
%

The rock wall in the existing tunnels
and sampled bore-hall cores are dominated by
Hornblende Biotite Gneiss and Migmatite
in the state of sound, intact rock mass.  This is desirable
for constructing such unprecedented large underground cavities.
The site has a neighboring mountain, Maru-yama,  just 2.3\,km away, whose
collapsed peak enables us to dispose of
more than one million m$^3$ of waste rock from the detector cavern excavation.
Based on the \textit{in situ} measurements of the rock quality and the rock stress,
it is confirmed that the Hyper-K caverns can be constructed with existing excavation techniques.

The Mozumi mine under Mt.\ Ikeno-yama, where the Super-K detector is located, is
another candidate site
which can provide more overburden ($\geq700$\,m) than the Tochibora site and reduced
background levels for low-energy physics, such as solar neutrinos and supernova relic neutrinos.
The geological surveys have been carried out at a vicinity of the candidate site, and
detailed stability analyses of the cavern construction and evaluation of the construction
period and cost are in progress.

\begin{figure}[tbp]
  \includegraphics[scale=1.0]{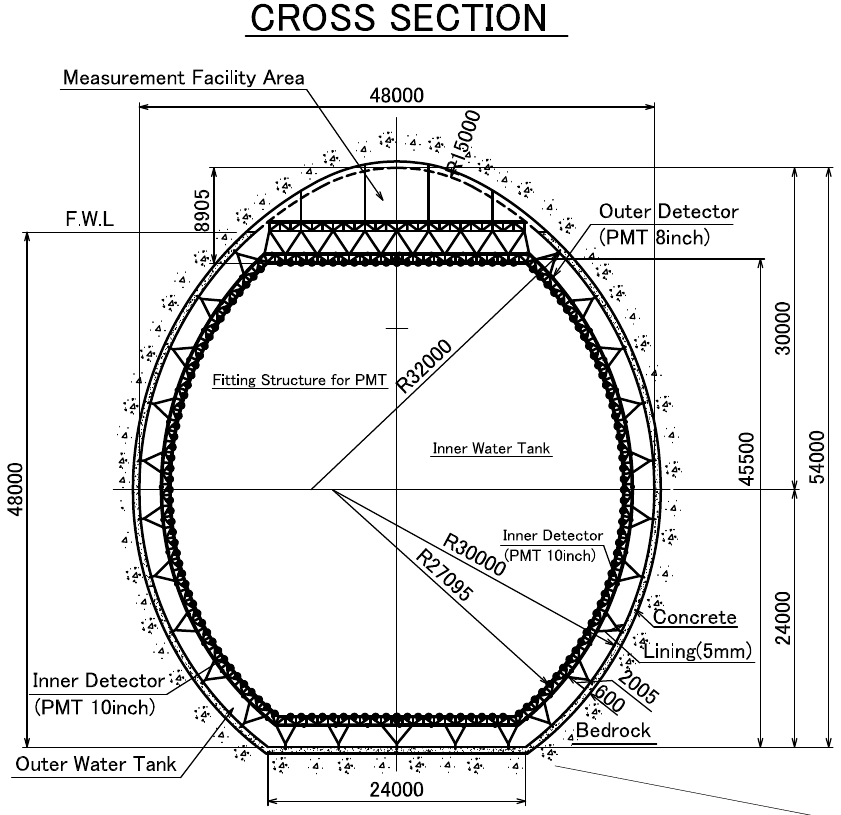}
  \caption{Cross section view of the Hyper-Kamiokande detector.\label{fig:crosssection}}
\end{figure}
%
%
\begin{figure}[htbp]
  \includegraphics[scale=0.65,angle=90]{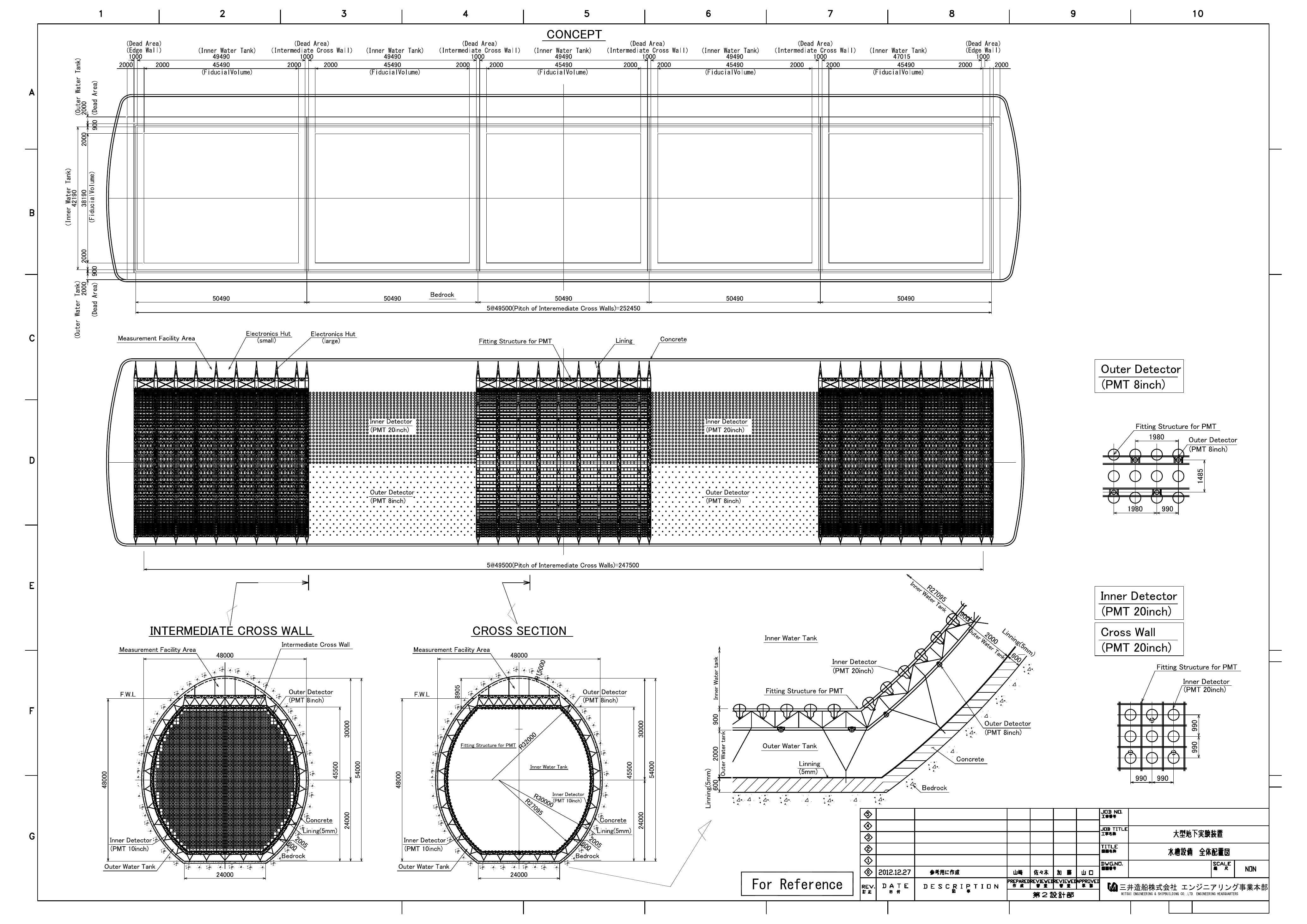}
  \caption{
    Profile of the Hyper-K detector.
    The left panel shows the detector segmentation.
    The right panel shows PMT arrays and the support structure for the inner
    and outer detectors.
    Each quasi-cylindrical tank
    lying horizontally is segmented by intermediate walls
    into five compartments.\label{fig:profile}}
\end{figure}
%
For the baseline design, the Hyper-K detector is composed of two separated caverns
as shown in Fig.~\ref{fig:hk_schematic}, each having an egg-shape cross section
48\,meters wide, 54\,meters tall, and 250\,meters long as shown
in Fig.~\ref{fig:crosssection} 
and \ref{fig:profile}.
The welded polyethylene tanks are filled up to a depth of 48\,m with
ultra-pure water: the total water mass equals 0.99 million tons.
%
%
A detailed design of the water containment system, e.g.\ concrete layers, polyethylene lining, 
and water leak detection/draining system, has been established.
Polyethylene lining sheet 
has been tested for pressure, tensile shear and tensile creep, and
has been confirmed to be sufficient for Hyper-K.

Each tank will be optically separated by segmentation walls
located every 49.5\,m to form 5 (in total 10) compartments as
shown in Fig.~\ref{fig:profile}, such that
event triggering and event reconstruction can be performed
in each compartment separately and independently.
Because the compartment dimension of 50 m is comparable with that of Super-K (36 m)
and is shorter than the typical light attenuation length in water achieved by the Super-K water
filtration system
($>100$\,m @ 400~nm),
we expect that the detector performance of Hyper-K for beam and atmospheric neutrinos will be effectively the same as
that of Super-K.
The water in each compartment is further optically
separated into three regions.
The inner region has a barrel shape of 42 m in height and width,
and 48.5 m in length,
and is viewed by an inward-facing array of 20-inch diameter photomultiplier tubes (PMTs).
The entire array consists of 99,000 Hamamatsu R3600  PMTs,
uniformly surrounding the region and giving a photocathode coverage of 20\%.
The PMT type, size, and number density are subject to optimization.
An outer region completely surrounds the 5 (in total 10) inner regions
and is equipped with 25,000 8-inch diameter PMTs.
This region is 2 m thick at the top, bottom,
and barrel sides, except at both ends of each cavern, where
the outer region is larger than 2 m due to rock engineering considerations.
A primary function of the outer detector is to reject
entering cosmic-ray muon backgrounds and to help in identifying
nucleon decays and neutrino interactions occurring in the inner detector.
The middle region or dead space is an uninstrumented, 0.9\,m thick
shell between the inner and outer detector volumes
where the stainless steel PMT support structure is located.
Borders of both inner and outer regions are lined with opaque sheets.
This dead space, along with the outer region, acts as a shield against
radioactivity from the surrounding rock.
The total water mass of the inner region is 0.74 million tons
and the total fiducial mass is 10 times 0.056 = 0.56 million tons.
The fiducial volume is defined as the region formed by
a virtual boundary located 2\,m away
from the inner PMT plane. 
%


The estimated cosmic-ray muon rate around the Hyper-K detector
candidate site is 
$\sim$ 8 $\times$ 10$^{-7}$ sec$^{-1}$cm$^{-2}$ which is
roughly 5 times larger than the flux at Super-K's location
($\sim$ 1.5 $\times$ 10$^{-7}$ sec$^{-1}$cm$^{-2}$).
The expected deadtime due to these muons is less than 1\% and
negligible for long baseline experiments, as well as nucleon decay searches and atmospheric neutrino studies.
\subsection{Water purification system} \label{sec:water}

Water is the target material and signal-sensitive medium of the detector, and thus its quality directly affects the physics sensitivity. In order to realize such a huge Cherenkov detector, achieving good water transparency is the highest priority. In addition, as radon emanating from the photosensors and detector structure materials is the main background source for low energy neutrino studies, an efficient radon removal system is indispensable.

In Super-Kamiokande the water purification system has been continually modified
and improved over the course of two decades from SK-I to SK-IV.  As a result, the transparency is now kept above 100~m and is very stable, and the radon concentration in the tank is held below 1~mBq/m$^3$.  Following this success, the Hyper-Kamiokande water system design will be based on the current Super-Kamiokande water system.

Naturally, ever-faster water circulation is generally more effective when trying to keep huge amounts of water clean and clear, but increasing costs limit this straightforward approach so a compromise between transparency and recirculation rate must be found.  In Super-Kamiokande, 50,000 tons of water is processed at the rate of 60~tons/hour in order to keep the water transparency (the attenuation length for 400--500~nm photons) above 100~m, and 20~m$^3$/hour of radon free air is generated for use as a purge gas in degas modules, and as gas blankets for both buffer tanks and the Super-Kamiokande tank itself. For the 0.99 million tons of water in Hyper-Kamiokande, these process speeds will need to be scaled-up to 1200~m$^3$/hour for water circulation and 400~m$^3$/hour for radon free air generation.

\begin{figure}
     \begin{center}
       \includegraphics[width=11cm]{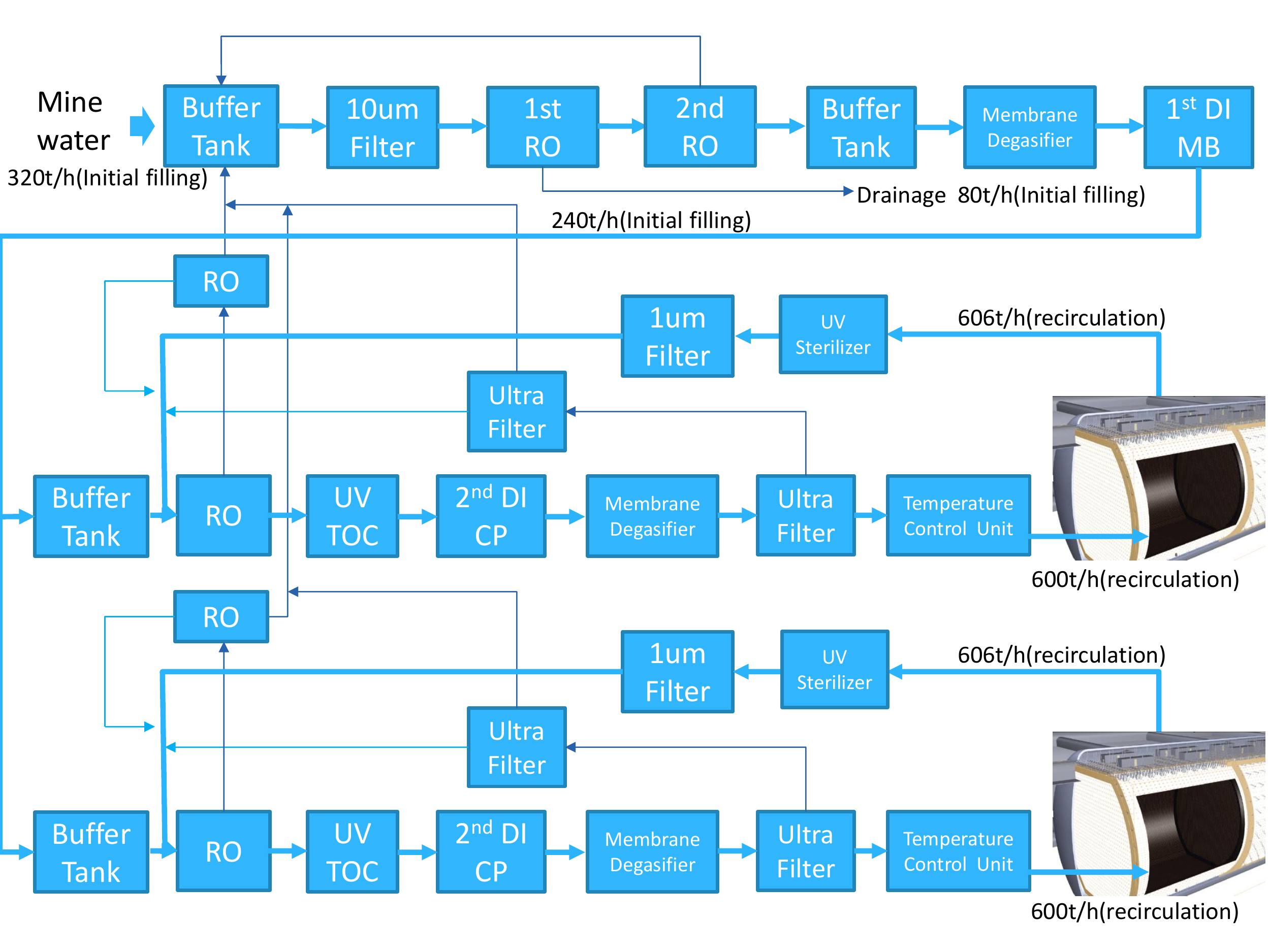}
       \caption{The water flow design of the Hyper-Kamiokande water system.}
       \label{water:water}
     \end{center}
\end{figure}

Figure \ref{water:water} shows the current design of the Hyper-Kamiokande water purification system. With these systems, the water quality in Hyper-Kamiokande is expected to be same as that in Super-Kamiokande.

Adding dissolved gadolinium sulfate for efficient tagging of neutrons has been studied as an option to enhance Hyper-K physics capability.
The feasibility of adding Gd to Super-K~\cite{Beacom:2003nk} is now under study with EGADS (Evaluating Gadolinium's Action on Detector Systems) project in Kamioka.
We have been careful to keep the possibility of gadolinium loading in mind when designing the overall Hyper-Kamiokande water system.

\subsection{Photosensors} \label{sec:photosensor}




In order to achieve Hyper-K's broad scientific goals, 
particles with a wide range of energies need to be reconstructed.
Depending on the energy of particle that emits Cherenkov photons,
the number of photons that hit each photosensor ranges from one to several hundred.
Thus, the photosensors are required to have a wide dynamic range
and good linearity.
%
The location of the interaction vertex is reconstructed using the
Cherenkov photon arrival timing information at each PMT.
Therefore, good timing resolution of the photosensors is essential,
and the jitter of the transit time is required to be 
less than 3~nsec (1\,$\sigma$) for a single photon.

For the baseline design, we have selected the 20-inch diameter PMT
(Hamamatsu R3600) that have been used successfully in Super-K
as the primary sensor candidate for the Hyper-K inner detector.
The R3600 is already known to satisfy the requirements above.
Moreover, they have been operating for more than 15 years in Super-K
and thus 
the long-term stability
is well understood.
The specifications of the 20-inch PMT is summarized in 
TABLE~\ref{pmt:pmtspec}. 
\begin{table}[tbp]
 \begin{center}
  \begin{tabular}{ll}
   \hline \hline
   Shape                  & Hemispherical \\
   Photocathode area      & 50\,cm diameter \\
   Quantum efficiency     & 22\,\% at $\lambda=390$\,nm \\
   Dynodes                & 11\,stage Venetian blind type \\
   Gain                   & 10$^7$ at $\sim2000$\,V \\
   Dark pulse rate        & 3\,kHz at $10^7$ gain \\
   Transit time           & 90\,nsec at 10$^7$ gain \\
   Transit time spread~~~ & 2.2\,nsec (1\,$\sigma$) for single photoelectron signals \\
   Weight                 & 13\,kg \\
   Pressure tolerance     & 6\,kg/cm$^2$\ water proof \\
   \hline \hline
  \end{tabular}
  \caption{Specifications of the 20-inch PMT (Hamamatsu R3600).}
  \label{pmt:pmtspec}
 \end{center}
\end{table}
%
%
The total number of inner detector PMTs in Hyper-K will be about 99,000.

For the 
outer detector, we have selected the same 
design as that of the Super-K outer detector.
The photosensors are Hamamatsu R1408 PMTs with an 8-inch diameter photocathode.
A total of 25,000 of these PMTs cover 1\% of the inner wall of the outer detector.
Like Super-K, an acrylic wavelength shifting plate of dimensions
60\,cm$\times$60\,cm is placed around the bulb of each of the 8-inch PMTs
to increase the photon detection efficiency.

\ \par
In order to further improve the performance and reduce the cost of the baseline design, 
we have been developing new photosensors as possible alternative options to the R3600.

Two types of new 20-inch sensors have been developed in cooperation with Hamamatsu Photonics, K.\ K.
One is a hybrid photo-detector (HPD), which uses an avalanche diode instead of a metal dynode for the multiplication of photoelectrons emitted from a photocathode. 
The other option is a PMT with a box-and-line dynode, which has a faster time response and a better collection efficiency compared to R3600. 
The specifications of three 20-inch photosensor candidates for Hyper-K are summarized in Table~\ref{tab:NewPhotosensors}.
The 8-inch HPDs and high QE, 20-inch box-and-line type PMTs are currently under test in a 200-ton water Cherenkov detector.
New 20-inch HPDs will be tested in the near future.

\begin{figure}[htbp]
 \includegraphics[width=0.5\textwidth]{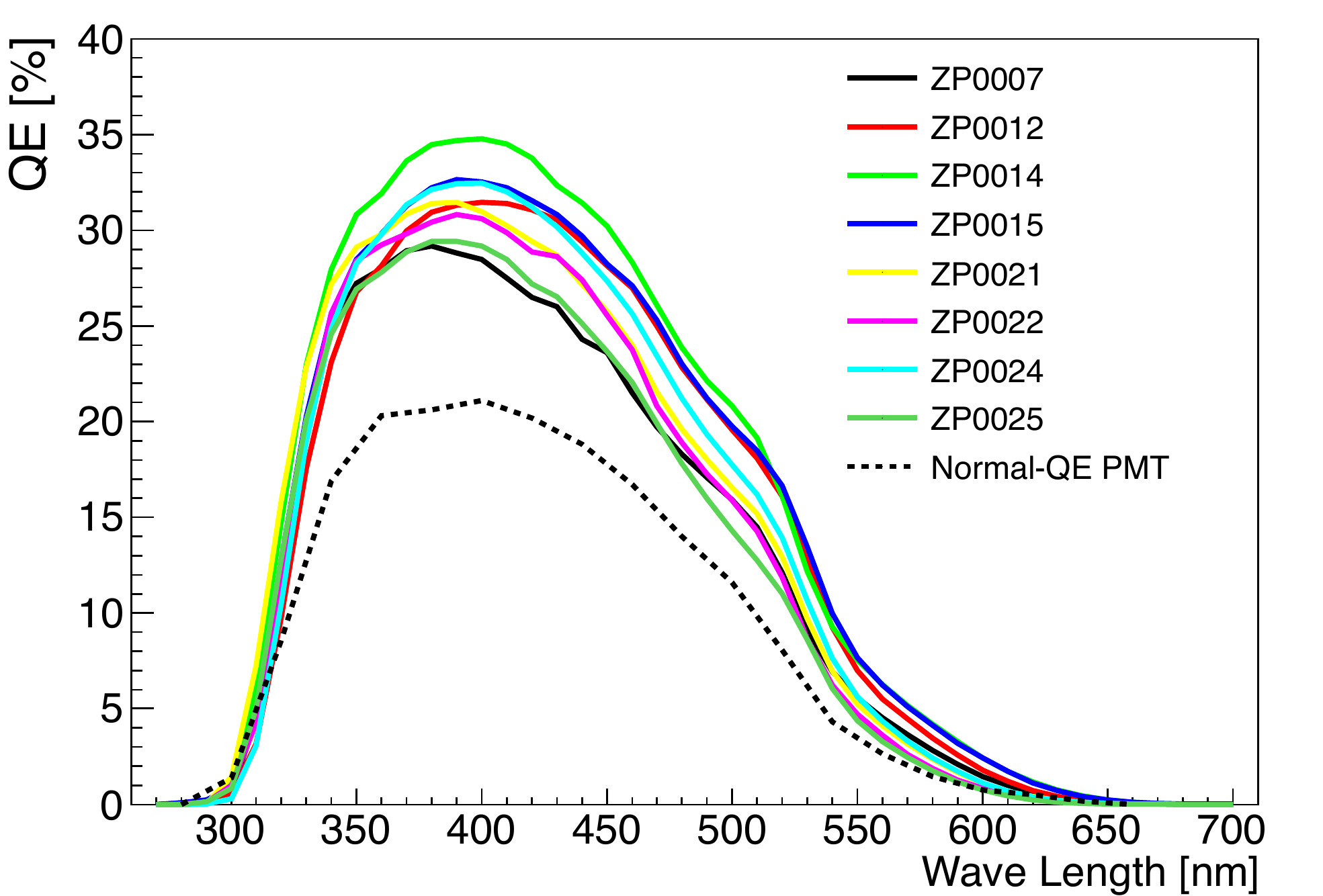}%
 \caption{Measured QE for eight high-QE R3600's (solid lines) and a normal R3600 (dashed line).\label{fig:HQE}}
 \end{figure}

As a common option for those large aperture photosensors, we have been developing a high quantum efficiency (QE) photocathode. 
The measured QE of eight high-QE R3600's and a typical QE of normal R3600 are shown in Fig~\ref{fig:HQE}.
%

 \begin{table}[tbp]
 \caption{Specification of three 20-inch diameter photosensors, which are candidates for Hyper-K. 
 \label{tab:NewPhotosensors}}
 \begin{ruledtabular}
 \begin{tabular}{l|c|c|c}
Type                       & PMT                         & PMT                       & Hybrid Photo-Detector \\
Amplification              & Venetian blind dynode       & Box-and-Line dynode       & Avalanche diode \\
&
   \begin{minipage}{0.25\hsize}
     \begin{center}
       \includegraphics[clip, width=4.0cm]{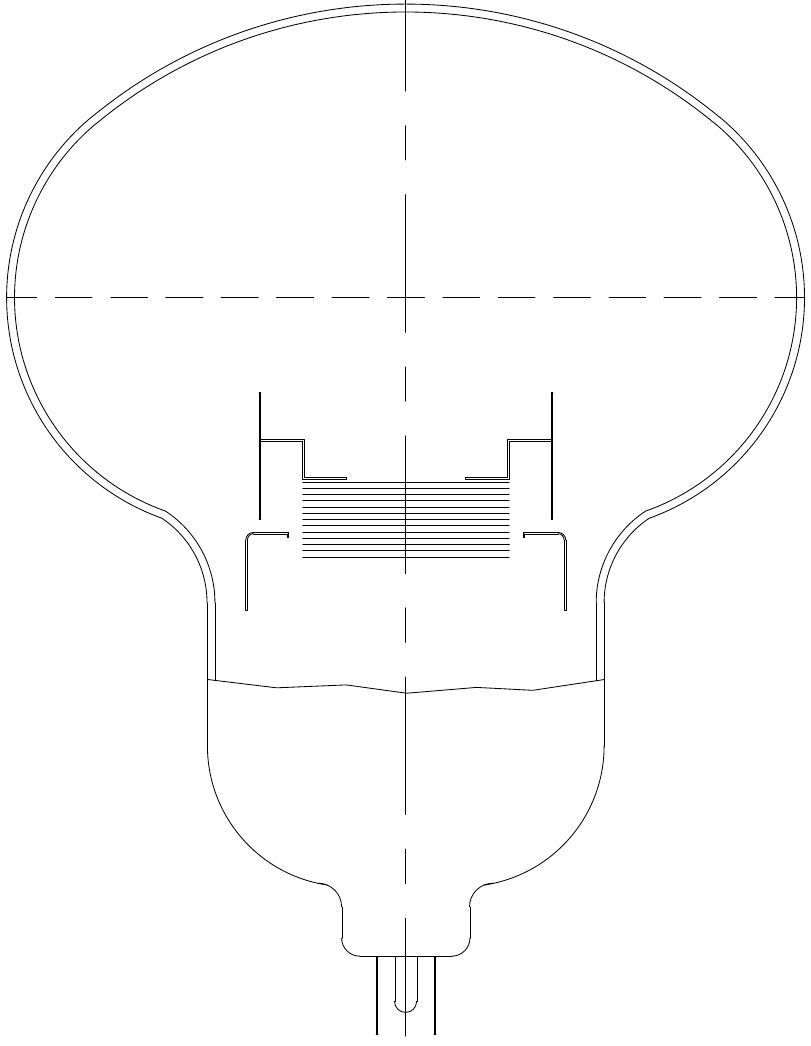}
     \end{center}
   \end{minipage}
&
   \begin{minipage}{0.25\hsize}
     \begin{center}
       \includegraphics[clip, width=4.0cm]{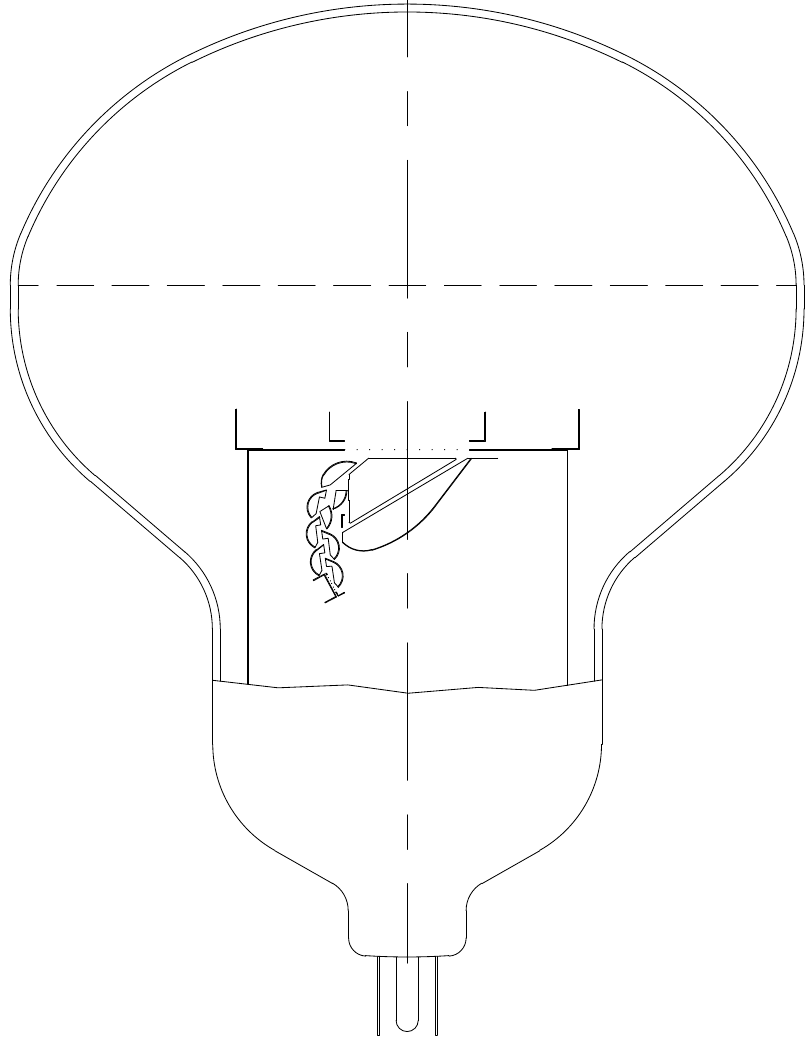}
     \end{center}
   \end{minipage}
&
   \begin{minipage}{0.25\hsize}
     \begin{center}
       \includegraphics[clip, width=4.0cm]{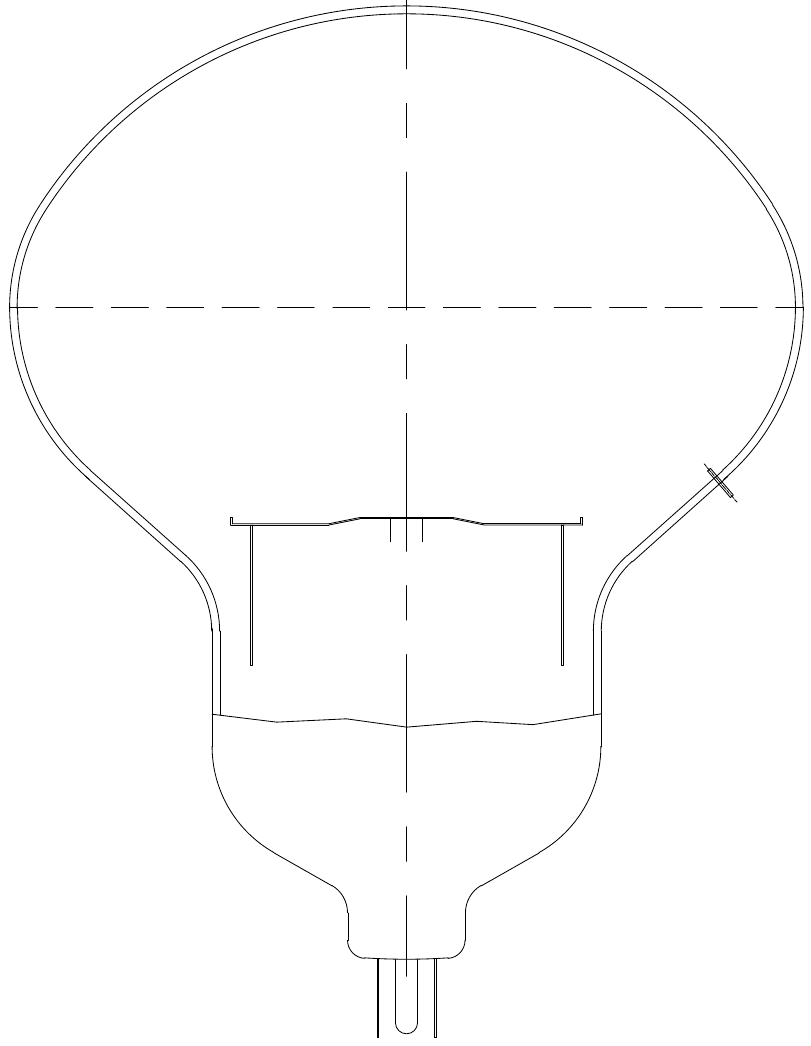}
     \end{center}
   \end{minipage}
\\
Model                      & R3600, HPK  & R12860, HPK & R12850, HPK \\
\hline
Collection efficiency      & 80\%        & 93\%        & 95\% \\
Transit time spread & \multirow{2}{*}{5.5\,nsec}   & \multirow{2}{*}{2.7\,nsec}   & \multirow{2}{*}{0.75\,nsec} \\
 ~~~(FWHM)  & & \\
Bias voltage               & 2\,kV       & 2\,kV       & 8\,kV \\
 \end{tabular}
 \end{ruledtabular}
 \end{table}


There are also other efforts to develop new photosensors that can potentially
be used for Hyper-K including the outer detector or new near detectors,
such as PMT by ETEL/ADIT~\cite{ADIT} and Large-Area Picosecond Photo-Detectors (LAPPD)~\cite{LAPPD}.

There have also been several attempts to improve the photon collection efficiency 
with special lens systems, wavelength shifters, or mirrors attached to the existing sensors. 
The effect of such additional system to the detector performance,
such as angular acceptance and timing resolution, needs to be carefully studied.

\subsection{Electronics and data acquisition system} \label{sec:daq}



In terms of the required specifications and the number of photosensors in one compartment, 
the baseline design of the Hyper-K detector is similar to that of the Super-K detector.
%
Therefore, it is possible for us to design the data acquisition 
system using the same concept as 
SK-IV,
reading out
all the hit information from the photosensors, including the dark noise
hits.

However, the egg-shape of Hyper-K detector makes
the cable routing and mechanical support difficult to design.
We are now planning to place the front-end electronics module and
the power supply for the photosensor in the detector water,
close to the photosensor.
The underwater front-end electronics will be enclosed
in a pressure tolerant water-tight housing. This approach has
been used in other experiments 
with
several established techniques.
%


\begin{figure}[tbp]
\begin{center}
  \includegraphics[width=0.6\textwidth]{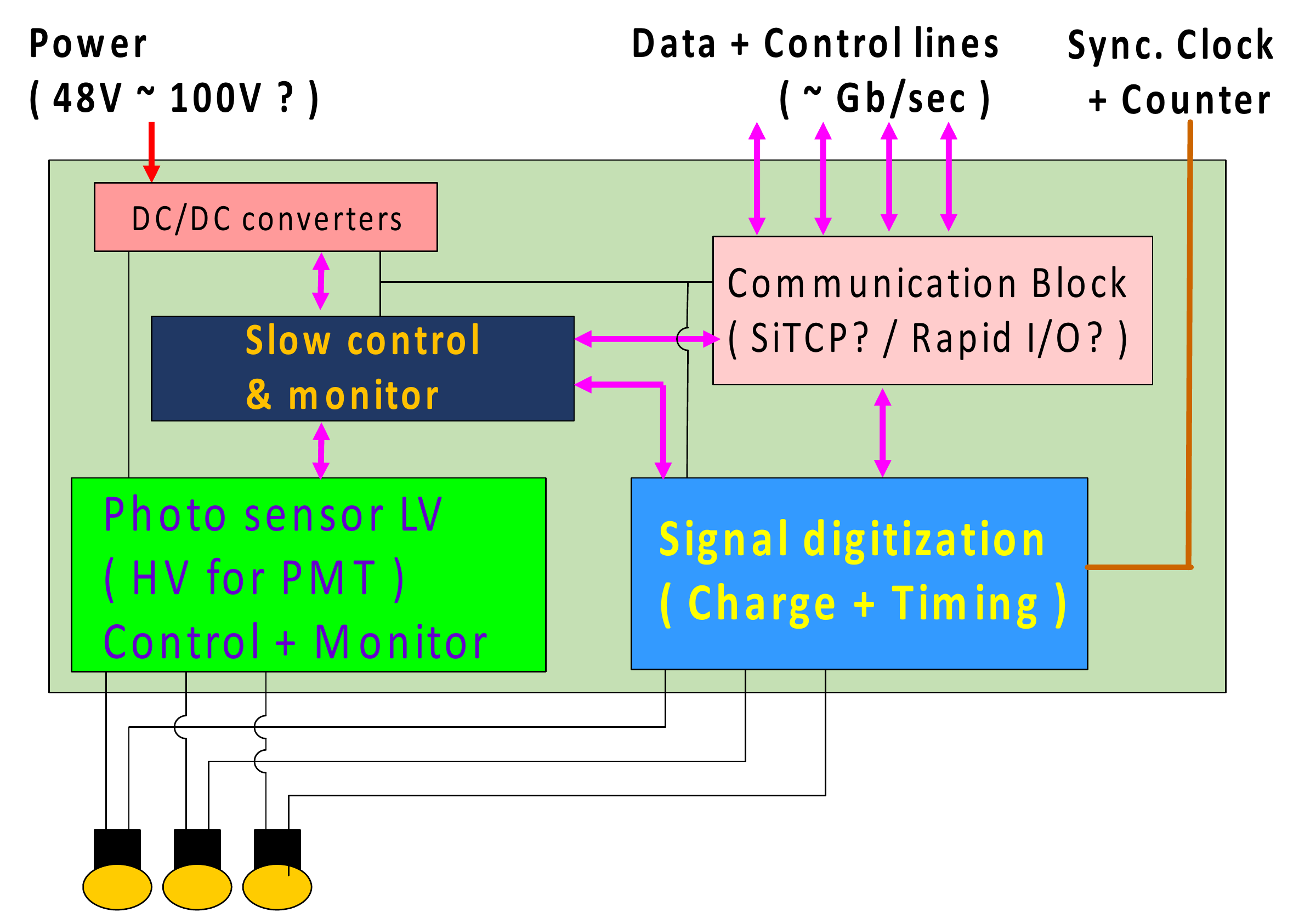}
  \caption{Schematic diagram of the front-end module.}
\label{schematic_frontend:daq}
\end{center}
\end{figure}
%
The schematic diagram of the front-end module is shown in 
Fig.~\ref{schematic_frontend:daq}.
There are four main functional blocks in the front-end board.
One module accepts signals from 24 photosensors.

%

The signal digitization block accepts the signals from the photosensors
and converts them into the digital timing and charge data. 
Because of similar requirements, the SK-IV front-end electronics using 
the charge to time conversion (QTC) chip~\cite{Nishino:2009zu} and 
ATLAS Muon TDC (AMT) chip~\cite{Arai:2002} is used as a reference for the baseline design.
The Hyper-K front-end is required to have equal or better performance than 
the specification of SK-IV electronics summarized in Table~\ref{SK-IV-elec:daq}.
\begin{table}[tb]
\caption{ Specification of the SK-IV front-end electronics }
\begin{tabular}{ll}
\hline \hline
Items & Required values \\
\hline
Built-in discriminator threshold~~ & 1/4 p.e ($\sim$ 0.3 mV)\\
Processing speed & $\sim$ 1$\mu$sec. / hit\\ 
Charge resolution & $\sim$ 0.05 p.e. (RMS) for $<$ 5 p.e. \\ 
Charge dynamic range & 0.2 $\sim$ 2500 pC (0.1 $\sim$ 1250 p.e.) \\ 
Timing resolution & 0.3 ns RMS (1 p.e.) \\ 
                  & $<0.3$ ns RMS ($>5$ p.e.) \\ 
Least significant bit resolution & 0.52 ns \\ 
\hline \hline
\end{tabular}\\
\label{SK-IV-elec:daq}
\end{table}
%
%
%
Because the relative timing is used to reconstruct the event 
vertex in the detector, all the modules have to be synchronized 
to the external reference clock. 


The photosensor power supply block 
controls the photosensor voltage supply.
For the HPD, we suppose that a voltage supply is embedded inside its housing.
If  standard PMTs are used as the photosensor, 
the high voltage modules will be put in the same enclosure as the front-end electronics.

%

The slow control block 
controls and monitors the status of the power supply 
for the photosensors. Also, the voltage, the current and the 
temperature in the front-end module have to be monitored. 

The communication block 
transports data from/to the other modules.
In order to reduce the amount of cables, 
the modules will be connected to each other to make a network of data transfer
lines. 
To avoid a single point failure, 
a module will be equipped with several communication ports and connected to multiple modules.
%



%
\begin{figure}[tb]
\begin{center}
  \includegraphics[width=0.8\textwidth]{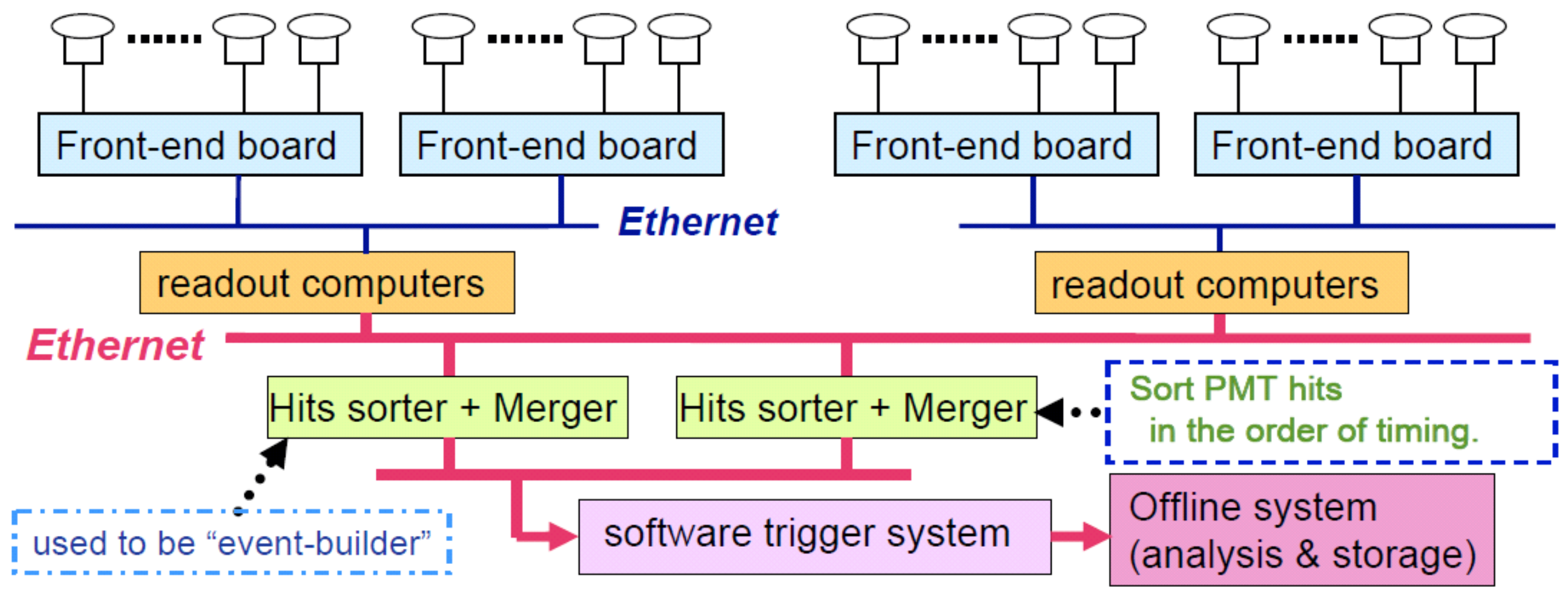}
  \caption{Schematic diagram of data readout and processing system.}
\label{online_schematics:daq}.
\end{center}
\end{figure}

The schematic diagram of the data readout and processing system
is shown in Fig.~\ref{online_schematics:daq}.
In the current baseline design, all the PMT signals above 
a certain threshold (e.g.\ $\sim$ 1/4 photoelectrons) 
are digitized and read out by the computer. 
The expected data rate from one front-end module is $\sim$ 2MB/sec and the total
data rate per compartment will be $\sim$ 1~GBytes/sec. 
Based on the experience with SK-IV,  about 20 computers will be necessary to read out the data from one compartment.
In order to select ``events'' to be transferred to the offline computers 
with the software trigger, $\sim$ 10 computers will be necessary.

A GPS-based  system will be used for the timing synchronization between a beam spill in J-PARC and a neutrino event at Hyper-K.
The GPS timestamp information for each beam spill is passed to the Hyper-K online system,
and used to define a software trigger to record all the hit information around the beam arrival time.
This technique has been well established in T2K.

\ \par
Although the Super-K experience shows that the baseline design will work, 
there are several ongoing R\&D activities to improve the performance of the electronics/DAQ for Hyper-K.
The current effort includes the development of a front-end electronics based on FADC, 
R\&D of an FPGA-based high precision TDC, 
and a more intelligent trigger for low energy events and/or events extending over multiple compartments.
It is planned to test multiple options with a prototype detector to evaluate their feasibility and performance.

\subsection{Detector calibration} \label{sec:calibration}
\begin{table}[td]
\caption{Calibration techniques used in Super-Kamiokande}
\begin{center}
\begin{tabular}{ll}
\hline \hline
Calibration source & Purpose \\ \hline
Nitrogen-dye laser & Timing response, charge linearity, OD \\
Laser with various wavelength & Water attenuation \& scattering \\
Xe lamp + scintillator ball & PMT gain, position dependence \\
Deuterium-tritium (DT) fusion generator [$^{16}$N] & Low energy response\\
Nickel + $^{252}$Cf \, [Ni($n$, $\gamma$)Ni] & Absolute gain, photo-detection efficiency \\
Cosmic ray muon / $\pi^{0}$ / decay electron & Energy calibration for high energy events\\
\hline \hline
\end{tabular}
\end{center}
\label{tab:SK-calib}
\end{table}%

In order to achieve the scientific goals of Hyper-K, 
precise calibration of the detector is indispensable.
Because the Super-K detector has been operating successfully for more than a decade
with many outstanding scientific achievements,
the Hyper-K detector calibration system will be designed based on the 
techniques established with the Super-K
calibration~\cite{Abe:2013gga}.

In Super-Kamiokande, various kind of calibrations have been carried out,
as summarized in Table~\ref{tab:SK-calib}.
Since 
Hyper-K has 
ten individual compartments, 
it is not realistic to perform the same calibration work with the same system as Super-K.
In addition, 
its egg-shaped cross section 
will make calibrations near the wall of PMTs difficult. 
Therefore, 
the detector should have dedicated, automated systems
for accurately placing various calibration sources at desired positions inside the tank.
Design of such deployment system is ongoing,
utilizing the experience in other experiment, such as Borexino, SNO, and KamLAND, in addition to Super-K.

In parallel, R\&D of advanced calibration sources, such as a light source using an LED and new neutron generators, is ongoing.
Also, a facility to characterize the response of a photosensor to light with various wavelength, incident angle, and location on the sensor, is being developed to provide more detailed information that can be used to improve the detector simulation.


\subsection{Expected detector performance} \label{sec:det_performance}

\begin{figure}[tb]
\begin{center}
  \hspace*{-0.5cm}\includegraphics[width=10cm,clip]{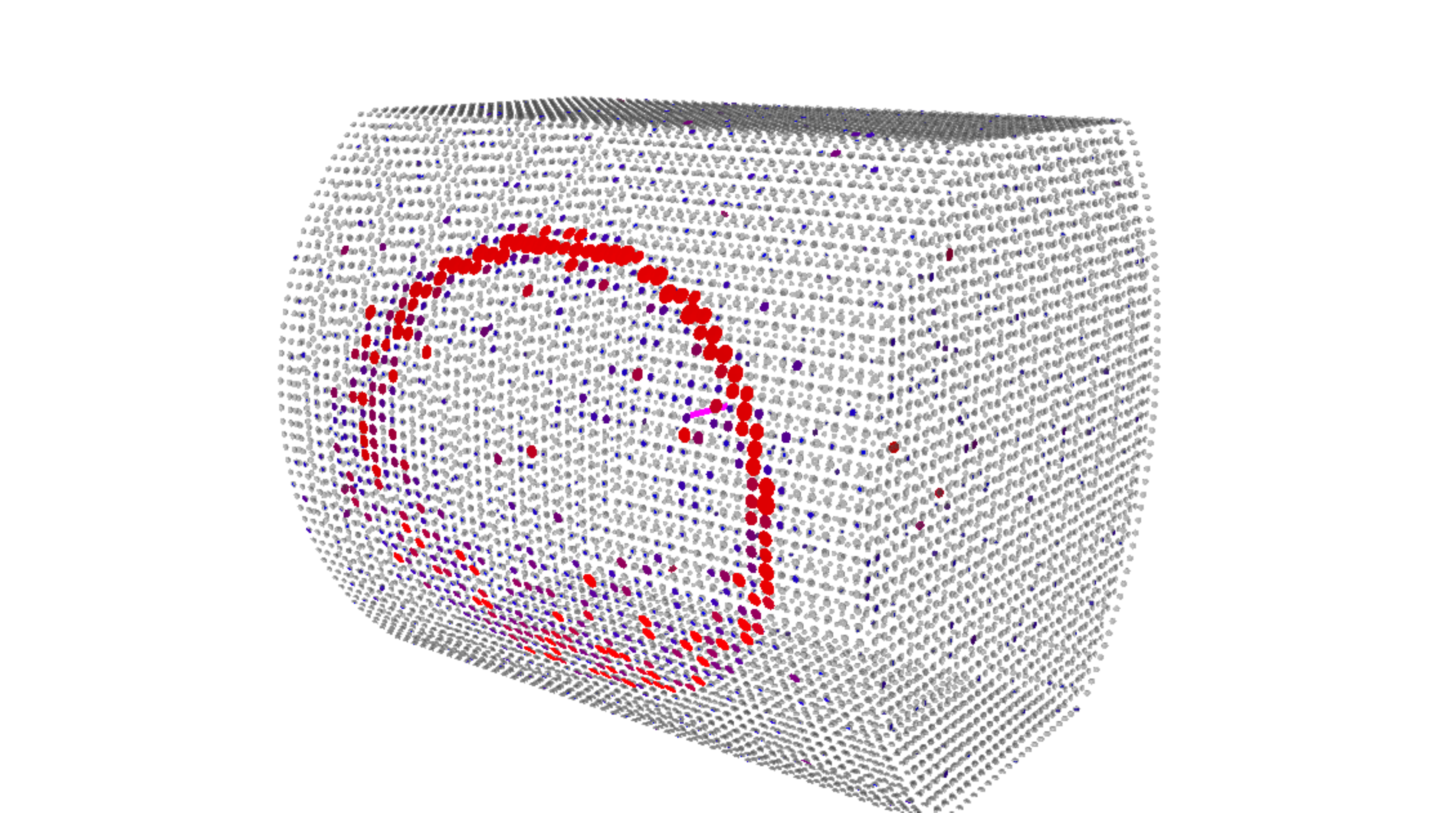}
  \caption
  {An event display of simulated event by WCSim. The ``egg-shape''
   cross section is implemented into WCSim. A muon is generated at the center of the detector 
   and is directed to the wall direction with 500\,MeV/c momentum. 
    }
  \label{evdisplay}
\end{center}
\end{figure}

We have been developing a detector simulation dedicated to Hyper-K based on
``WCSim,''~\cite{WCsim} which is an open-source water
Cherenkov detector simulator based on the GEANT4 library~\cite{Agostinelli:2002hh,Allison:2006ve}.
First, the simulation model of WCSim was validated by implementing the Super-K detector geometry in WCSim and
comparing the detector responses with those by the official Super-K MC simulation based on GEANT3~\cite{Brun:1994zzo} 
and tuned with the Super-K data.
%
Then, the detailed Hyper-K detector geometry 
was implemented in WCSim. 
An example event display is shown in Fig.~\ref{evdisplay}.

A new reconstruction algorithm developed for Super-K/T2K~\cite{Abe:2013hdq}, named ``fiTQun,'' has been adopted for the Hyper-K analysis.
It uses a maximum likelihood fit with charge and time probability density functions constructed for every PMT hit
assuming several sets of physics variables (such as vertex, direction, momentum, and particle type)~\cite{Patterson:2009ki, Abe:2013hdq}.
In the conventional 
event reconstruction in Super-K, physics variables are determined step-by-step,
while fiTQun can determine all physics variables at a time. 
In addition, fiTQun uses information from not only fired PMTs but also from
PMTs which have not fired, 
utilizing more information than 
the conventional method which uses only fired PMTs.

\begin{figure}[tb]
\begin{center}
  \hspace*{-0.5cm}\includegraphics[width=8cm,clip]{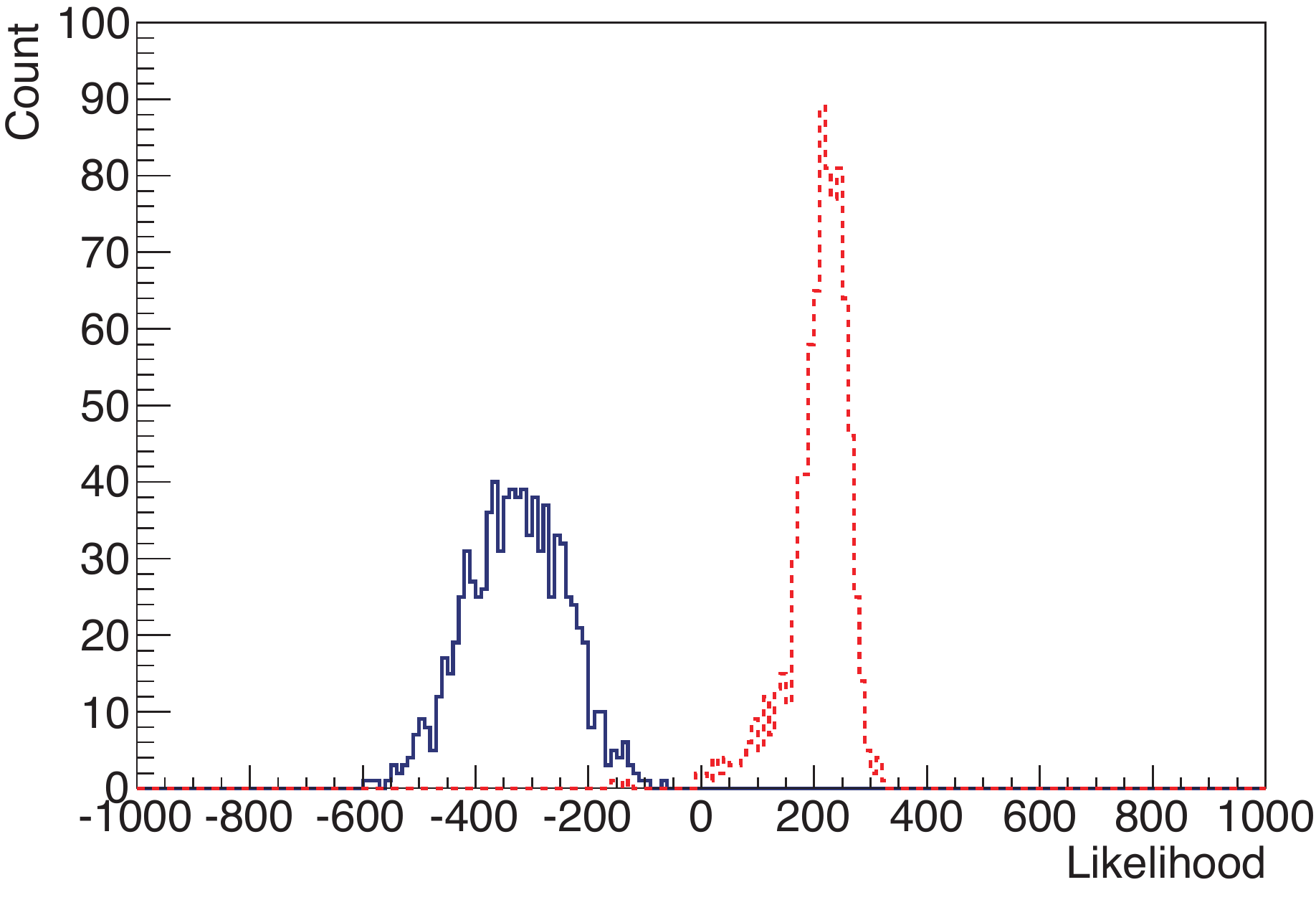}
  \caption
  { PID likelihood functions for electron (blue solid histogram) and $\mu$ (red dashed histogram) with 500\,MeV/$c$ momentum. A negative (positive) value indicates electron-like ($\mu$-like) particle.
    }
  \label{pid}
\end{center}
\end{figure}

\begin{table}
  \begin{center}
   \caption{
Comparison of performance of SK-II (20\% photo-coverage), SK-IV (40\% photo-coverage), 
and the expected performance of Hyper-Kamiokande baseline design (20\% photo-coverage) with preliminary Hyper-K simulation and reconstruction.
}
  \begin{tabular}{l|cc|cc|cc}
      \hline \hline
      & \multicolumn{2}{c|}{SK-II}  & \multicolumn{2}{c|}{SK-IV}  & \multicolumn{2}{c}{Hyper-K}\\
      \hline
      Particle type ($p=$500\,MeV/$c$)  & $e$     & $\mu$   & $e$     & $\mu$   & $e$       & $\mu$\\
      \hline 
      Vertex resolution                &  28~cm  &  23~cm  &  25~cm  &  17~cm  & 27~cm     &  30~cm \\
      Particle identification          &  98.5\% &  99.0\% &  98.8\% &  99.5\% & $>$99.9\% &  99.2\%  \\
      Momentum resolution              &  5.6\%  &  3.6\%  &  4.4\%  &  2.3\%  & 4.0\%     &  2.6\%   \\
      \hline \hline
    \end{tabular}
  \label{tab:performance}
  \end{center}
\end{table}

In the baseline design of Hyper-K, 20-inch PMTs cover 20\% of the inner detector surface.
This is the same setup as Super-K from 2002 to 2005 (the SK-II period). 
Based on the experience with SK-II,
the effect of the photo-coverage difference between 20\% and 40\% is known to be small
for neutrino events with an energy relevant to the long baseline oscillation experiment.
Therefore, the Hyper-K detector is expected to have similar performance as Super-K.

We have evaluated the expected performance of the Hyper-K detector using 
the MC simulation and reconstruction tools under development.
Electrons and muons with 500\,MeV/$c$ are generated with a fixed vertex (at the center of
the tank) and direction (toward the barrel of the tank) 
in the Hyper-K detector simulation,  
and the fiTQun reconstruction is applied. 
Figure~\ref{pid} shows the likelihood function for the particle identification.
A negative (positive) value indicates electron-like ($\mu$-like) particle.
It demonstrates a clear separation of electrons and muons.
The obtained performance of Hyper-Kamiokande is compared with the performance of 
SK-II (20\% photo coverage, old electronics) and SK-IV (40\% photo coverage, new electronics) in Table~\ref{tab:performance}.
The vertex resolution for muon events will be improved to the same level as
Super-K with an update of the reconstruction program.
From the preliminary studies, 
the performance of Hyper-K is similar to or possibly better than
SK-II or SK-IV with the new algorithm.
In the physics sensitivity study described in Section~\ref{sec:physics_sensitivities},
a Super-K full MC simulation with the SK-IV configuration is used because 
it includes the simulation of new electronics and is tuned with the real data,
while giving similar performance with Hyper-K as demonstrated above.


%
%
%

%
%
\section{Neutrino Beam at J-PARC} 
\label{sec:jparcbeam}
This section describes the J-PARC accelerators/neutrino beamline and 
planned operational parameters for the design beam power of 750\,kW. 
The work necessary to ramp up to this beam power from the current 
level of 240\,kW is well underway, and is to be accomplished 
considerably earlier than the start of Hyper-K data taking.
The prospects for realizing future multi-MW beam powers with the 
existing facility are then outlined.
A state-of-the-art prediction of the neutrino flux that will be generated 
by the facility has been examined in detail by the current T2K experiment. 
This is described together with the expected uncertainties.
%
\subsection{J-PARC accelerator cascade and the neutrino experimental 
            facility} \label{sec:jparc}
%
The J-PARC accelerator cascade~\cite{JPARCTDR} 
consists of a normal-conducting LINAC as an injection 
system, a Rapid Cycling Synchrotron (RCS), and a Main Ring synchrotron (MR). 
H$^-$ ion beams, with a peak current of 50 mA and pulse width of 500 $\mu$s, 
are accelerated to 400\,MeV by the LINAC. 
Conversion into a proton beam is achieved by charge-stripping 
foils at injection into the RCS ring, which accumulates and accelerates
two proton beam bunches up to 3\,GeV 
at a repetition rate of 25 Hz. Most of the bunches are extracted to the 
Materials and Life science Facility (MLF) to generate intense 
neutron/muon beams. 
The beam power of RCS extraction is rated at 1\,MW. 
With a prescribed repetition cycle, four successive beam pulses 
are injected from the RCS 
into the MR at 40 ms (= 1/25 Hz) intervals to 
form eight bunches in a cycle, and accelerated up to 30\,GeV. 
In fast extraction (FX) mode operation, 
the circulating proton beam bunches are extracted within a single turn 
into the neutrino primary beamline by a kicker/septum magnet system. 

\begin {figure}[htbp]
  \begin{center}
    \includegraphics[width=0.8\textwidth]
     {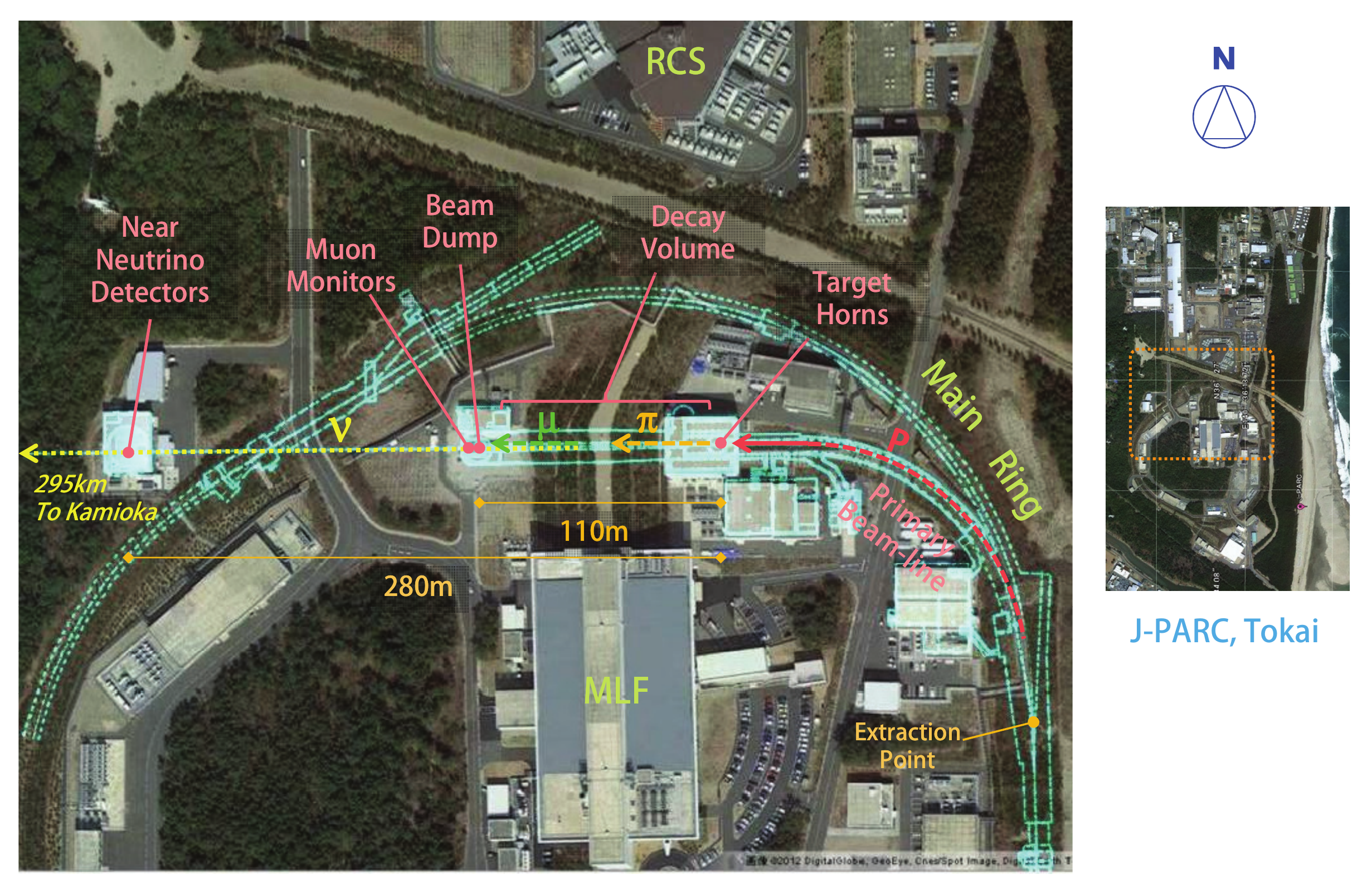}
    \caption{The neutrino experimental facility 
             (neutrino beamline) at J-PARC.}
    \label{fig:beamline}
  \end{center}
\end {figure}
Fig.~\ref{fig:beamline} shows an overview of the neutrino experimental 
facility~\cite{Abe:2011ks}. The primary beamline guides the extracted 
proton beam to a production target/pion-focusing horn system
in a target station (TS). The pions decay into muons and neutrinos during 
their flight in a 110 m-long decay volume. A graphite beam dump 
is installed at the end of the decay volume, and muon monitors 
downstream of the beam dump monitor the muon profile. 
A neutrino near detector complex is situated 280 m 
downstream of the target to monitor neutrinos at production.
To generate a narrow band neutrino beam, 
the beamline utilizes an off-axis beam configuration~\cite{OffAxisBeam} 
for the first time ever, with the capability to vary the off-axis angle 
in the range from 2.0$^\circ$ to 2.5$^\circ$. 
The latter value has been used for the T2K experiment and 
is assumed also for the proposed project. 
The centreline of the beamline extends 295 km to the west, 
passing midway between Tochibora and Mozumi, so that both 
sites have identical off-axis angles.  

\begin {figure}[htbp]
  \begin{center}
    \includegraphics[width=0.8\textwidth]
     {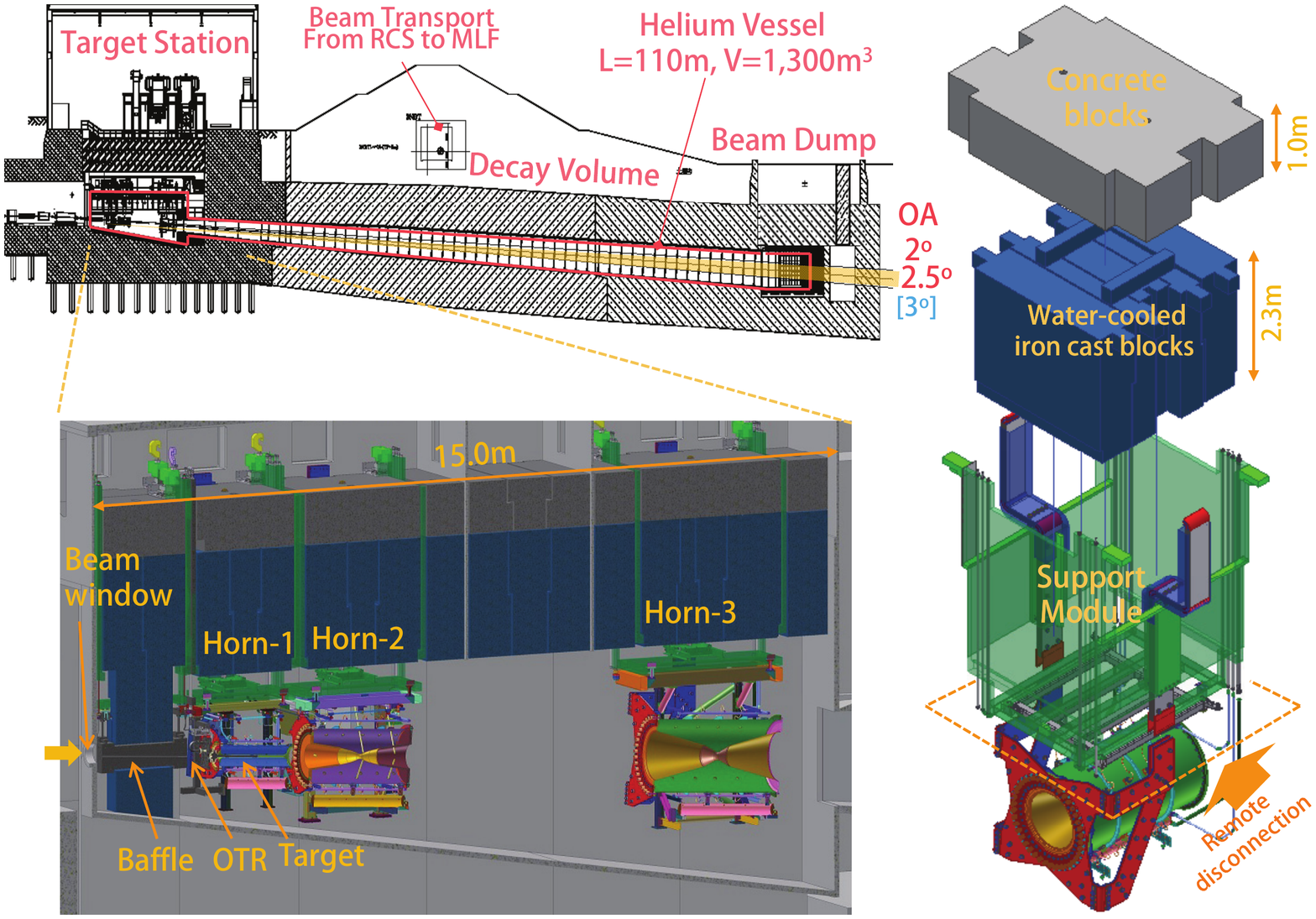}
    \caption{(Left) Side view of the secondary beamline, 
             with a close up of the target station helium vessel.
             (Right) A schematic view of a support module and shield blocks 
             for horn-3. If a horn fails, the horn together with its 
             support module is transferred remotely to a purpose-built 
             maintenance area, disconnected from the support module 
             and replaced. 
             }
    \label{fig:secondaryBL}
  \end{center}
\end {figure}
Fig.~\ref{fig:secondaryBL} shows a cross section of the secondary beamline,
and a close-up of the TS helium vessel. 
A helium-cooled graphite production target is inserted within the bore 
of the first of a three-horn pion-focusing system. 
At 750\,kW operation, $\sim$20\,kW heat load is generated 
in the target.\footnote{
  The beam size on the target (r= $\sim$4 mm) should be strictly controlled 
  with beam monitors at a final focusing section of the primary 
  beam-line and an optical transition radiation monitor (OTR)\cite{OTR} 
  in front of the target.
}
The horns are suspended from the lid of the TS helium vessel. 
Each horn comprises two co-axial cylindrical conductors which carry 
a 320 kA pulsed current. This generates a peak toroidal magnetic field 
of 2.1 Tesla which focuses one sign of pions.
The heat load generated in the inner conductors by secondary particles 
and by joule heating is removed by water spray cooling.\footnote{ 
 Gaseous hydrogen and oxygen are generated from the cooling water 
 by radiolysis, which could limit the beam power. 
 A recombination catalyst
 is installed to prevent the risk of explosion.
}
A helium cooled, double skin titanium alloy beam window separates 
the helium environment in the TS vessel ($\sim$1 atm pressure) 
from the vacuum of the primary beamline.
All secondary beamline components 
become highly radioactive during operation and replacements require handling 
by a remotely controlled overhead crane in the target station.
Failed targets can be replaced within horn-1 using a bespoke target 
installation and exchange mechanism.
Both the decay volume and the beam dump dissipate 
$\sim$1/3 of the total beam power, respectively. 
The steel walls of the decay volume and the
graphite blocks of the hadron absorber (core of the beam dump) 
are water cooled and both 
are designed to deal with 
3$\sim$4\,MW beam since neither can be upgraded nor maintained 
after irradiation by the beam.\footnote{
 The water cooling systems in the utility buildings 
 only have capacity for 750\,kW operation, thus need 
 to be upgraded for multi-MW beam operation.
} 

%
\subsection{Power upgrade of Main Ring synchrotron and the neutrino beamline} 
\label{sec:powerup-nu}
%
\begin{table}[tbp]
  \begin{center}
    \caption{Planned parameters of the J-PARC Main Ring for fast extraction. 
             Numbers in parentheses are those achieved up until May 2013.}
    \begin{tabular}{lcc}
      \hline \hline
      parameter &  \multicolumn{2}{c}{value} \\
      \hline
      circumference          & \multicolumn{2}{c}{1567.5\,m } \\
      beam kinetic energy    & \multicolumn{2}{c}{30\,GeV   } \\
      beam intensity         & $2.0\times 10^{14}$\,ppp~\tablenote{
           Most recent studies on space-charge tracking simulation
           show~\cite{jparc-midterm2} that 
           2.3$\times$10$^{14}$ ppp (2.9$\times$10$^{13}$ ppb, 
           equivalent to RCS 700\,kW operation) is achievable by 
           introducing 2nd harmonic RF during injection. } 
      & ($1.24\times 10^{14}$) \\
      ~                      & $2.5\times 10^{13}$\,ppb
      & ($1.57\times 10^{13}$) \\
      $[$ RCS equivalent power $]$   &   $[$ 610\,kW $]$   &  (377)  \\
      RF frequency      & \multicolumn{2}{c}{1.67$-$1.72\,MHz } \\
      harmonic number        & \multicolumn{2}{c}{9} \\
      number of bunches      & \multicolumn{2}{c}{8~/~spill} \\
      spill width            & \multicolumn{2}{c}{$\sim$~5\,$\mu$s} \\
      bunch full width       & 150$-$$\sim$400\,ns & ($\sim$160)  \\
      maximum RF voltage             & 560\,kV   & (280)\   \\
      repetition period      & 1.3\,sec & (2.48) \\
      ~ & 0.12${}_{\rm inj}$+0.5${}_{\rm acc}$+0.68${}_{\rm decel}$
        & ~~(0.14+1.4+0.94)~~ \\
      beam power             & 750\,kW &  (240)   \\
      \hline \hline
    \end{tabular}
    \label{jparc:MRFXpara}
  \end{center}
\end{table}
In the MR FX mode operation, so far 1.24$\times$10$^{14}$ protons per pulse
(ppp) beam intensity 
has been achieved, a world record for extracted ppp for any 
synchrotron and equating to an average beam power of 240\,kW. 
The accelerator team is following 
a concrete upgrade scenario~\cite{jparc-midterm1,jparc-midterm2}
to reach the design power of 750\,kW in forthcoming years, 
with a typical planned parameter set as listed in Table~\ref{jparc:MRFXpara}.
This will double the current repetition rate by 
(i) replacing the magnet power supplies, 
(ii) replacing the RF system, and 
(iii) upgrading injection/extraction devices. 
Furthermore, conceptual studies on how to realize 1$\sim$2\,MW 
beam powers and even beyond are now underway~\cite{jparc-longterm}, 
such as by raising the RCS top energy, enlarging the MR aperture, or 
inserting an ``emittance-damping'' ring between the RCS and MR. 

%
%
%
The neutrino production target and the beam window are designed 
for 750\,kW operation with 3.3$\times$10$^{14}$ ppp 
(equivalent to RCS 1\,MW operation) and 2.1 sec cycle. In the target,
the pulsed beam generates an instantaneous temperature rise per pulse 
of 200 C$^\circ$ and a thermal stress wave of magnitude 7 MPa, 
giving a safety factor of $\sim$5 against the tensile strength. 
Although this safety factor will be reduced by cyclic fatigue, 
radiation damage\footnote{Graphite loses integrity at proton fluences 
  of around 10$^{22}$ protons/cm$^2$, which would be reached after 
  around 5 years operation at 750\,kW. 
  The target has been designed to operate at a maximum 
  temperature of around 700$^\circ$C, which from neutron irradiation data 
  should minimise any dimensional changes and reduction 
  in thermal conductivity.
}
and oxidization of the graphite,
%
a lifetime of 2$-$5 years is expected.\footnote{
 By adopting the double rep-rate scenario, the number of protons 
 per pulse will be reduced, and hence the thermal shock per pulse 
 will be reduced. 
}
In order to both increase lifetimes and to realize multi-MW beam operation, 
the beamline team intends to investigate modifications to 
the existing design, and even to develop 
a new concept that can dissipate a higher heat load and 
may be more resilient to radiation damage.

So far the horns were operated with a 250 kA pulsed current and 
a minimum repetition cycle of 2.48 sec. 
To operate the horns at a doubled repetition rate of $\sim$1 Hz  
requires new individual power supplies for each horn utilizing an energy 
recovery scheme and low inductance/resistance striplines. 
These upgrades will reduce the charging 
voltage/risk of failure, and, as another benefit, increase 
the pulsed current to 320 kA.
The horn-1 water-spray cooling system 
has sufficient capacity to keep the conductor 
below the required 80$^\circ$C at up to 2\,MW.  

\begin{table}[btp]
  \begin{center}
    \caption{Acceptable beam power and achievable parameters 
             for each beamline component~\cite{IFW-nu750kW,IFW-numultiMW}.
             Limitations as of May 2013 are also given in parentheses.}
    \begin{tabular}{lcc}
      \hline \hline
      component   & \multicolumn{2}{c}{beam power/parameter} \\ 
      \hline
      target      & \multicolumn{2}{c}{3.3$\times$10$^{14}$ ppp } \\
      beam window & \multicolumn{2}{c}{3.3$\times$10$^{14}$ ppp } \\
      horn        &   ~  & ~ \\
      \multicolumn{1}{c}{cooling for conductors} & 
      \multicolumn{2}{c}{2\,MW 
      } \\
      \multicolumn{1}{c}{stripline cooling}   
      & 1$\sim$2\,MW & ( 400\,kW )\\
      \multicolumn{1}{c}{hydrogen production} 
      & 1$\sim$2\,MW & ( 300\,kW ) \\ 
      \multicolumn{1}{c}{horn current} & 320 kA & ( 250 kA ) \\
      \multicolumn{1}{c}{power supply repetition} &  1 Hz  &  ( 0.4 Hz ) \\
      decay volume   &  \multicolumn{2}{c}{4\,MW} \\
      hadron absorber/beam dump  &  \multicolumn{2}{c}{3\,MW} \\
      \multicolumn{1}{c}{water cooling facilities}  
       & $\sim$2\,MW & ( 750\,kW ) \\                  
      radiation shielding   & 4\,MW & ( 750\,kW )\\
      radioactive air leakage to the TS ground floor 
      & $\sim$2\,MW &  ( 500\,kW ) \\ 
      radioactive cooling water treatment 
      & $\sim$2\,MW &  ( 600\,kW ) \\
      \hline \hline
    \end{tabular}
    \label{jparc:BLupgrade}
  \end{center}
\end{table}
Considerable experience has been gained on the path to achieving 240\,kW 
beam power operation, and the beamline group is promoting upgrades
to realize 750\,kW operation, 
such as by improving the activated air confinement in TS, and is proposing 
to expand the facilities for the treatment of activated water.
Table~\ref{jparc:BLupgrade} gives a summary of acceptable beam power and/or 
achievable parameters for each 
beamline component~\cite{IFW-nu750kW,IFW-numultiMW}, 
after the proposed upgrades in forthcoming years. 

\subsection{The neutrino flux calculation}
\label{sec:nuflux}

The T2K flux~\cite{Abe:2012av} is estimated by simulating the 
J-PARC neutrino beam line while tuning the modeling 
of hadronic interactions using data from NA61/SHINE~\cite{Abgrall:2011ae,Abgrall:2011ts}
 and other experiments measuring hadronic interactions on nuclei. To date, NA61/SHINE has
provided measurements of pion and kaon production multiplicities for proton interactions on a 
0.04 interaction length graphite target, as well as the inelastic cross section for protons
on carbon.  Since ``thin" target data are used, the secondary interactions of hadrons 
inside and outside of the target are modeled using other data or scaling the NA61/SHINE data to different center of mass
energies or target nuclei.

\begin {figure}[htbp]
  \begin{center}
    \includegraphics[width=0.45\textwidth]{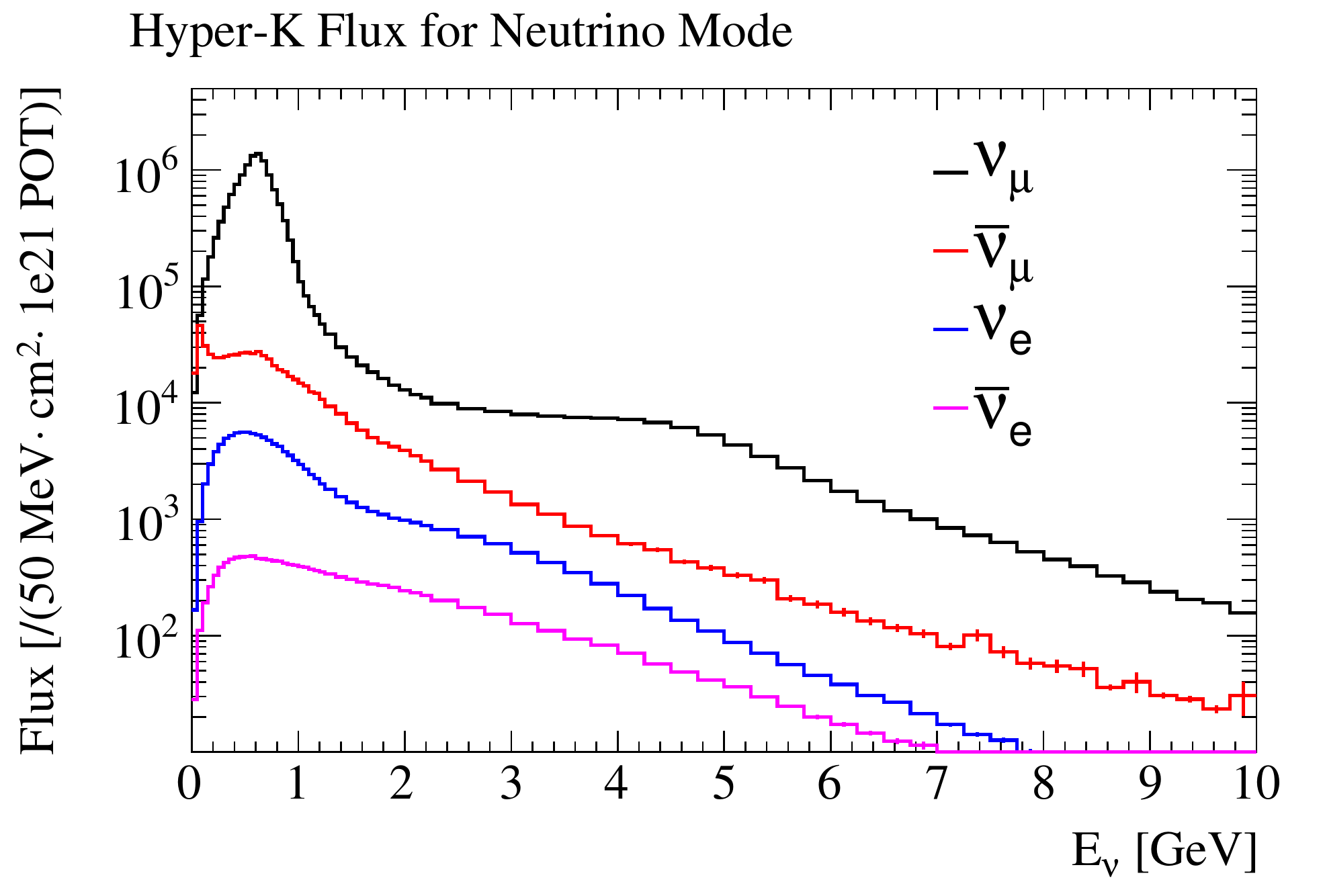}
    \includegraphics[width=0.45\textwidth]{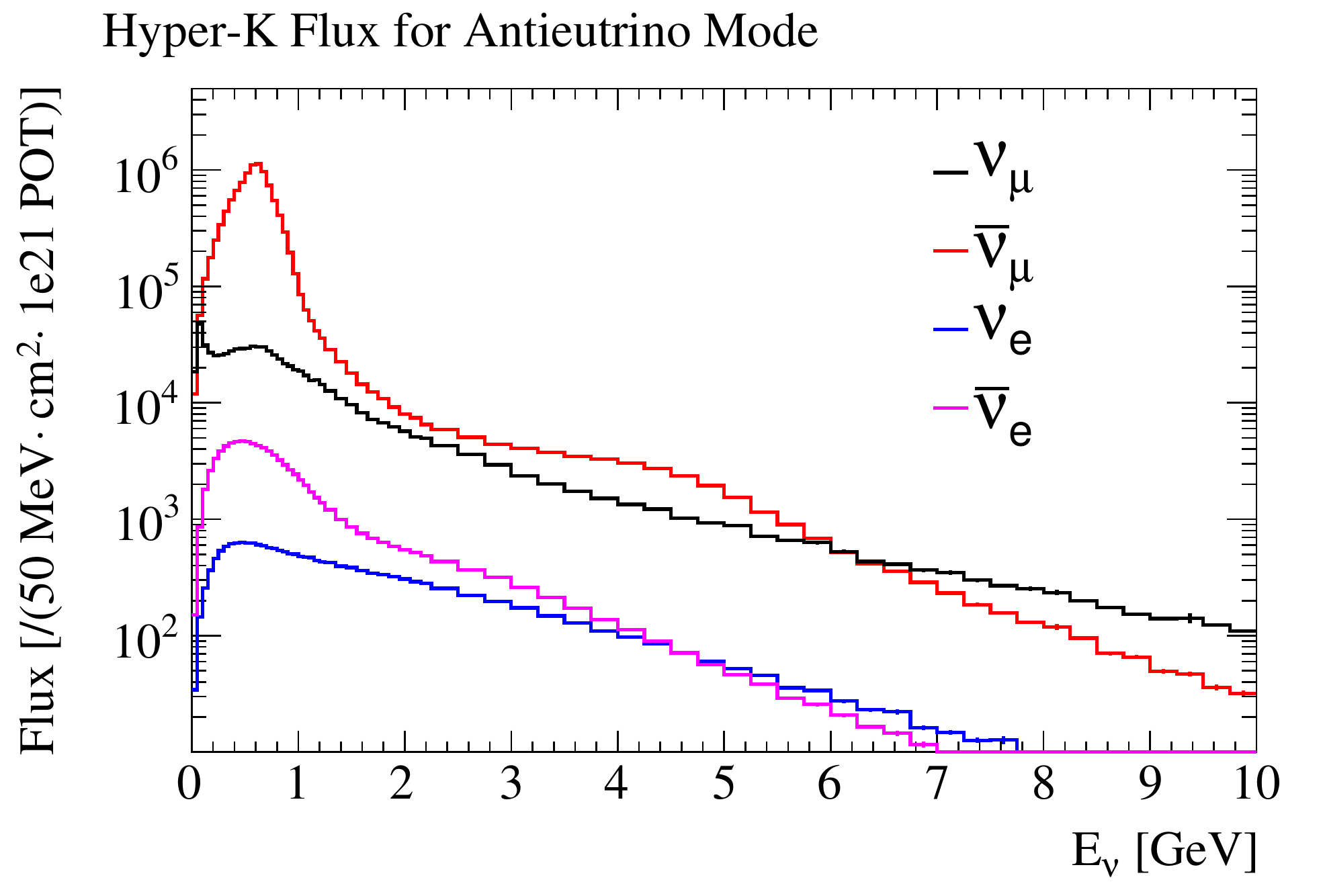}
    \caption{The predicted Hyper-K neutrino fluxes from the J-PARC beam without oscillations. The
neutrino enhanced beam is shown on the left and the antineutrino enhanced beam is shown on the right.}
    \label{fig:flux_pred}
  \end{center}
\end {figure}

 For the studies presented in this
document, the T2K flux simulation has been used with the horn currents raised from 250 kA to 320 kA. 
The flux is estimated for both
polarities of the horn fields, corresponding to neutrino enhanced and antineutrino enhanced fluxes. 
The calculated fluxes at Hyper-K, without oscillations, are shown in Fig.~\ref{fig:flux_pred}.  

\subsubsection{The flux uncertainties}

The sources of uncertainty in the T2K flux calculation include:
\begin{itemize}
\item{Uncertainties on the primary production of pions and kaons in proton on carbon collisions.}
\item{Uncertainties on the secondary hadronic interactions of particles in the target or 
beam line materials after the initial hadronic scatter.} 
\item{Uncertainties on the properties of the proton beam incident on the target, including the absolute
current and the beam profile.}
\item{Uncertainties on the alignment of beam line components, including the target and magnetic horns.}
\item{Uncertainties on the modeling of the horn fields, including the absolute field strength and 
asymmetries in the field.}
\end{itemize}

The uncertainties on the hadronic interaction modeling are the largest contribution to the flux
uncertainty and may be reduced by using replica target data.  
A preliminary analysis using a subset of the replica target data from NA61/SHINE
has shown that it can be used 
to predict the T2K flux~\cite{Abgrall:2012pp}.  Since it is expected that replica target data will
be available for future long baseline neutrino experiments, the Hyper-K flux uncertainty is 
estimated assuming the expected uncertainties on the measurement of particle multiplicities from the replica
target.  Hence, uncertainties related to the modeling of hadronic interactions inside the target are 
no longer relevant, however, uncertainties for interactions outside of the target are considered.  
The uncertainties on the measured replica target multiplicities are estimated by applying the same uncertainties
that NA61/SHINE has reported for the thin target multiplicity measurements.

\begin {figure}[htbp]
  \begin{center}
    \includegraphics[width=0.45\textwidth]{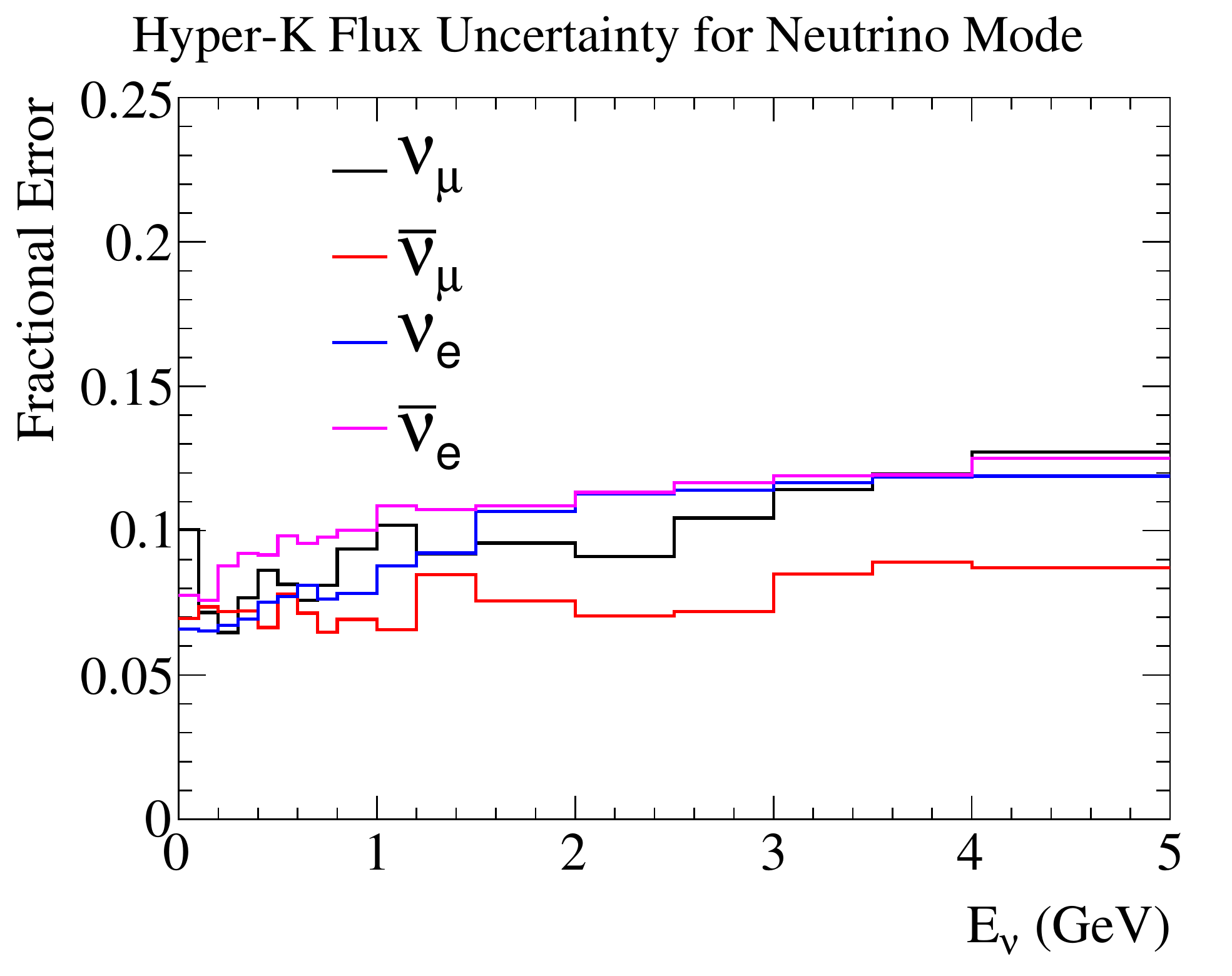}
    \includegraphics[width=0.45\textwidth]{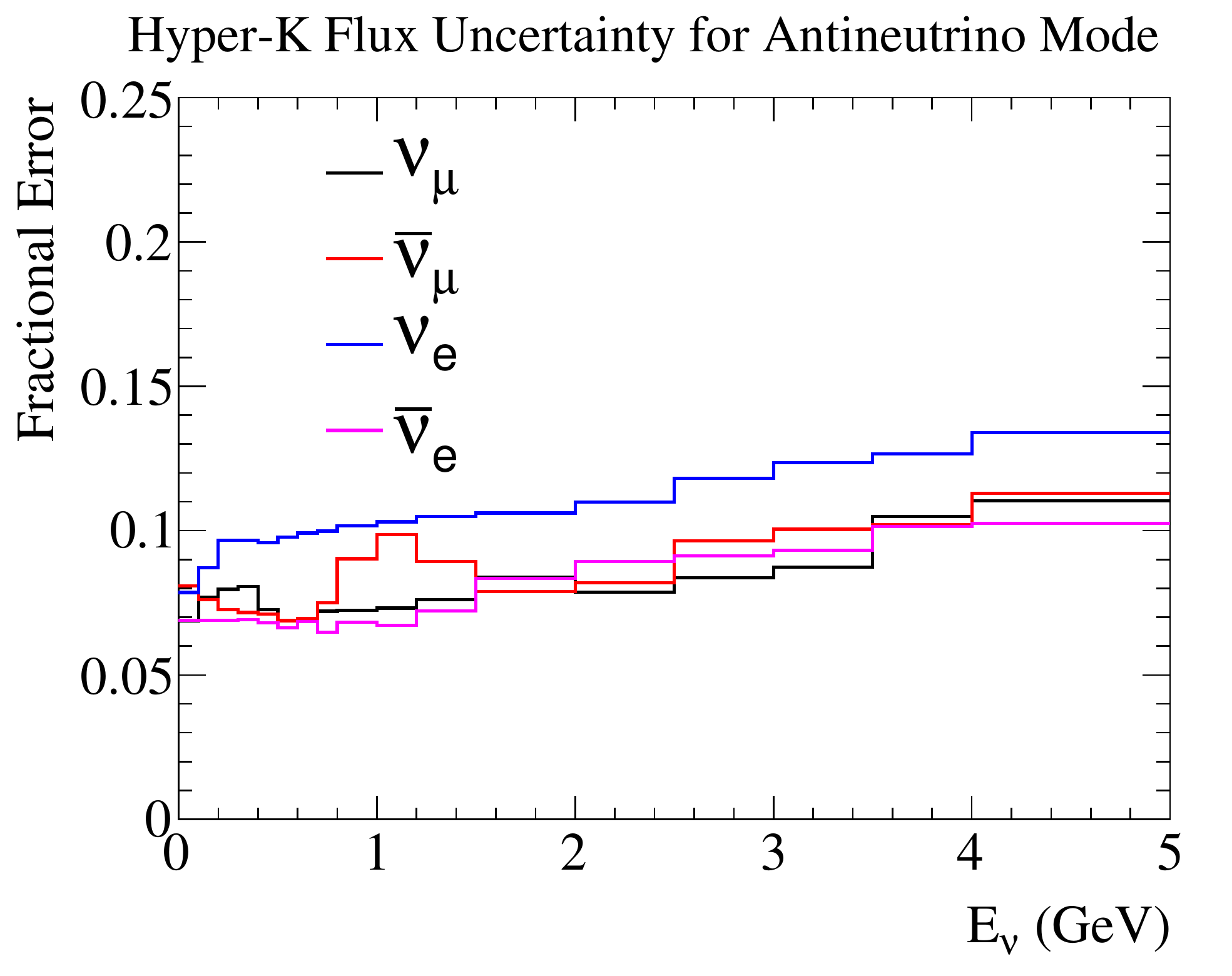}
    \caption{The predicted uncertainty on the neutrino flux calculation assuming replica target hadron production data
are available.}
    \label{fig:flux_unc}
  \end{center}
\end {figure}

\begin {figure}[htbp]
  \begin{center}
    \includegraphics[width=0.45\textwidth]{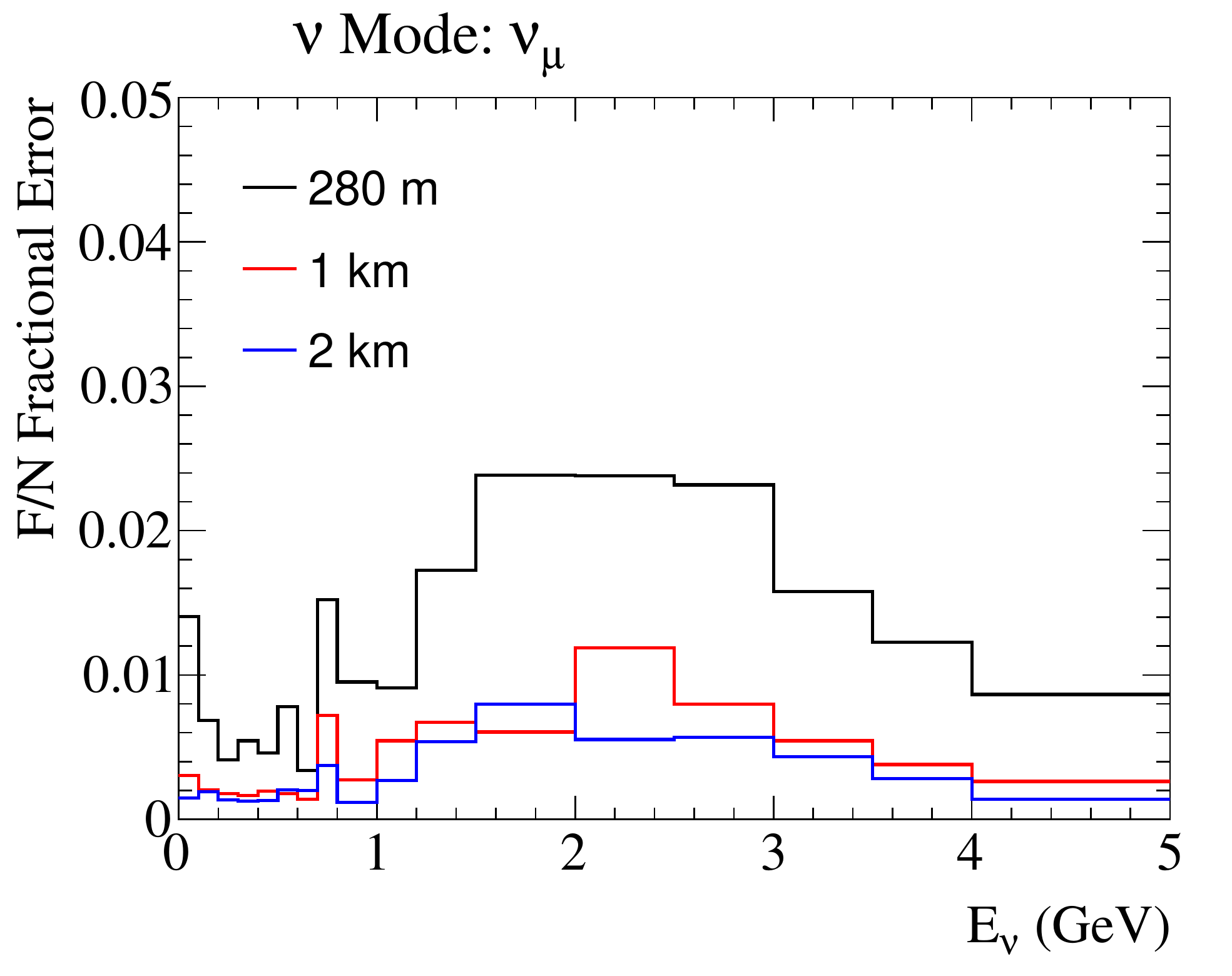}
    \includegraphics[width=0.45\textwidth]{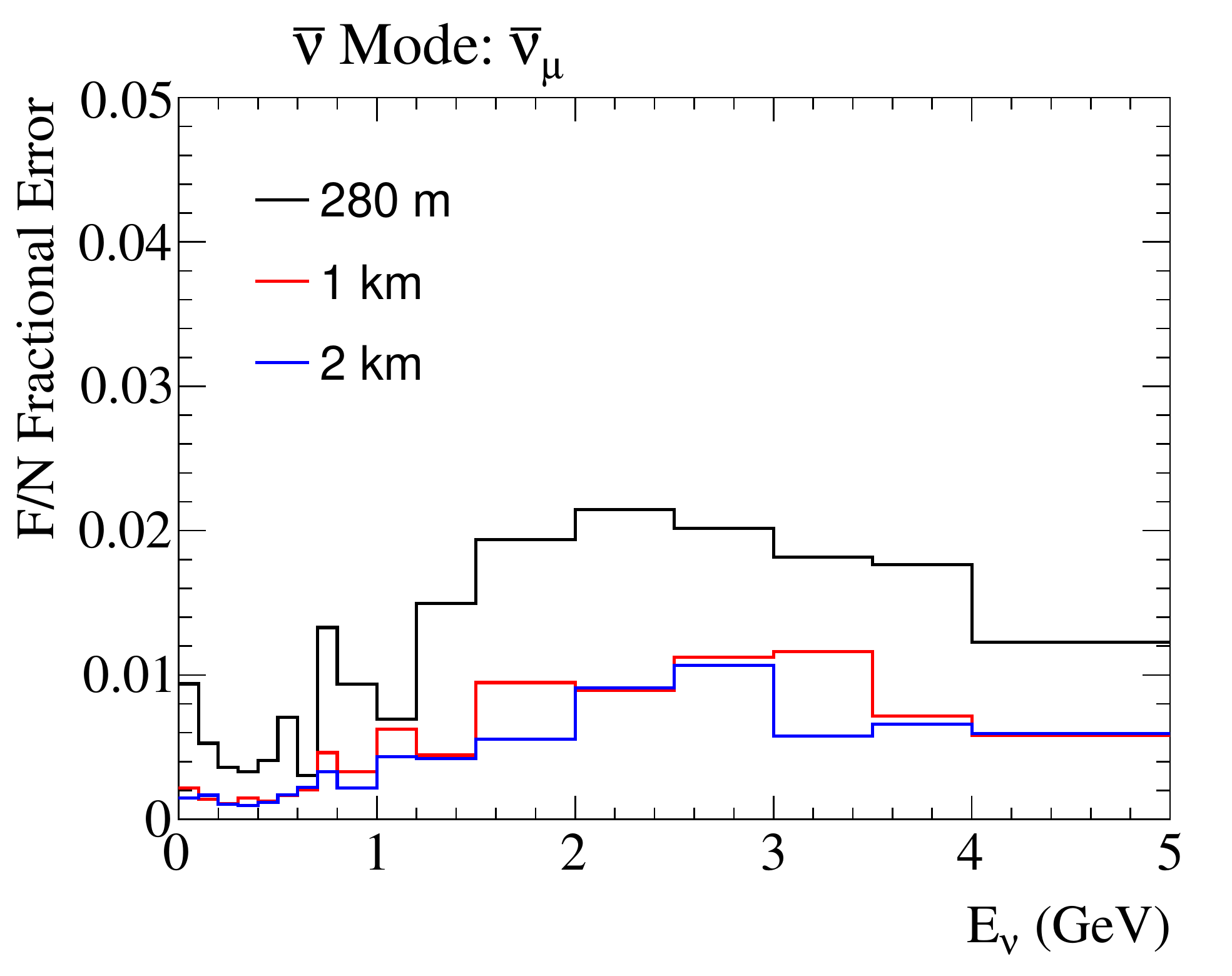}
    \includegraphics[width=0.45\textwidth]{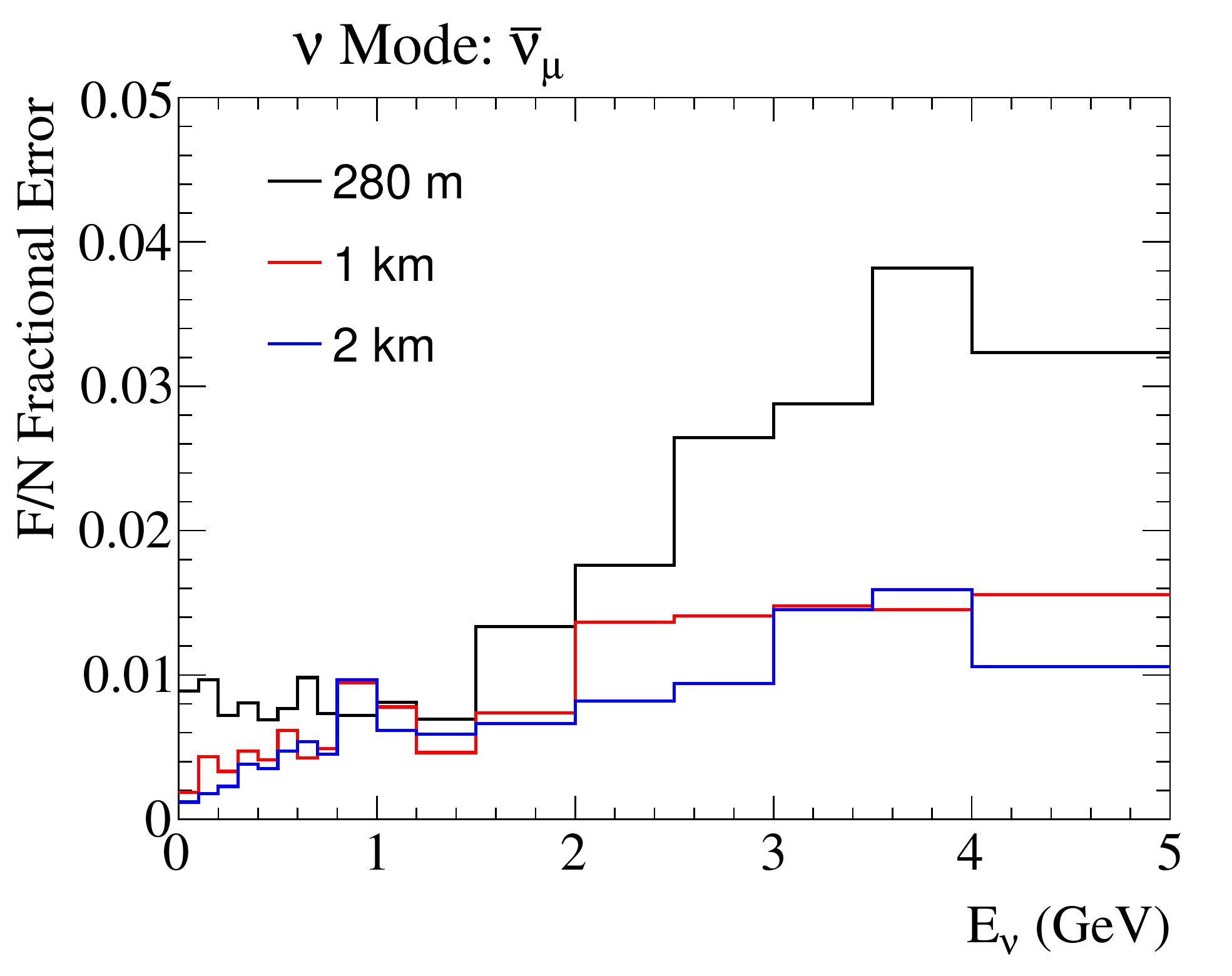}
    \includegraphics[width=0.45\textwidth]{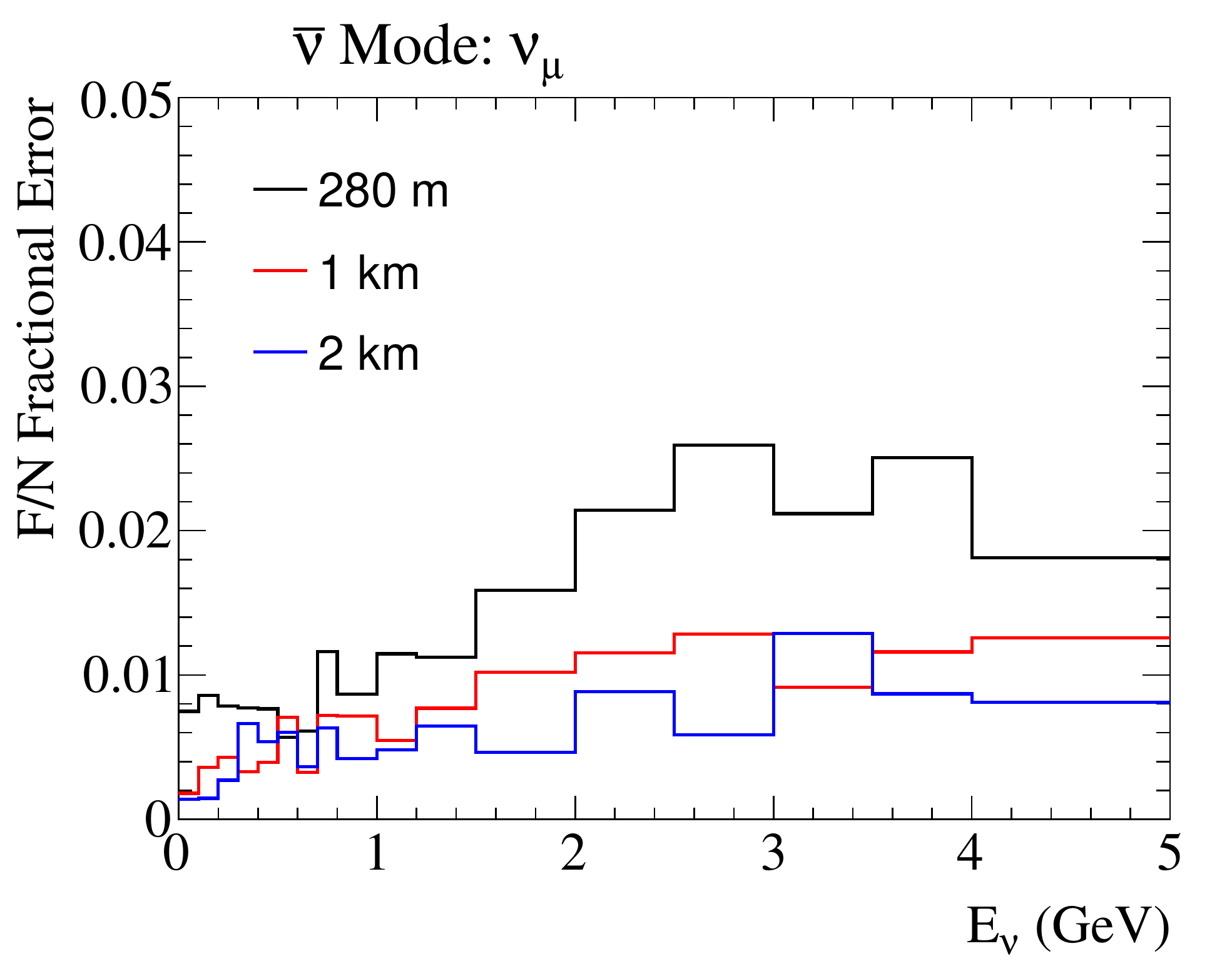}
    \caption{The uncertainty on the far-to-near flux ratio for near detectors at 280 m, 1 km and 2 km.  Left: neutrino 
enhanced beam.  Right: antineutrino enhanced beam.  Top: the focused component of the beam.  Bottom: the defocused component of
the beam.}
    \label{fig:fn_unc}
  \end{center}
\end {figure}

The total uncertainties on the flux as function of the neutrino energy are shown in Fig.~\ref{fig:flux_unc}.
In oscillation measurements, the predicted flux is used in combination with measurements of the neutrino
interaction rate from near detectors.  Hence, it is useful to consider the uncertainty on the ratio of the flux
at the far and near detectors: 
\begin{equation}
\delta_{F/N}(E_{\nu}) = \delta \left (\frac{\phi_{HK}(E_{\nu})}{\phi_{ND}(E_{\nu})} \right )
\end{equation}
Here $\phi_{HK}(E_{\nu})$ and $\phi_{ND}(E_{\nu})$ are the predicted fluxes at Hyper-K and the near detector
respectively.  T2K uses the ND280 off-axis detector located 280 m from the T2K target.  At that distance,
the beam-line appears as a line source of neutrinos, compared to a point source seen by Hyper-K, and the 
far-to-near ratio is not flat.  For near detectors placed further away, at 1 or 2 km for example, the 
far-to-near flux ratio becomes more flat and there is better cancellation of the flux uncertainties 
between the near and far detectors.  Fig.~\ref{fig:fn_unc} shows how the uncertainty on the far-to-near
ratio evolves for baselines of 280 m, 1 km and 2 km.  While this extrapolation uncertainty is reduced for 
near detectors further from the production point, even the 280 m to Hyper-K uncertainty is less than $1\%$ near
the flux peak energy of 600\,MeV.

\subsubsection{The neutrino beam direction\label{sec:beam_dir}}
The previously described uncertainties assume that the near detector is located on the line from the
average neutrino production point to Hyper-K.  
 This is expected to be the optimal configuration for uncertainties on the beam direction.  If the off-axis
near detector only covers a small solid angle, it is only sensitive to changes in the off-axis angle,
and cannot distinguish between vertical or horizontal shifts of the neutrino beam direction.  For 
a near and far detector on the same line, the effects of vertical or horizontal shifts are the same
and uncertainties on the beam direction cancel in the far-to-near ratio.  The T2K ND280 detector is situated on
the line to Super-K.

\begin {figure}[htbp]
  \begin{center}
    \includegraphics[width=0.45\textwidth]{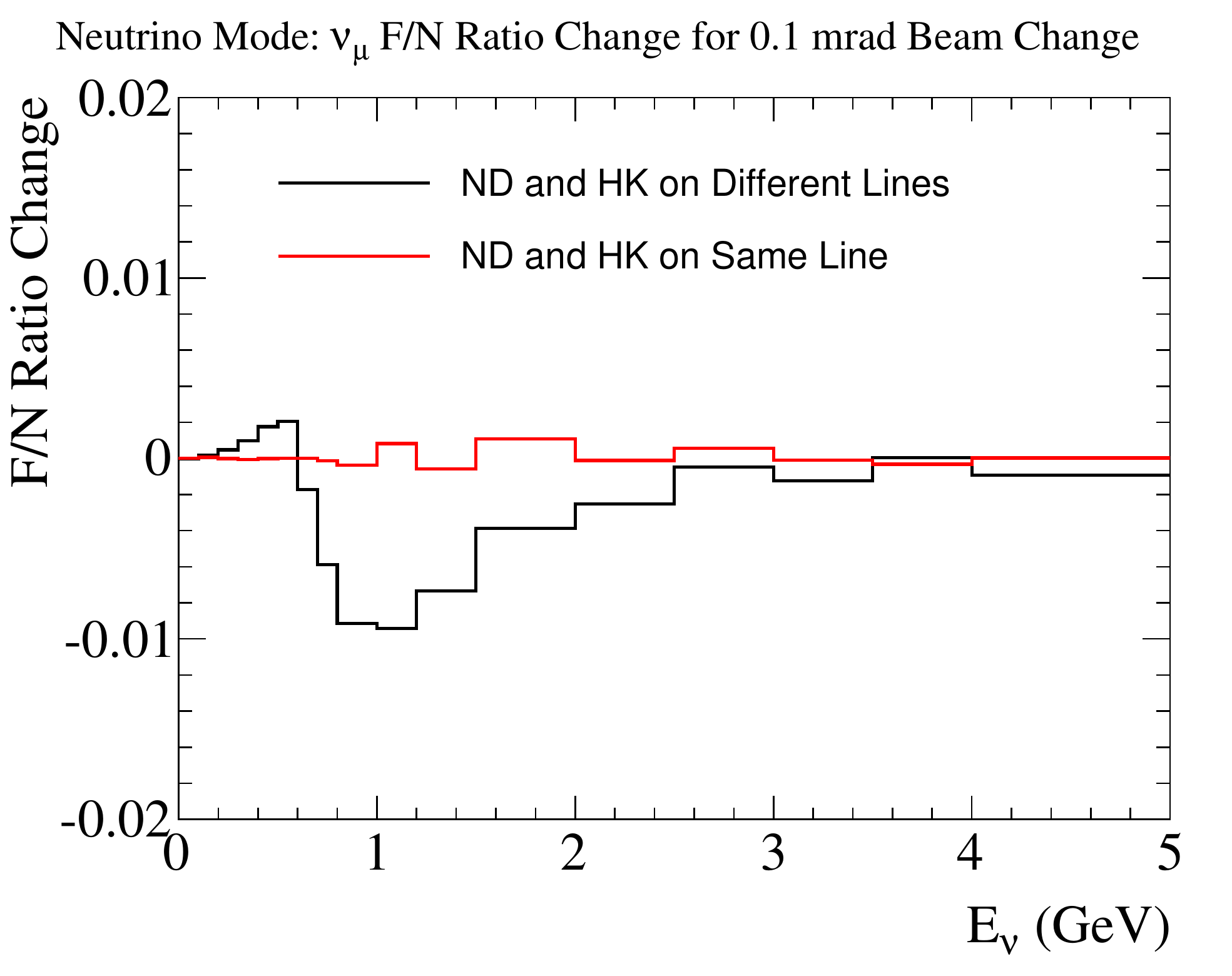}
    \includegraphics[width=0.45\textwidth]{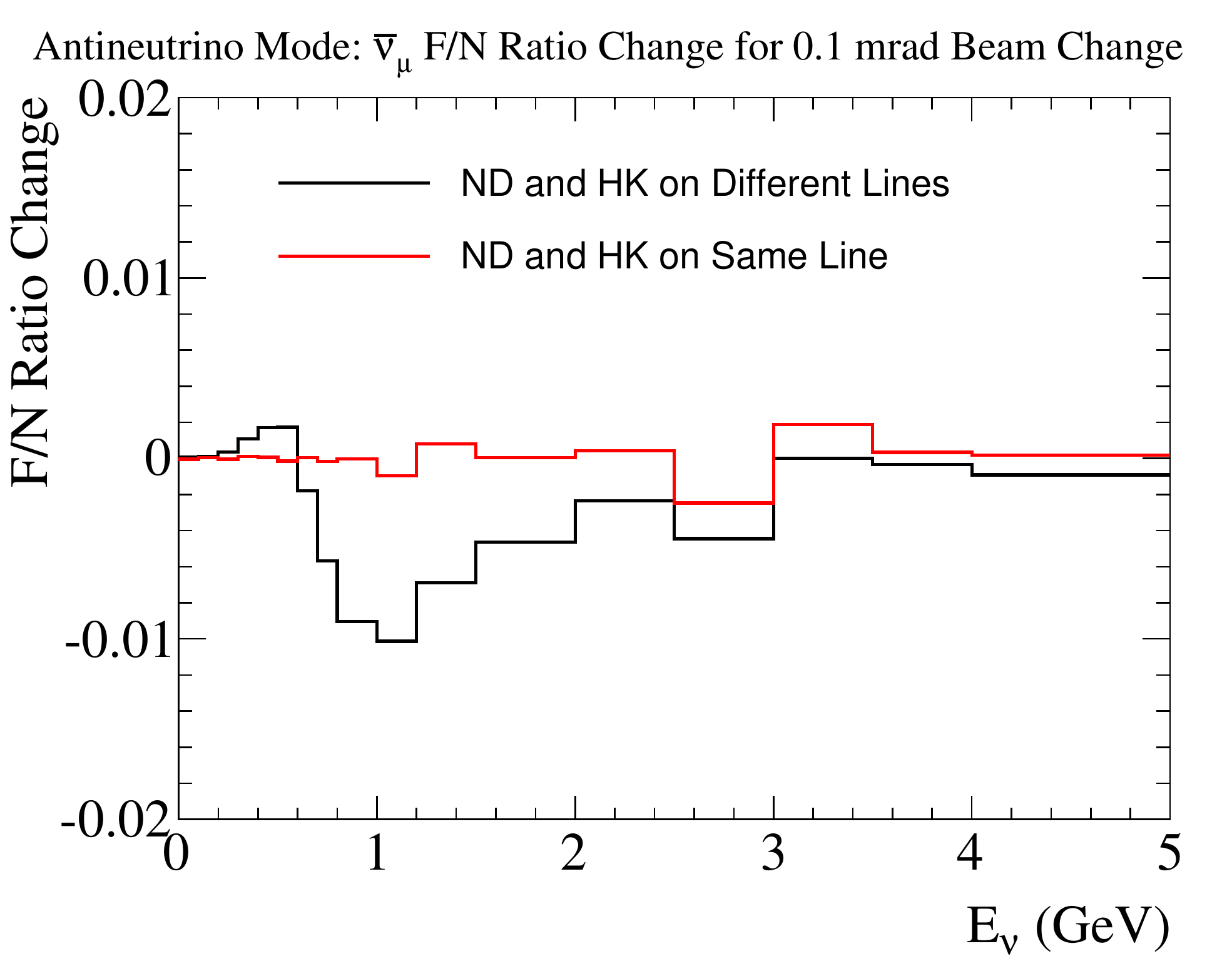}
    \caption{The change to the far-to-near ratio when the beam is shifted in the horizontal direction
by 0.1 mrad toward the far detector.  The ratio is unchanged when the near and far detector are in the
same direction (red), and changed by up to 1\% when the near and far detector are not in the same direction. }
    \label{fig:offaxis_xshift}
  \end{center}
\end {figure}

Two sites are being considered for Hyper-K, the Mozumi site near Super-K and the Tochibora
site.  As illustrated in Fig.~\ref{fig:map}, the horizontal displacement from the beam direction for 
these two sites is opposite, hence ND280 is not situated on the line to the Tochibora site. The
bias on the far-to-near ratio when the horizontal displacement of the near detector is opposite to the
far detector is estimated when the beam is shifted by 0.1 mrad in the horizontal direction, the current uncertainty 
on the beam direction measurement by the T2K INGRID detector.  As Fig~\ref{fig:offaxis_xshift} shows that
the far-to-near ratio can be distorted by 1\% for a 0.1 mrad horizontal shift when the near detector and far
detector are not on the same line.  This may be a significant source of uncertainty for estimation of the
flux at Hyper-K and must be considered when designing the near detectors for Hyper-K.

%

\section{\label{sec:ND} Near Detectors} 

The accelerator neutrino event rate observed at Hyper-K depends on the oscillation probability, neutrino flux, neutrino
interaction cross-section, detection efficiency, and the detector fiducial mass of Hyper-K.  To 
extract estimates of the oscillation parameters from data, one must model the neutrino flux, cross-section and 
detection efficiency with sufficient precision.  In the case of the neutrino cross-section, the model must describe the exclusive differential
cross-section that includes the dependence on the incident neutrino energy, $E_{\nu}$, the kinematics of the outgoing
lepton, $p_{l}$ and $\theta_{l}$, and the kinematics of final state hadrons and photons.  
In our case,
the neutrino energy
is inferred from the lepton kinematics, while the modeling of reconstruction efficiencies depends on the hadronic 
final state as well.


The neutrino flux and cross-section models can be constrained by data collected at near detectors, situated close enough
to the neutrino production point so that oscillation effects are negligible.  
Our approach to using near detector
data will build on the experience of T2K while considering new near detectors that may address important uncertainties in the
neutrino flux or cross-section modeling.

The conceptual design of the near detectors is being developed based on the physics sensitivity studies described in Section~\ref{sec:physics_sensitivities}.
In this section, we present basic considerations on the near detector requirements and conceptual designs.
More concrete requirements and detector design will be presented in future.
We first discuss the current understanding of neutrino cross section based on the T2K experience and issues relevant for the near detector requirements.
Then, the design, performance, and future prospects of T2K near detectors are described as a reference.
In order to further reduce the uncertainty and to enhance the physics sensitivity of the project, we have been studying a possibility of building new detectors.
As examples of such new detectors, two possible design of 
new intermediate water Cherenkov detectors are presented.

\subsection{Neutrino cross section uncertainties relevant for near detector requirements}
T2K has successfully applied a method of fitting to near detector
data with parameterized models of the neutrino flux and interaction cross-sections.  The model parameters in the flux and 
nucleon level cross-section description are constrained by the near detector data so that their contribution to the uncertainty 
on the Super-K event rate predictions is reduced to only $\sim3\%$, as discussed in Section~\ref{sec:nd280_perf}.  However,
additional uncertainties on the modeling of nuclear effects and the modeling of the $\nu_e$ 
interaction cross section relative to the $\nu_{\mu}$ cross section contribute an uncertainty of 5-10\% on the Super-K
event rate predictions.  

The use of near detector data
is complicated by the fact that the neutrino beam's energy dependence and flavor content at the near and far detectors can be different due to
the neutrino oscillations.  This complication introduces critical sources of uncertainty for future long baseline
experiments measuring CP violation:
\begin{itemize}
\item The relative cross-section for $\nu_{\mu}$ and $\nu_{e}$ interactions.
\item The relationship between the incident neutrino energy and the final state kinematics used to 
estimate the true neutrino energy.  In our case, this is the charged lepton four momentum.
\item The difference in the reaction cross-sections on different nuclei in the near and far detectors 
in the case that those target nuclei are different.  
\end{itemize}

Since the intrinsic $\nu_e$ contribution in the beam is $\sim1\%$, using the near detector data to constrain the $\nu_e$ 
interaction cross-section is challenging. Recent work has shown that theoretical uncertainties on the cross section ratio
$\sigma_{\nu_{e}}/\sigma_{\nu_{\mu}}$ can be a few percent at the relevant energy and mimic a CP violation effect with opposite
sign for neutrinos and antineutrinos~\cite{Day:2012gb}.  The potential to measure $\nu_e$ interactions in the T2K 
near detectors and new near detectors is discussed in the following sections.  

The oscillation probability depends on the neutrino energy, while we estimate the neutrino energy from the
observed four momentum of the final state charged lepton.  Correctly modeling the relationship between neutrino
energy and final state lepton kinematics is essential to correctly applying the oscillation probability, even when there is a 
constraint on the event rate from the near detector data.  The signal modes are a charge lepton in the final state,
with no detected pion.  The main contribution to this topology is charge current quasi-elastic (CCQE) scatters, where the neutrino
energy can be estimated from final state lepton momentum and scattering angle.  In recent years, much theoretical work
has been done to calculate contributions to this topology from non-CCQE processes such as, two body currents or final state
interactions that can absorb a pion~\cite{marteau,Martini:2009uj,Carlson:2001mp,Shen:2012xz,Bodek:2011ps,Martini:2010ex,Martini:2013sha,
Nieves:2005rq,PhysRevC.83.045501}.  These nuclear effects often lead to the ejection of multiple nucleons in the final state
and are referred to as multinucleon processes here.  The additional final state nucleons can carry away energy, leading to kinematics
that are different from CCQE scatters.  Currently, the theoretical calculations do not all agree with one another and do not include all 
processes leading to such invisible energy loss.  

Even if models disagree,  these effects may be constrained by data from 
near detectors and dedicated cross-section experiments~\cite{AguilarArevalo:2010zc,AguilarArevalo:2013hm,Fields:2013zhk,Fiorentini:2013ezn}.
However, such measurements of the reaction rate or of the nucleon content of the final state can only test one of these
calculations within a model, and do not directly probe the relationship of final state kinematics to neutrino energy.  
Since the multinucleon processes arise from nuclear effects, near detector measurements with the same nuclear target as the far
detector are preferred.  These may be made with the existing (Section~\ref{sec:nd280_perf}) or upgrade (Section~\ref{sec:nd280_upgrade}) of T2K near detectors, or new near detectors (Section~\ref{sec:int_wc}).
Additionally, the relationship between the incident neutrino energy and final state lepton four momentum
can be studied in more detail by using multiple neutrino spectra with different peak energies, as discussed in Section~\ref{sec:nuprism}.

\subsection{The T2K near detectors}

\subsubsection{The T2K INGRID and ND280 detectors}

\begin {figure}[htbp]
  \begin{center}
    \includegraphics[width=0.45\textwidth]{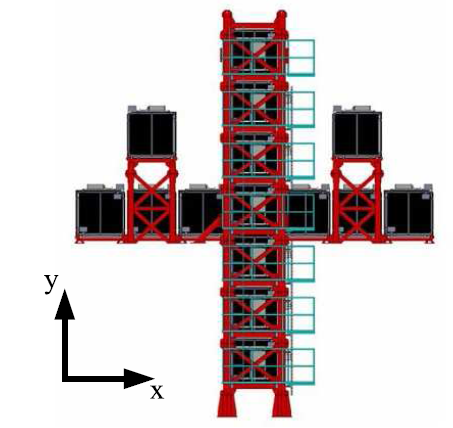}
    \includegraphics[width=0.45\textwidth]{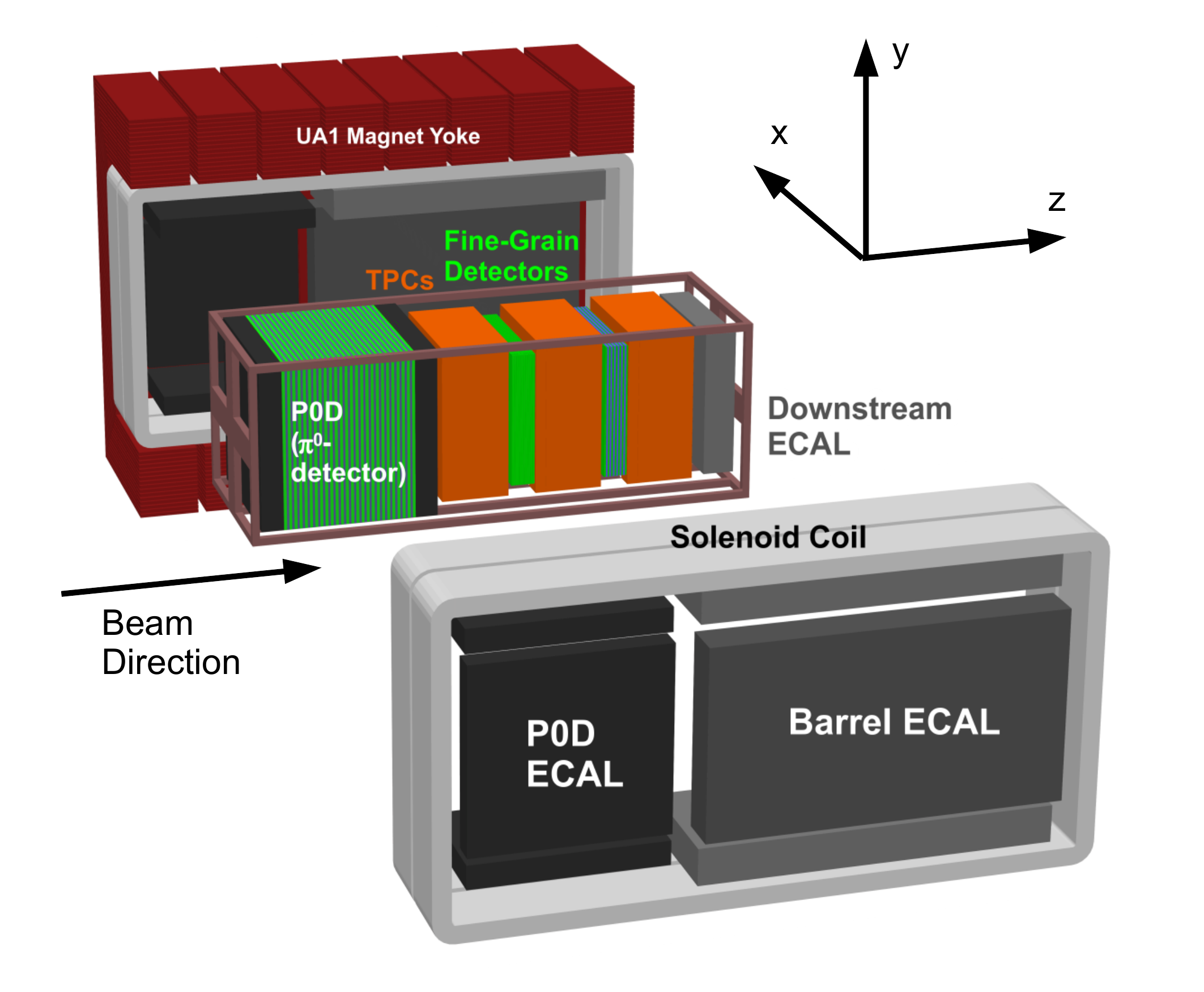}
    \caption{The INGRID detector modules (left) and the exploded view of the ND280 off-axis detector (right).}
    \label{fig:nd280_exploded}
  \end{center}
\end {figure}

The INGRID detector~\cite{Otani:2010zza} consists of 16 iron-scintillator modules configured in a cross pattern centered on the beam axis 280 m downstream
from the T2K target, as shown in Fig.~\ref{fig:nd280_exploded}.
The rate of interactions in each module is measured and a profile is constructed to 
constrain the neutrino beam direction.  The ND280 off-axis detector is located 280 m downstream from the T2K target as well, but
at an angle of 2.5 degrees away from the beam direction.
Fig.~\ref{fig:nd280_exploded} shows the components of ND280:
the P0D $\pi^{0}$ detector~\cite{Assylbekov201248}, time projection chambers (TPCs)~\cite{Abgrall:2010hi},
fine grain scintillator bar detectors (FGDs)~\cite{Amaudruz:2012pe} and surrounding
electromagnetic calorimeters (ECALs).  The detectors are immersed in a 0.2 T magnetic field and the magnetic yoke is instrumented with
plastic scintillator panels for muon range detection~\cite{Aoki:2012mf}.  The magnetic field allows for momentum measurement and 
sign selection of charged particles.  The magnetization of ND280 is particularly important for operation in antineutrino mode where
the neutrino background is large.  In that case, ND280 is able to separate the ``right-sign" $\mu^{+}$ from the ``wrong-sign" $\mu^{-}$. The 
P0D and FGDs act as the neutrino targets, while the TPCs provide measurements of momentum and ionizing energy loss for particle
identification.  The P0D and one of the FGDs include passive water layers that allow for neutrino interaction rate measurements on 
the same target as Super-K.
ND280 has been employed to measure the rates of charged current $\nu_{\mu}$ and $\nu_{e}$ interactions, as well as
NC$\pi^0$ interactions.

\subsubsection{The INGRID beam direction measurement}  

The INGRID detector is used to measure the neutrino beam direction.  Neutrino interactions originating
in each INGRID module are detected and corrections including those for events
originating in the surrounding sand and inefficiencies in event pile-up detection, {\it i.e.} more than
one event for beam bunch, are applied.  The profiles of event rates across the vertical and horizontal arrays of 
modules are fit to extract the beam center.  Systematic uncertainties on the beam center 
measurement are 0.094 mrad and 0.104 mrad for the horizontal and vertical respectively.  As discussed
in Section~\ref{sec:beam_dir}, a precise knowledge of the beam direction is important if the off-axis
near detector and far detector are not situated along the same direction.  The 0.1 mrad systematic error
from INGRID is sufficient to control the flux extrapolation uncertainty due to the uncertainty on the 
beam direction to $<1\%$ in that scenario, under the 
assumption that the beam direction constraint from INGRID can be extrapolated to 2.5 degrees off-axis.

\subsubsection{ND280 measurements \label{sec:nd280_perf}}

The ND280 detector is used to measure charge current (CC) interaction
rates binned by lepton kinematics and hadronic final states, as well
as neutral current (NC) interactions with detected $\pi^{0}$, $\pi^{\pm}$ or
protons in the final state.  These measurements are used to constrain
the neutrino flux and cross-section models, including the $\nu_e$ contamination 
of the beam.

\paragraph{ND280 $\nu_{\mu}$ CC measurement}

The $\nu_{\mu}$ charged current interactions in ND280 are used to
constrain the neutrino energy spectrum and cross-section model parameters.
CC events are selected with a vertex in the most upstream FGD (FGD1) with a track passing through
the second TPC and having an energy loss consistent with a muon. The
selected CC candidate events are divided into several samples to help
constrain the cross-sections: CC-0$\pi$, with no identified pions;
CC-1$\pi^+$, with exactly one $\pi^+$ and no $\pi^-$ or $\pi^0$; and
CC-other, with all the other CC events. The current analysis uses interactions
in FGD1, which consists only of plastic scintillator targets.  

\begin{figure} 
   \centering
   \includegraphics[width=0.45\textwidth]{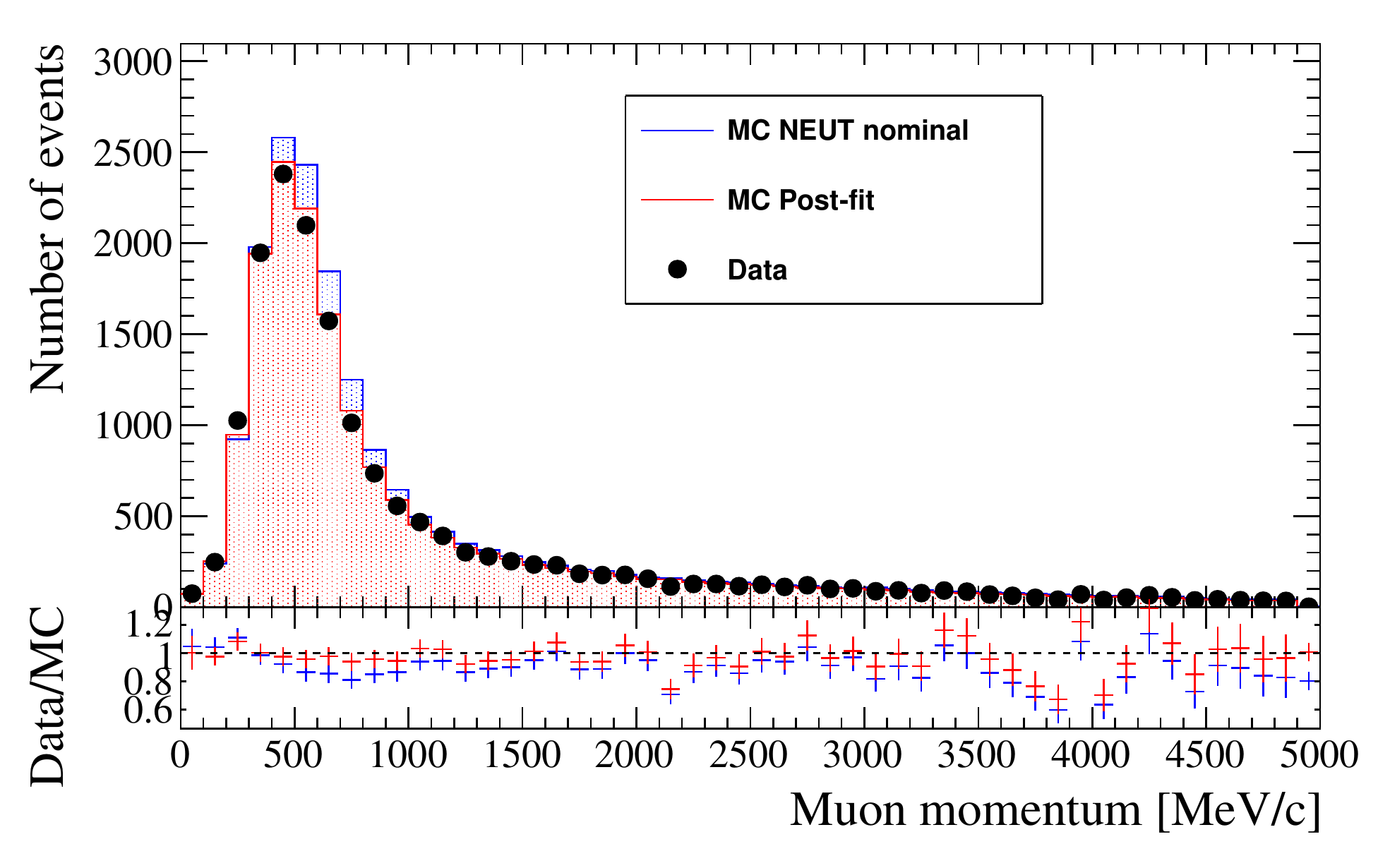} 
   \includegraphics[width=0.45\textwidth]{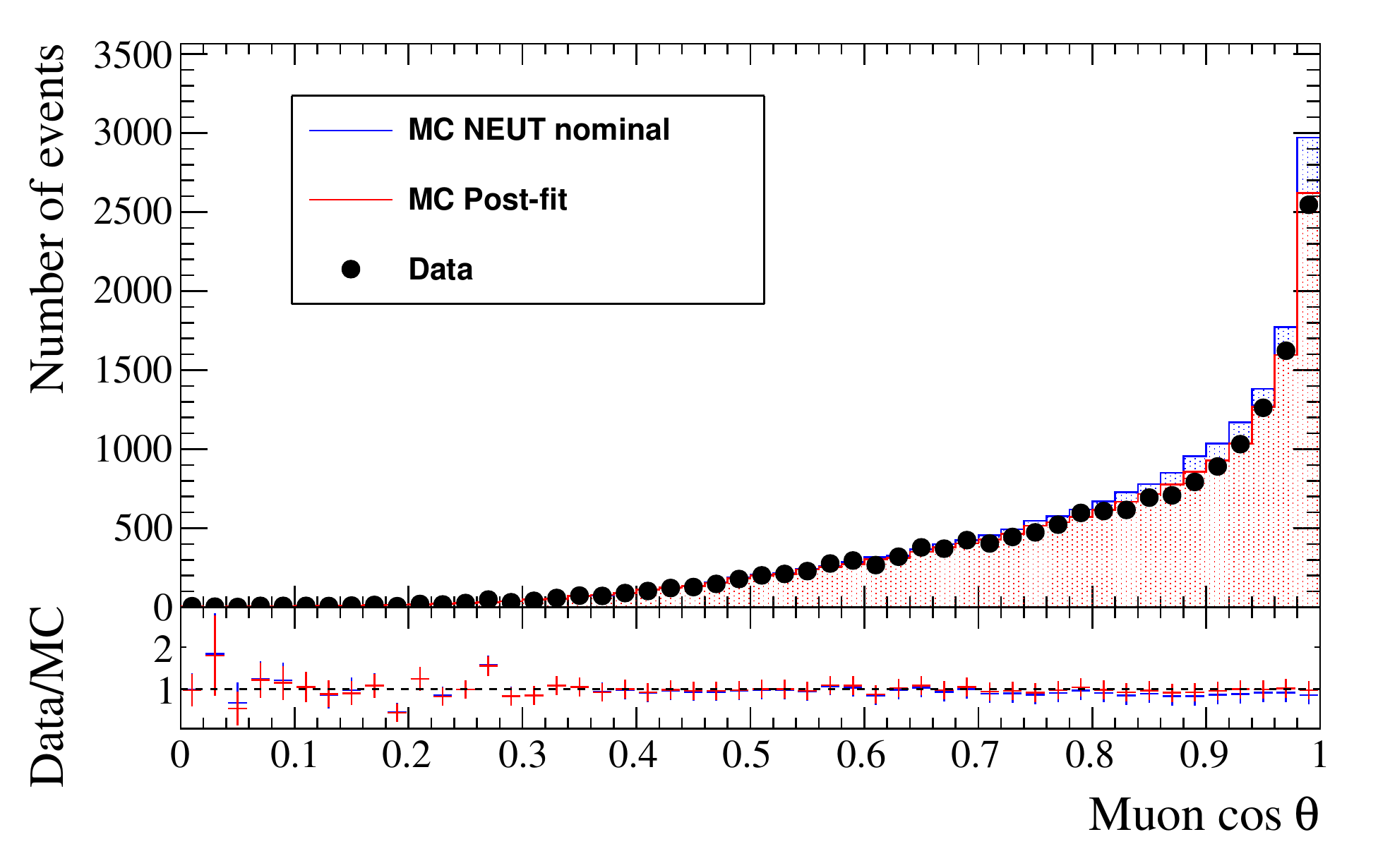} 
   \caption{The momentum (left) and angle (right) distributions for the CC0$\pi$ data set.  The blue histograms are the predicted
distribution before constraining the flux and cross-section models with a fit to the data.  The red histograms are the predicted
distributions after the constraint from the data.}
   \label{fig:fitted_cc0pi}
\end{figure}

The numbers of selected events for $5.9\times10^{20}$ protons on target are shown
in Table~\ref{tab:tracker_numu_events}. These data are fit while allowing the
flux and cross-section model parameters to vary, and the improved agreement in the
modeled event rates and muon kinematic distributions can be seen in Table~\ref{tab:tracker_numu_events}
and Fig.~\ref{fig:fitted_cc0pi}.

\begin{table}
\begin{center}
\caption{The measured and predicted number of events in the ND280 $\nu_{\mu}$ CC enhanced samples.}
\label{tab:tracker_numu_events}
\begin{tabular}{lcccc}
\\ \hline \hline
                                &  CC0$\pi$ & CC1$\pi$  & CC Other  & CC Inclusive  \\  \hline
Data                            &  17369    & 4047      & 4173      &  25589      \\
Model before data constraint    &  19980    & 5037      & 4729      &  29746      \\ \hline
Model after data constraint     &  17352    & 4110      & 4119      &  25581      \\ \hline \hline
\end{tabular}
\end{center}
\end{table}

The constrained flux and cross-section parameters related to the nucleon level cross-section are used
to predict the $\nu_{\mu}$ and $\nu_{e}$ interaction rates at Super-K.  The uncertainties on these 
parameters are reduced by the fit to ND280 data, hence reducing the uncertainties on the predicted
Super-K event rates, as shown in Table~\ref{tab:sk_unc}.  The overall event rate uncertainties for
the $\nu_{\mu}$ and $\nu_{e}$ candidate predictions are reduced from 23.4\% and 27.5\% to 8.1\% and 8.9\% respectively. 

\begin{table}
\begin{center}
\caption{The uncertainties on the SK $\nu_{\mu}$ and $\nu_{e}$ candidate rate predictions.}
\label{tab:sk_unc}
\begin{tabular}{l|cc|cc} \hline \hline
           & \multicolumn{2}{c|}{Before ND280 Data Constraint} & \multicolumn{2}{c}{After ND280 Data Constraint} \\      
Source     & $\nu_{\mu}$ Candidates & $\nu_{e}$ Candidates & $\nu_{\mu}$ Candidates & $\nu_{e}$ Candidates \\ \hline           
Constrained Flux \& Cross-section Param. & 21.6\% & 26.0\%   & 2.7\% & 2.9\% \\
Unconstrained Cross-Section Param.       & 5.9\%  & 7.6\%   & 4.9\% & 7.6\% \\ 
Super-K Modeling Uncertainties               & 6.3\%  & 3.5\%   & 5.6\% & 3.5\% \\ \hline
Total Error                                  & 23.4\% & 27.5\%   & 8.1\% & 8.9\% \\ \hline \hline
\end{tabular}
\end{center}
\end{table}

The ``Unconstrained Cross-section Param." uncertainty in Table~\ref{tab:sk_unc} is dominated by uncertainties in the modeling of the target oxygen nucleus. 
Thus far, the ND280 analyses used in the oscillation measurement have only used interactions in FGD1, which is composed 
entirely CH scintillator bars with no oxygen targets. 
The downstream FGD2 contains layers of water interspersed within its scintillator layers. A simultaneous fit of the interactions in both FGDs 
can provide a constraint on nuclear effects in oxygen, and may potentially reduce the corresponding nuclear model uncertainties. The ultimate 
event samples in both FGDs are shown in Table~\ref{tab:futurerates}. The statistical precision of a subtraction of interactions on scintillator 
from interactions on water is better than 1\%, which is more precise than current detector systematic uncertainties ($\sim$3\%).  Implementing the FGD2 data
to reduce the cross-section modeling uncertainties is a high priority for T2K.  

Additionally, the P0D is capable of operating with and without water targets dispersed throughout its active volume, measuring the event rates separately 
in these two configurations.  A CC $\nu_{\mu}$ event selection with the P0D and downstream TPC can produce samples of forward muons produced in the 
P0D water layers. The expected P0D event rates are given in Table~\ref{tab:futurerates}.

\begin{table}
\caption{The number of selected CC-Inclusive events in FGD1, FGD2, and the P0D are given for the ultimate expected T2K POT ($7.8\times10^{21}$) assuming 50\% $\nu$-mode horn operation and 50\% $\bar{\nu}$-mode. The subset of events that are right-sign interactions (i.e. $\nu$-interactions in $\nu$-mode and $\bar{\nu}$ interactions in $\bar{\nu}$-mode) on water are shown separately.\label{tab:futurerates}}
\begin{tabular}{cccc} \hline \hline
\multicolumn{2}{c}{\multirow{3}{*}{Event Sample}} & Total & Right-Sign Event \\
\multicolumn{2}{c}{} & Event Rate & Rate on Water \\ \hline
\multirow{4}{*}{$\nu$-mode~~} & FGD1 & 169,000 & -- \\
& FGD2 & 166,000 & 84,000   \\
& P0D Water Out & 144,000 & --   \\
& P0D Water In & 210,000 & 66,000 \\ \hline
\multirow{4}{*}{$\bar{\nu}$-mode~~} & FGD1 & 57,000 & --   \\
& FGD2 & 56,000 & 28,000   \\
& P0D Water Out & 63,000 & --   \\
& P0D Water In & 93,000 & 30,000   \\ \hline \hline
\end{tabular}
\end{table}

\begin {figure}[htbp]
  \begin{center}
    \includegraphics[width=0.45\textwidth]{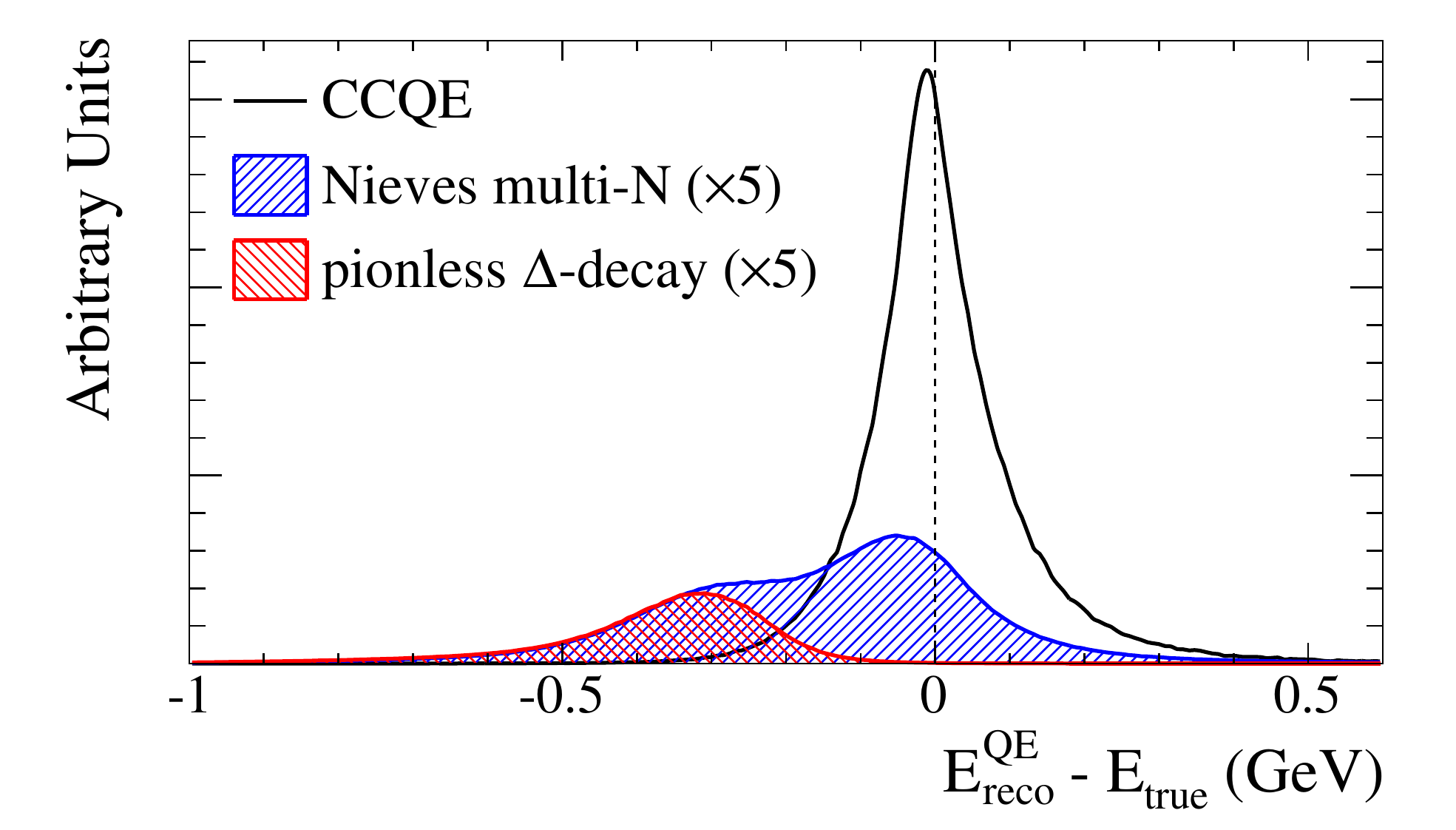}
    \caption{The difference between the energy reconstructed assuming quasi-elastic kinematics and the true energy for
events with no pions in the final state. Black are the NEUT simulation of CCQE events and red are the NEUT simulation of
resonant production where the $\Delta$ is absorbed.  The blue is the from the two-body current calculation of
Nieves {\it et al.}~\cite{PhysRevC.83.045501}.}
    \label{fig:t2k_mec}
  \end{center}
\end {figure}

As discussed in the Section~\ref{sec:ND}, there are large uncertainties in the theoretical modeling of interactions that
involve the ejection of no pions and multiple nucleons.  T2K has studied the potential biases from the mismodeling of these
nuclear effects on the measurements of
$\theta_{23}$ and $\Delta m^{2}_{32}$ in fits to $\nu_{\mu}$ candidates at SK~\cite{Abe:2014ugx}.
Toy data for both ND280 and SK are generated using NEUT with additional
two body current interactions based on the model of Nieves {\it et al.}~\cite{PhysRevC.83.045501}.  In addition to
the Nieves model a second ad-hoc simulation of two-body currents was studied.  This ad-hoc model was chosen to
cover the range of two-body current calculations in the literature.
Fig.~\ref{fig:t2k_mec} illustrates the energy reconstruction
bias from two-body currents in the calculation of Nieves {\it et al.}.

Both the ND280 and SK toy data are fit assuming the NEUT model,
as they are fit in the T2K oscillation analyses, and the biases on $\theta_{23}$ and $\Delta m^{2}_{32}$ are evaluated
when there are additional two-body current contributions in the toy data.  The average and RMS of the oscillation
parameter biases for many toy experiments are shown in Table.~\ref{tab:mec_bias}.  The average bias can be as large as
2.9\% on $\theta_{23}$ while there is an additional variation of the bias by as much as 3.6\%, for individual toy experiments.
This study indicates that the modeling of two-body currents and other nuclear effects can be a dominant systematic effect, even
when ND280 data are used to constrain the neutrino interaction model.  Addressing uncertainties in the modeling of two-body
currents with additional data from the hadronic final states or novel approaches using the final state lepton kinematics will
be necessary to control these uncertainties for future experiments.

\begin{table}
\begin{center}
\caption{The oscillation parameter average bias and RMS bias of toy experiments with
two body current models. }
\label{tab:mec_bias}
\begin{tabular}{l|cc|cc}
\hline \hline
      & \multicolumn{2}{c|}{$\theta_{23}$} & \multicolumn{2}{c}{$\Delta m^{2}_{32}$} \\ 
Model & Bias Mean & Bias RMS & Bias Mean & Bias RMS   \\ \hline
Nieves {\it et al.}~\cite{PhysRevC.83.045501}  & 0.3\%  & 3.6\% & -0.2\% & 0.6\% \\
Ad-hoc Model & -2.9\% & 3.2\% &  0.5\% &  0.6\% \\ \hline \hline
\end{tabular}
\end{center}
\end{table}

\paragraph{ND280 cross checks of CC $\nu_e$ and NC$\pi^{0}$ rates}

Currently, T2K also uses ND280 to make cross-checks on the rate of $\nu_{e}$ CC interactions~\cite{Abe:2014usb} and 
NC$\pi^{0}$ interactions, both important backgrounds for the $\nu_{e}$ appearance measurement at Super-K.

\begin{figure} 
   \centering
   \includegraphics[width=0.45\textwidth]{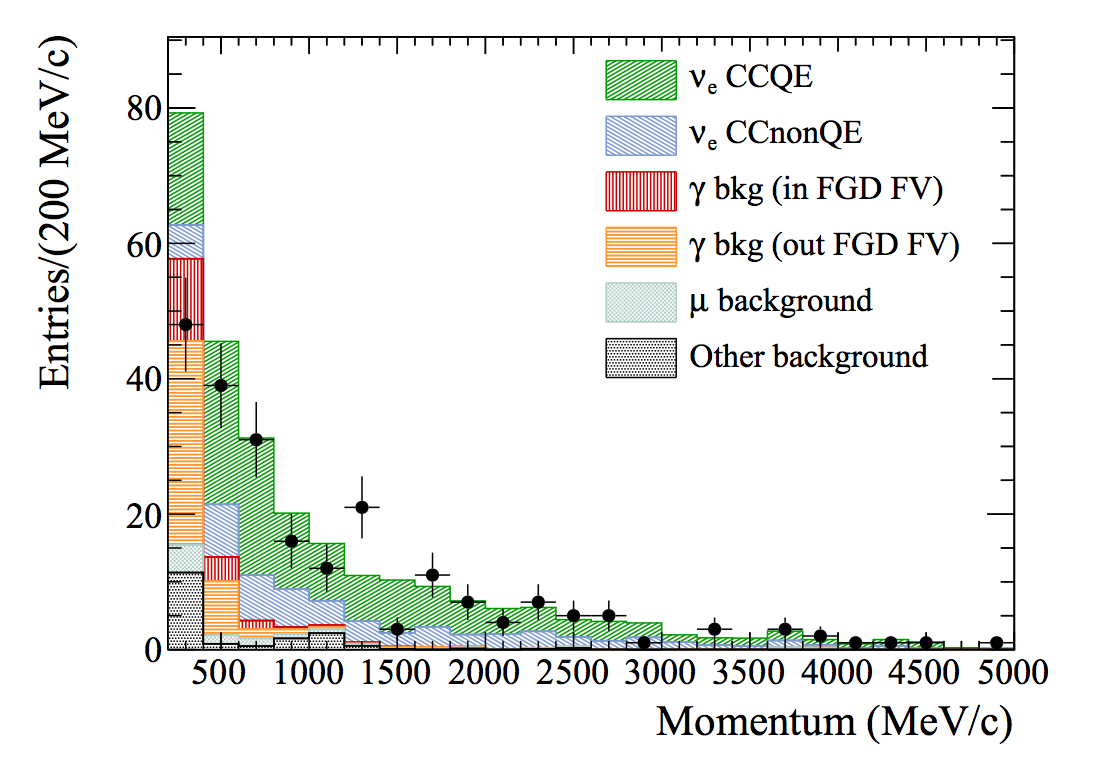}
   \includegraphics[width=0.45\textwidth]{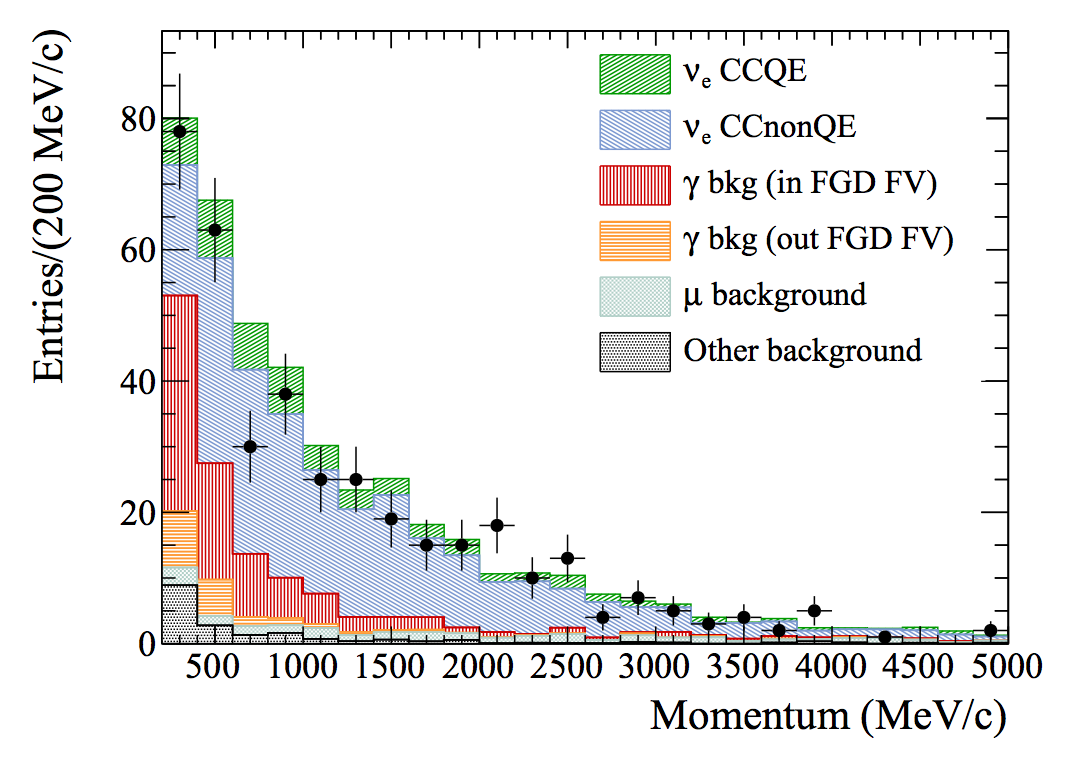}
   \caption{The ND280 CC $\nu_{e}$ candidates in the CCQE-like (left) and CCnonQE-like (right) sub-samples. The predicted rates are
shown in the stacked color histograms.}
   \label{fig:nd280_nue}
\end{figure}

The ND280 $\nu_{e}$ CC candidates are selected in a similar manner to the $\nu_{\mu}$ candidates with the following
changes: candidates from interactions in the downstream FGD2 are included, the ionizing energy loss must be consistent
with an electron, ECAL showers are used in the particle identification when present, a veto on $e^{+}e^{-}$ pairs is applied, and
a veto on events with reconstructed upstream objects is applied to reduce the $\gamma$ background.  The data are broken
into CCQE-like and CCnonQE-like sub-samples, shown in Fig.~\ref{fig:nd280_nue}.  A fit of the flux and cross-section models to the
data provides a ratio of measured $\nu_{e}$ CC interactions to the model prediction, $1.01\pm0.10$ (syst+stat.).  The most relevant
data for the T2K oscillation measurements are interactions of $\nu_{e}$ with $E_{\nu}<1.5$\,GeV. These tend to populate the low 
momentum region where there is a large background from converting photons. Reducing this background to improve the constraint for
the T2K oscillation analysis is a challenge and high priority for ND280.  

ND280 measures the rate of NC$\pi^{0}$ with the P0D detector from a
data set corresponding to $8.55\times10^{19}$ POT.  When normalized to the
ND280 CC $\nu_{\mu}$ data, the measured ratio of the data rate over the model prediction is $0.81 \pm{} 0.15 \mathrm{(stat.)} \pm{} 0.14 \mathrm{(syst.)}$.
This selection of $\pi^{0}$ candidates is limited to the forward region, $cos(\theta_{\pi})>0.6$, whereas Super-K can detect
photons from $\pi^0$ decays with 4$\pi$ coverage.  The current measurement does not separate events on water or other materials in 
the P0D.  Future analysis will attempt to address these limitations in the current measurement.

\subsubsection{Potential ND280 upgrades \label{sec:nd280_upgrade}}

The T2K collaboration is in the process of discussing various upgrade possibilities at the ND280 site. These include the deployment of heavy water ($\mbox{D}_2\mbox{O}$) within the passive water targets in FGD2 that would allow the extraction of neutrino interaction properties on the quasi-free neutron in deuterium via a subtraction with data taken with light water $\mbox{H}_2\mbox{O}$. The use of a water-based liquid scintillator (WbLS) developed at BNL is  being explored in the context of a tracking detector with comparable or finer granularity than the FGD to allow the detailed reconstruction of hadronic system emerging from the neutrino interactions or a larger detector with coarser segmentation that would allow high statistics studies. Either would significantly enhance the study of neutrino interactions on water by reducing the reliance on subtraction and enhancing the reconstruction capabilities relative to the currently deployed passive targets. Finally, a high pressure TPC that can contain various noble gases (He, Ne, Ar) to serve both as the target and tracking medium is being studied. Such a detector would allow the ultimate resolution of the particles emitted from the target nucleus while allowing a study of the $A$-dependence of the cross-sections and final state interactions to rigorously test models employed in neutrino event generators.

While the above options would be deployed within the UA1 magnet, another proposal would place a scintillating tracking detector outside of the magnet on the B2 floor of the NU1 building surrounded by range detectors to measure the muon momentum over a large range of angles. The inner tracking detector would allow passive water and plastic targets to be deployed in order to measure water and CH cross-sections.

\subsection{\label{sec:int_wc} Intermediate water Cherenkov detectors}

Since many of the uncertainties on the modeling of neutrino interactions arise from uncertainties on nuclear effects, the ideal 
near detector should include the same nuclear targets as the far detector.  The ND280 P0D and FGD detectors include passive water 
layers, however extracting water only cross sections requires complicated analyses that subtract out the interactions on other 
materials in the detectors.  An alternative approach is to build a water Cherenkov (WC) near detector to measure the cross section 
on H$_2$O directly and with no need for a subtraction analysis.  This approach was taken by K2K~\cite{Ahn:2006zza} and was proposed 
for T2K~\cite{t2k2km}.   The MiniBooNE experiment has also employed a mineral oil Cherenkov detector at a short baseline to great 
success~\cite{AguilarArevalo:2008qa}.  A WC near detector design is largely guided by two requirements:
\begin{enumerate}
\item The detector should be large enough to contain muons up to the momentum of interest for measurements at the far detector.
\item The detector should be far enough from the neutrino production point so that there is minimal pile-up of interactions in the same beam timing bunch.
\end{enumerate}
These requirements lead to designs for kiloton size detectors located at intermediate distances, 1-2 km from the target, for the J-PARC neutrino beam.  

The main disadvantage of the WC detector is the inability to separate positively and negatively charged leptons, and hence 
antineutrino and neutrino interactions.   This ability is especially important for a CP violation measurement where the wrong 
sign contribution to the neutrino flux should be well understood.  Hence, the WC detector will most likely be used in conjunction 
with a magnetized tracking detector such as ND280.  Recent developments in the addition of Gadolinium (Gd) and Water-based Liquid 
Scintillator (WbLS) compounds to water do raise the possibility to separate neutrino and antineutrino interactions by detecting the 
presence of neutrons or protons in the final state.

Two conceptual designs for possible intermediate WC detectors have been studied and are described in this section.  The Tokai Intermediate Tank for
Unoscillated Spectrum (TITUS) is a 2 kiloton WC detector located about 2 km from the target at the same off-axis angle as the far detector.  
At this baseline the detector sees fluxes for the neutral current and $\nu_e$ backgrounds that are nearly identical to the Hyper-K fluxes.  The 
detector geometry and the presence of a muon range detector are optimized to detect the high momentum tail of the muon spectrum.  The use of 
Gd in TITUS to separate neutrino and antineutrino interactions is being studied.  The $\nu$PRISM detector is located 1 km from the target and is 
50 m tall, covering a range of off-axis angles from 1-4 degrees.  The $\nu$PRISM detector sees a range of neutrino spectra, peaked at energies 
from 0.4 to 1.0\,GeV.  The purpose of $\nu$PRISM is to use these spectra to better probe the relationship between the incident neutrino energy and 
final state lepton kinematics, a part of the interaction model with larger uncertainties arising from nuclear effects.

\subsubsection{The TITUS water Cherenkov detector}

As discussed in the previous text, the challenges in the use
of ND280 data include the measurement of a different flux than at SK, the
limited phase space coverage, and the implementation of analyses to
extract cross-sections on water.  These limitations can be addressed
with a complementary water Cherenkov detector strategically located at
an intermediate distance of about 2~km from the neutrino
production point.

\begin{figure}[htb]
\begin{center}
\includegraphics[height=0.2\textheight]{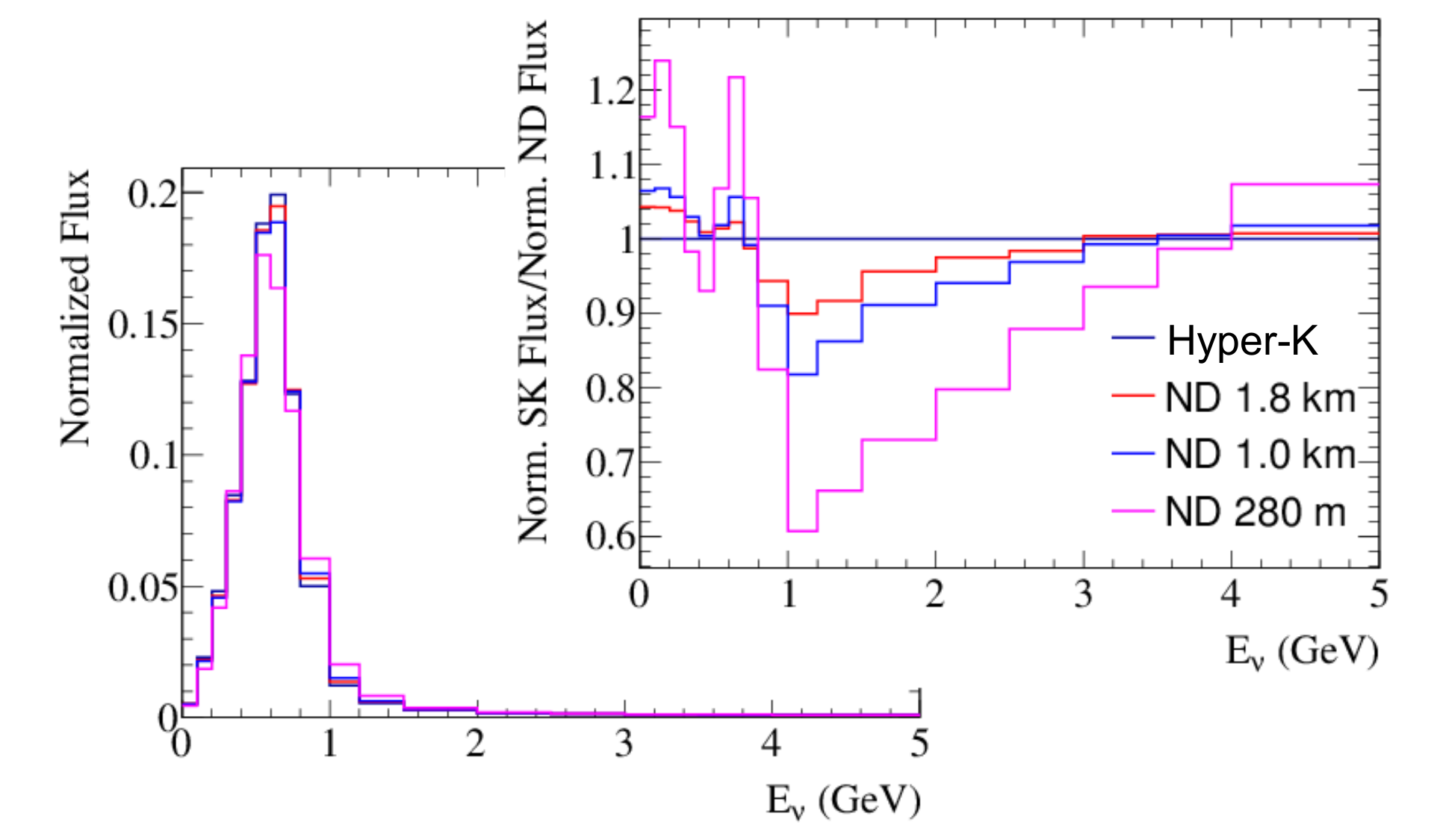}
\includegraphics[height=0.2\textheight]{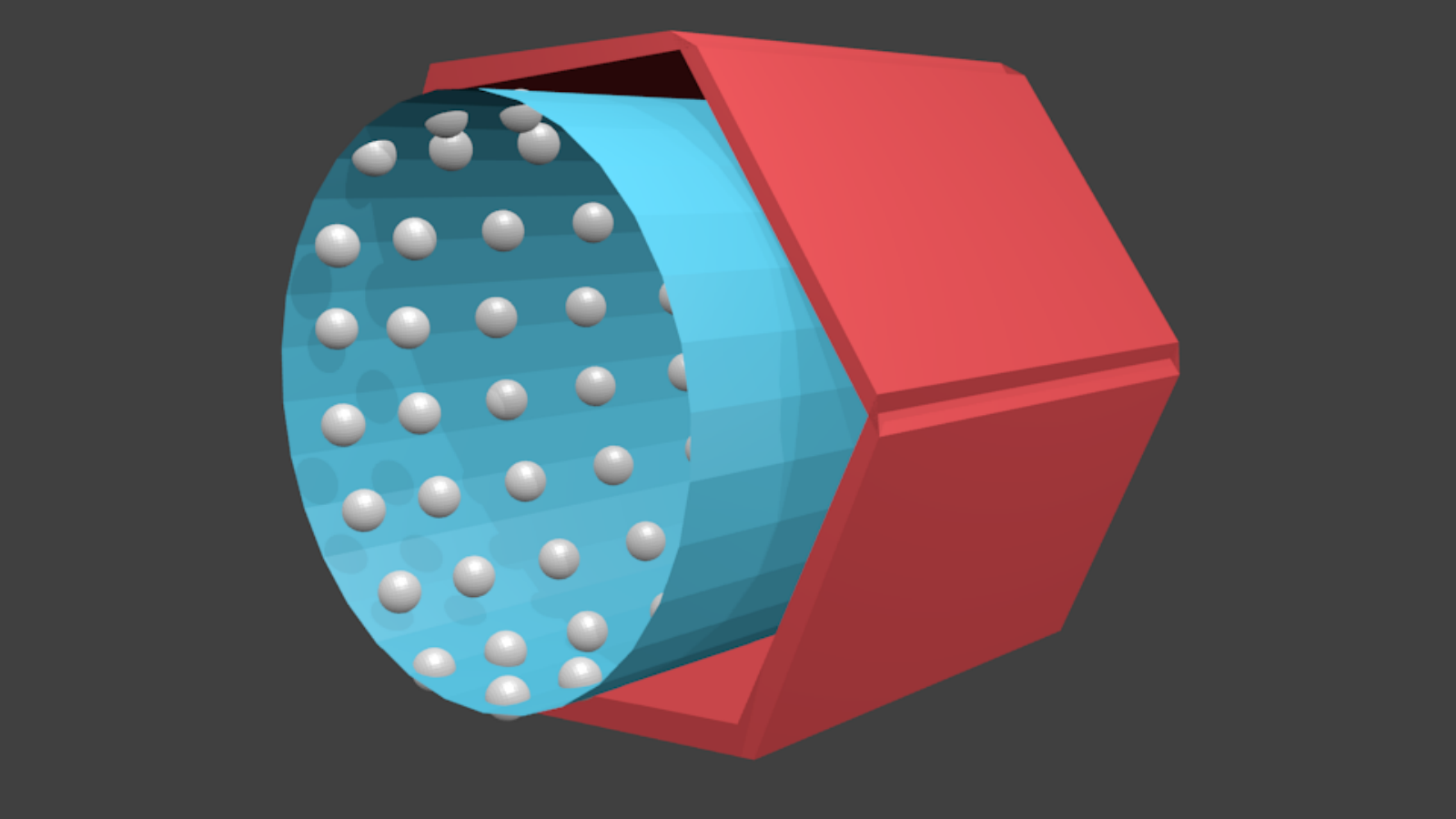}
\end{center}

\caption{\label{fig:titus} Left: neutrino energy spectrum at different
  baselines, and corresponding far-to-near ratio.
  Right: Detector design, consisting of a WC Gd-doped tank (blue) surrounded by a MRD or MIND detector (red).}
\end{figure}

At this distance, the TITUS 
detector sees almost the same spectrum as at Hyper-K. The
maximum difference in shape is $\sim$5\% at the peak energy instead
of almost 20\% with ND280, see Fig.~\ref{fig:titus} (left) for the
ratio. 

The TITUS detector consists of a 2~kton Gd-doped WC tank
(about 5.5~m radius for about 22~m length)
surrounded by a Muon Range
Detector (MRD) or a Magnetized Iron Neutrino Detector (MIND) covering 3/4 of the length of the sides and the
downstream of the detector (see Fig.~\ref{fig:titus}).
Optimization of the dimensions and
shape of the tank and the MRD or MIND is in progress.  The current detector
size was chosen using two main criteria: muon containment and pile-up.

The photosensors planned to be used are both LAPPDs and HPDs.
The LAPPDs will provide excellent time and spatial resolution that
will greatly aid the reconstruction, and will provide very good
identification of the NC$\pi^0$ events that are a major background
for electron neutrino appearance at Hyper-K. 

We seek to take advantage of the ANNIE
experiment~\cite{Anghel:2014sia}, should it be funded and running in
the next few years, to provide an additional environment for testing.  The
ANNIE experiment has a similar configuration, but smaller size, than
TITUS.

%

%

The number of events observed by the detector at 2 km are shown in Table~\ref{tab:TITUSevtrt}.

\begin{table}
\begin{center}
\caption{Expected true neutrino interaction event rates at TITUS for fiducial volumes (FV) of 2~kton and 1.17~kton for the full proposed beam run in neutrino- and anti-neutrino-enhanced modes (3.9$\times 10^{21}$ and 11.7$\times 10^{21}$ POT, respectively).  Coherent pion production is in the ``Other'' category; resonant pion production is in 1$\pi$.  The category labeled MEC are multinucleon ejection events modeled based on the calculations of Nieves {\it et. al.}~\cite{PhysRevC.83.045501}}
\begin{tabular} { l|cccc|ccc|cc|cc|cc}
\hline \hline
             & \multicolumn {4}{|c|}{$\nu_{\mu}$ CC ($10^{4}$) } &  \multicolumn {3}{|c|}{$\overline{\nu}_{\mu}$ CC ($10^{4}$)} & \multicolumn {2}{|c|}{$\nu_{e}$ CC ($10^{4}$) }  & \multicolumn {2}{|c|}{$\overline{\nu}_{e}$ CC ($10^{4}$) }  & \multicolumn {2}{|c}{NC ($10^{4}$)}  \\
Interaction: & QE & MEC & 1$\pi$ & Other & QE & MEC &  Other & QE & Other & QE & Other & $\pi^{0}$ & Other \\\hline 
+320kA (FV = 2~kton) & 428 & 72.5 & 236 & 143 & 8.37 & 2.30 & 12.4 & 5.26 & 13.8 & 0.411 & 1.07 & 55.8 & 249 \\
+320kA (FV = 1.17~kton) & 240 & 40.6 & 132 & 79.8& 4.68 & 1.29 & 6.95 & 2.95 & 7.75 & 0.230 & 5.97 & 31.3 & 139 \\\hline
-320kA (FV = 2~kton) & 93.0 & 20.5 & 99.9 & 122 & 276 & 62.8 & 184 & 3.41 & 13.1 & 3.95 & 7.15 & 59.8 & 307 \\
-320kA (FV = 1.17~kton) & 52.0 & 11.5 & 55.9 & 68.5 & 154 & 35.1 & 103& 1.91 & 7.35 & 2.21 & 4.00 & 33.5 & 172 \\
\hline
\end{tabular}
\label{tab:TITUSevtrt}
\end{center}
\end{table}

\begin{figure}[htb]
\begin{center}
\includegraphics[width=0.45\textwidth]{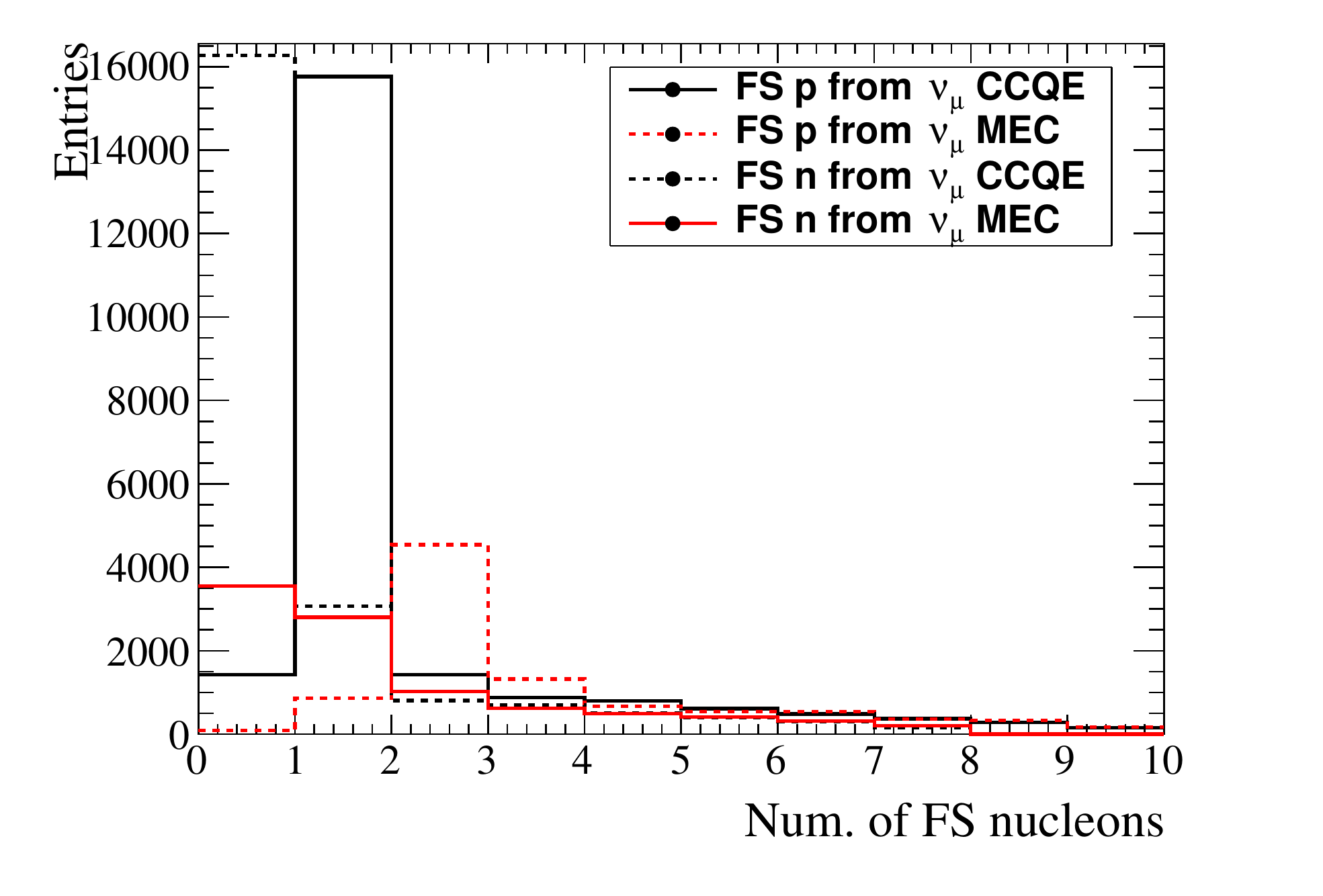}
\includegraphics[width=0.45\textwidth]{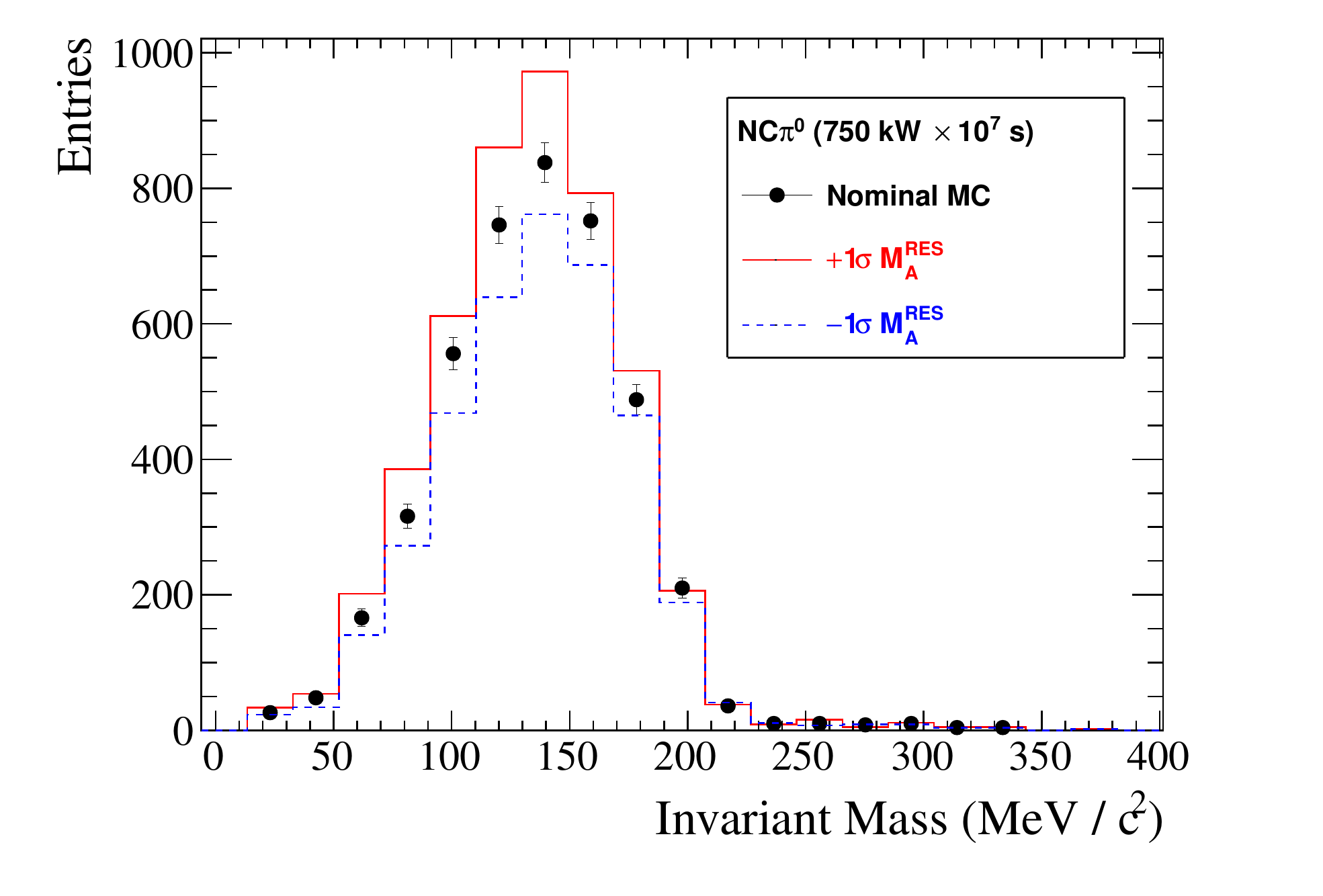}
\end{center}
\caption{\label{fig:titussens} Left: nucleon multiplicity for interaction final states. 
Right: reconstructed $\pi^0$ mass from NC$\pi^0$ events. }
\end{figure}

Adding Gd to the water~\cite{Beacom:2003nk} provides TITUS with excellent neutron tagging
capabilities.
With a 0.1\% concentration, $\sim$90\% of neutrons will
capture on Gd, producing a 8\,MeV gamma cascade of typically 2-3 gammas
from neutrino capture, resulting in sufficient optical light to be
detected in the volume.  Tagging events by the presence and number
of final-state neutrons provides a unique capability to discriminate
between different species of neutrino interactions (e.g. CCQE vs MEC
separation, NC versus CC separation, $\overline{\nu}/\nu$,
$\nu_{e}/\nu_{\mu}$). 
For instance, Fig.~\ref{fig:titussens} (left)
shows the current nucleon multiplicity prediction after FSI, 
assuming that the $n$-$p$ pair is 80\% and dominant, 
where the $n$-$n$ and $p$-$p$ pairs are 10\% each, 
as nuclear theorists speculate~\cite{Martini:2009uj} and
partially supported by electron scattering data~\cite{Piasetzky_pair}.
The error on the FSI neutrons is of the order of 3\%(33\%) for CCQE(MEC) interactions on water.
As Fig.~\ref{fig:titussens} shows, different interaction types have different nucleon multiplicities and 
counting nucleons gives an additional handle to study them.  
This would improve our knowledge on neutrino cross-sections, 
and eventually reduce the error on the far detector measurements 
coming from neutrino cross-sections.

The thermal neutrons, in particular the spallation neutrons from
cosmic rays, are potential backgrounds that can change the neutron
counting scheme. From an initial investigation based on the Super-K analysis,
they can be strongly suppressed by tagging the parent cosmic
rays. The neutron background coming from the beam totals
about one per particle/bunch (including also the interactions with the
surrounding rock), that can be further reduced by a selection.

The Gd-doped design of TITUS makes possible the
characterization of $\nu$ versus $\overline{\nu}$, significantly reducing the error on the $\overline{\nu}/\nu$ ratio.
The anti-neutrino interactions have higher neutron multiplicities than neutrino interactions.  

The detector will allow NC$\pi^0$ events to be clearly
identified. Within one year of running the uncertainty on the NC$\pi^0$ rate
will be known better than the current error on the axial mass, see
Fig.~\ref{fig:titussens} (right).

The observed $\nu_e/\nu_{\mu}$ candidate ratio in TITUS is used to estimate the
$\nu_{e}$ beam contamination and to constrain uncertainties on the
relative reaction cross-sections of  $\nu_e$ and $\nu_{\mu}$.
Using a selection similar to the current T2K selection
at SK for $\nu_e$ and $\nu_{\mu}$ shows that a statistical precision of
1-2\% on the measurement of the $\nu_e$ rate can be achieved.  The complete
uncertainty on the interaction rate will depend on the uncertainties in 
the muon and neutral current backgrounds.  The uncertainty on the reaction
cross-section measurement will also depend on the flux uncertainty that can be achieved. 

Finally, there are a range of further important studies that the
detector will be able to address. In particular, the measurement of the
neutron rate, that is a crucial background in the proton decay search;
supernova neutrinos, TITUS can be included in the SNEWS (SuperNova
Early Warning system); 
reactor neutrinos, under the assumption of a
reactor operating close-by, this detector would be able to measure the
reactor rate.

\subsubsection{\label{sec:nuprism} The $\nu$PRISM detector}

\begin {figure}[htbp]
  \begin{center}
    \includegraphics[width=0.35\textwidth]{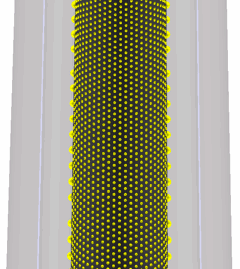}
    \caption{An illustration of a $\nu$PRISM detector segment showing the vertical cylinder geometry with inward facing photo-multiplier tubes
viewing an inner detector and outward facing photo-multiplier tubes viewing an outer detector.  At a baseline of 1 km, the $\nu$PRISM cylinder would
be 50 m tall.}
    \label{fig:nuprism_draw}
  \end{center}
\end {figure}

\begin {figure}[htbp]
  \begin{center}
    \includegraphics[width=0.43\textwidth]{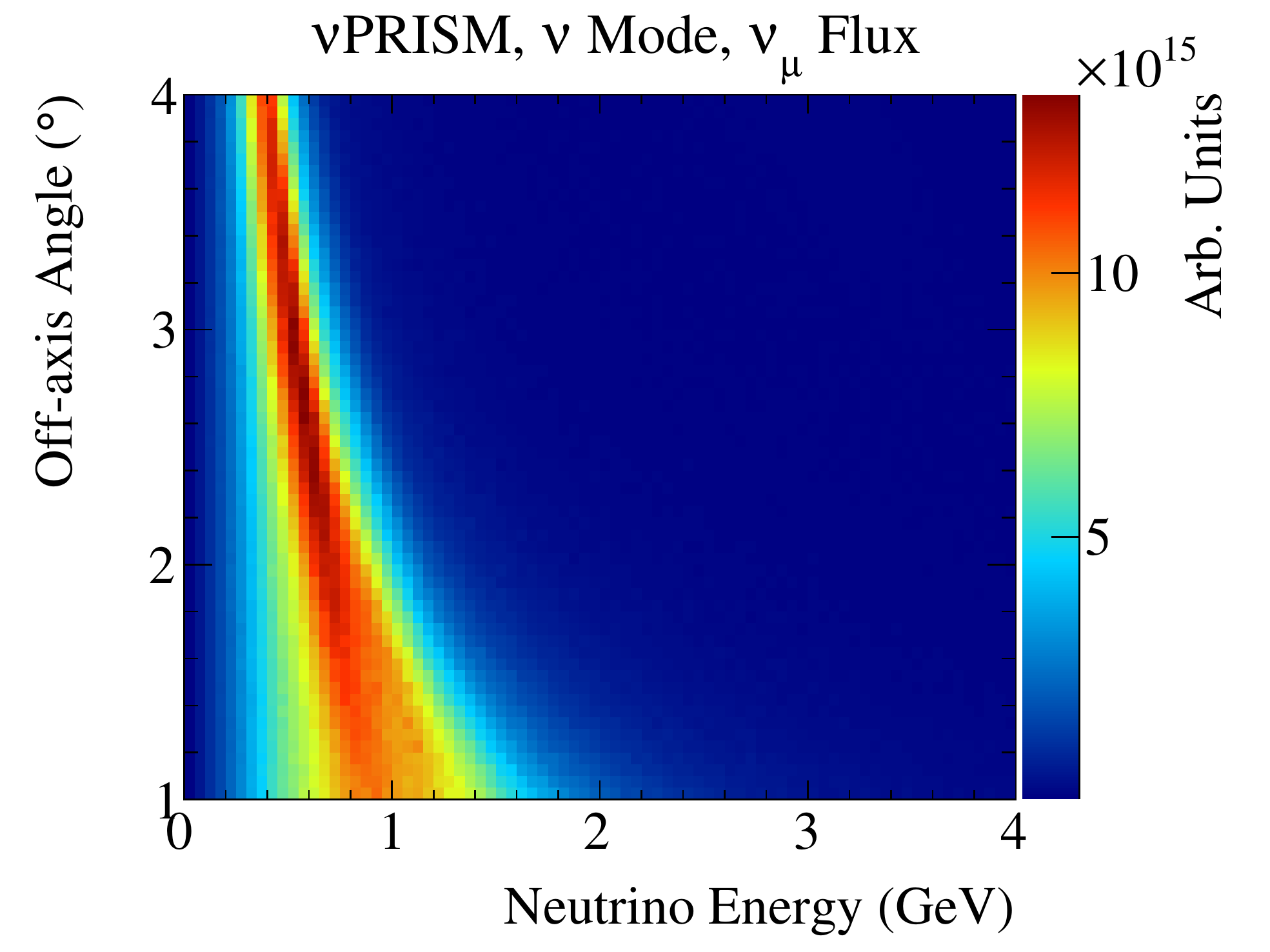}
    \includegraphics[width=0.43\textwidth]{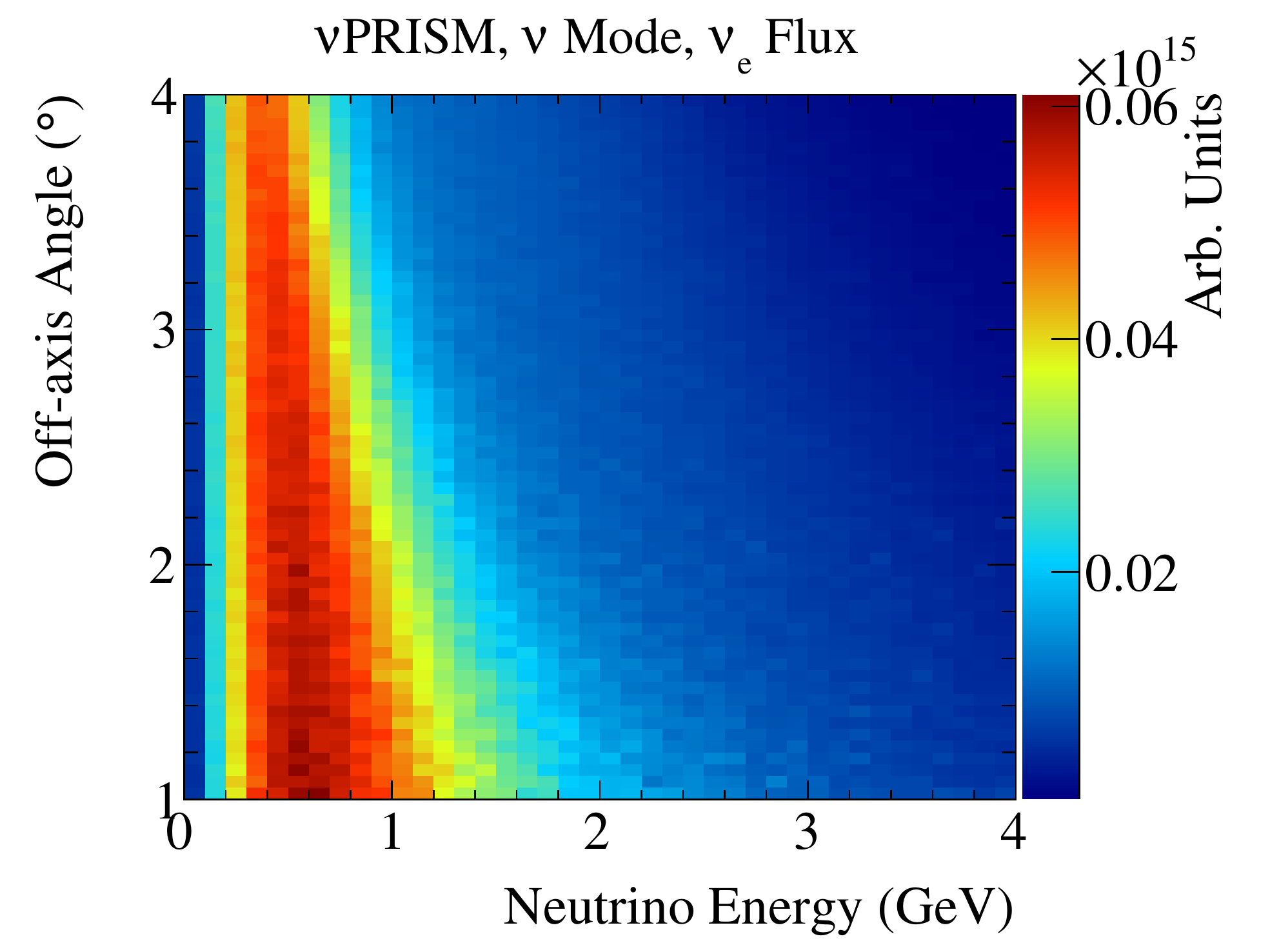}
    \caption{The predicted $\nu_{\mu}$ (left) and $\nu_{e}$ (right) flux for the neutrino enhanced beam as a function of the off-axis angle at the $\nu$PRISM detector.}
    \label{fig:offaxis_flux}
  \end{center}
\end {figure}

The problem of determining the relationship between neutrino energy and lepton kinematics in CC0$\pi$
interactions could be easily solved if mono-energetic beams of neutrinos could be produced at 
$\mathcal{O}(1\text{\,GeV})$.
While mono-energetic beams cannot be produced, beams of varying peak energy can be produced by changing 
the off-axis angle of the beam.  Fig.~\ref{fig:offaxis_flux} shows how the neutrino flux varies from an off-axis 
angle of 1$^{\circ}$ to 4$^{\circ}$.  The $\nu$PRISM detector, as illustrated in Fig.~\ref{fig:nuprism_draw},  would consist of a vertical column
water cherenkov detector located $\sim1$ km from the neutrino production point, and extending over a 
3-4$^{\circ}$ range of off-axis angles.  Using $\nu$PRISM and the neutrino flux prediction, it is
possible to detect interactions from a variety of neutrino spectra by identifying the off-axis angles using 
the location of the interaction
vertices in the detector.  Hence the dependence of final state lepton kinematics on neutrino energy 
can be studied with a single detector and a single neutrino beam.  

The expected detected event rates for single-ring lepton candidates in $\nu$PRISM with 
neutrino mode and antineutrino mode beams are shown in Table~\ref{tab:nuprism_rates}.  Pure, high statistics samples of charged current 
$\nu_{\mu}$ candidates can be detected.  When operating with the antineutrino enhanced beam, the purity is
reduced due to the neutrino background, however the properties of this background can be well constrained by the
neutrino flux prediction and $\nu$PRISM measurements made with the neutrino enhanced beam.

The $\nu$PRISM detector is also well optimized to study the $\nu_{e}$ contamination in the beam from muon and
kaon decays.  The $\nu$PRISM measurement of $\nu_{e}$ candidates at 2.5$^{\circ}$ off-axis angle can be used to 
predict the expected $\nu_{e}$ background rate at Hyper-K.  The $\nu_{e}$ candidates in $\nu$PRISM can also be used
to make measurements of the $\nu_e$ cross-section at $\mathcal{O}(1\text{\,GeV})$.
Given recent improvements to the SK reconstruction
that reduce the misidentification of muons or $\pi^{0}$s as electrons, it is possible to select $\nu_{e}$ candidate
samples in $\nu$PRISM with $>70\%$ purity of $\nu_{e}$ charged current interactions.  Even higher purities may be achieved
by optimizing the granularity of the PMTs used in $\nu$PRISM and optimizing the event reconstruction and selection.
 As indicated in Table~\ref{tab:nuprism_rates},
the highest purity can be achieved at larger off-axis angles, where the background of NC$\pi^0$ reactions is reduced
due to the decrease in the high energy $\nu_{\mu}$ flux.  With high purity samples of $1\times10^3-1\times10^4$ events,
$\nu$PRISM has the potential to measure the $\nu_e$ interaction cross-section relative to the $\nu_{\mu}$ interaction cross-section
to better than $10\%$ precision, depending on the flux and reconstruction uncertainties that can be achieved.

\begin{table}
\begin{center}
\caption{The event rates and purities for single muon-like ring and single electron-like ring selections
for $3.9\times 10^{21}$ ($11.7\times 10^{21}$) POT in the $\nu$PRISM detector with neutrino (antineutrino) mode.}
\label{tab:nuprism_rates}
\begin{tabular}{ccccc}
\hline \hline
      & \multicolumn{2}{c}{1 Ring $\mu$} & \multicolumn{2}{c}{1 Ring e} \\ 
~~Off-axis Angle ($^{\circ}$)~~ & Candidates & CC $\nu_{\mu}$($\bar{\nu}_{\mu}$) Purity & Candidates & CC $\nu_{e}$($\bar{\nu}_{e}$) Purity   \\ \hline
1.0-2.0                        & $3.42\times10^6(3.06\times10^{6})$  & 97.5\%(84.7\%) & $2.56\times10^4(2.95\times10^4)$ &  45.8\%(27.1\%) \\
2.0-3.0                        & $1.76\times10^6(1.65\times10^{6})$  & 97.7\%(81.8\%) & $1.36\times10^4(1.66\times10^4)$ &  67.2\%(38.0\%) \\
3.0-4.0                        & $7.85\times10^5(8.02\times10^{5})$  & 97.2\%(76.2\%) & $7.91\times10^3(1.09\times10^4)$ &  74.9\%(40.1\%) \\ \hline \hline
\end{tabular}
\end{center}
\end{table}

\section{Physics Sensitivities} 
\label{sec:physics_sensitivities}

\subsection{Overview}
A comparison of muon-type to electron-type transition probabilities between neutrinos and anti-neutrinos is 
one of the most promising methods to observe the lepton $CP$ asymmetry.
Recent observation of a nonzero, rather large value of $\theta_{13}$~\cite{Abe:2011sj, Abe:2011fz,Ahn:2012nd,An:2012eh} makes 
this exciting possibility more realistic.

\begin{figure}[b]
\includegraphics[width=0.45\textwidth]{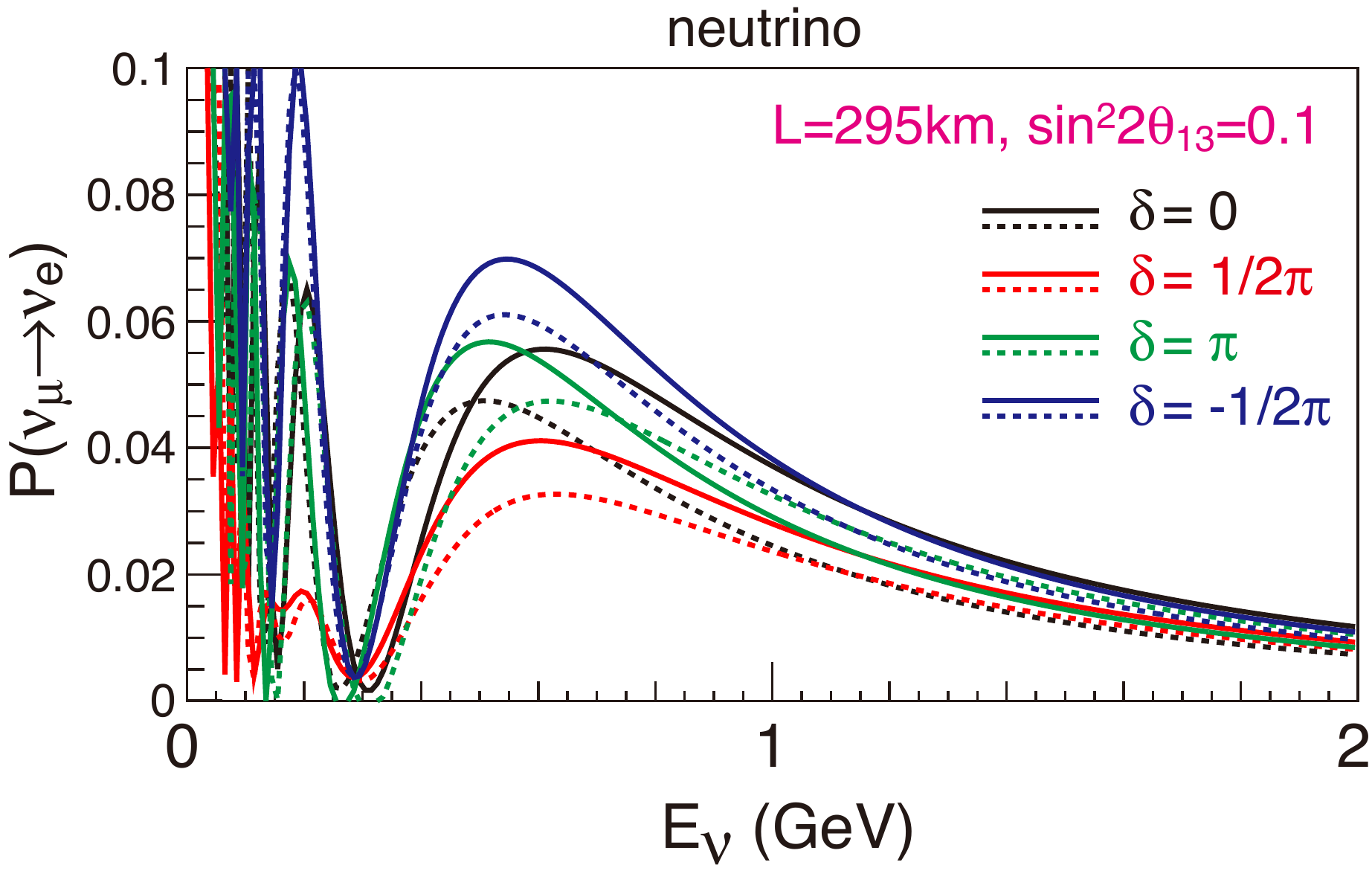}
\includegraphics[width=0.45\textwidth]{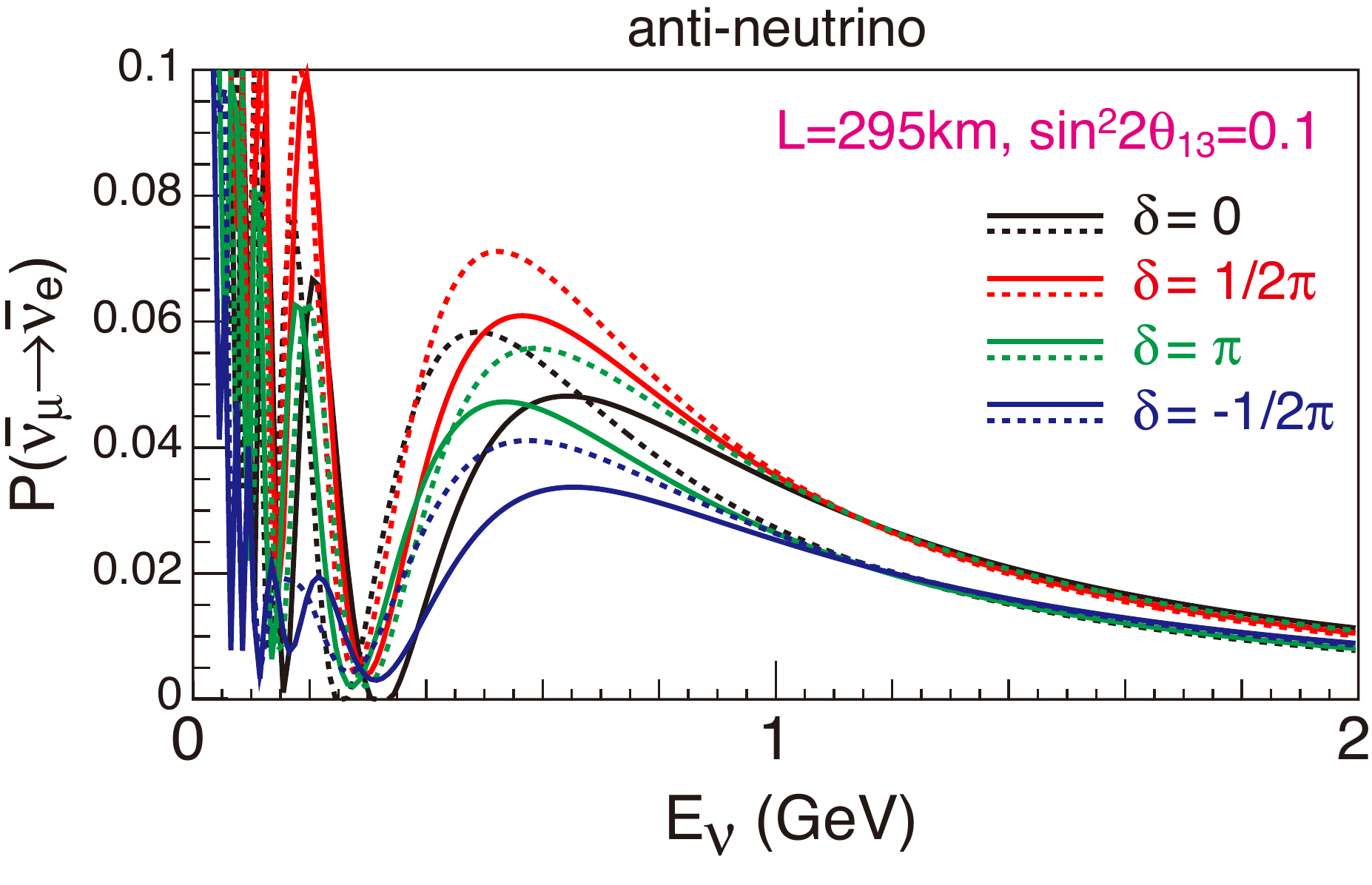}
\caption{Oscillation probabilities as a function of the neutrino energy for $\numu \to \nue$ (left) and $\numubar \to \nuebar$ (right) transitions with L=295~km and $\sin^22\theta_{13}=0.1$. 
Black, red, green, and blue lines correspond to $\deltacp = 0, \frac{1}{2}\pi, \pi$, and $-\frac{1}{2}\pi$, respectively.
Solid (dashed) line represents the case for a normal (inverted) mass hierarchy.
\label{fig:cp-oscpob}}
\end{figure}

Figure~\ref{fig:cp-oscpob} shows the $\numu \to \nue$ and $\numubar \to \nuebar$ oscillation probabilities as a function of the true neutrino energy for a baseline of 295~km.
The cases for $\deltacp = 0, \frac{1}{2}\pi, \pi$, and $-\frac{1}{2}\pi$, are overlaid. 
Also shown are the case of normal mass hierarchy ($\Delta m^2_{32}>0$) with solid lines and inverted mass hierarchy ($\Delta m^2_{32}<0$) with dashed lines.
The oscillation probabilities depend on the value of $\deltacp$, and by comparing the neutrinos and anti-neutrinos,
one can see the effect of $CP$ violation.
There are sets of different mass hierarchy and values of $\deltacp$ which give similar oscillation probabilities.
This is known as the degeneracy due to unknown mass hierarchy and may introduce an ambiguity 
if we do not know the true mass hierarchy.

\begin{figure}[tbp]
\includegraphics[width=0.45\textwidth]{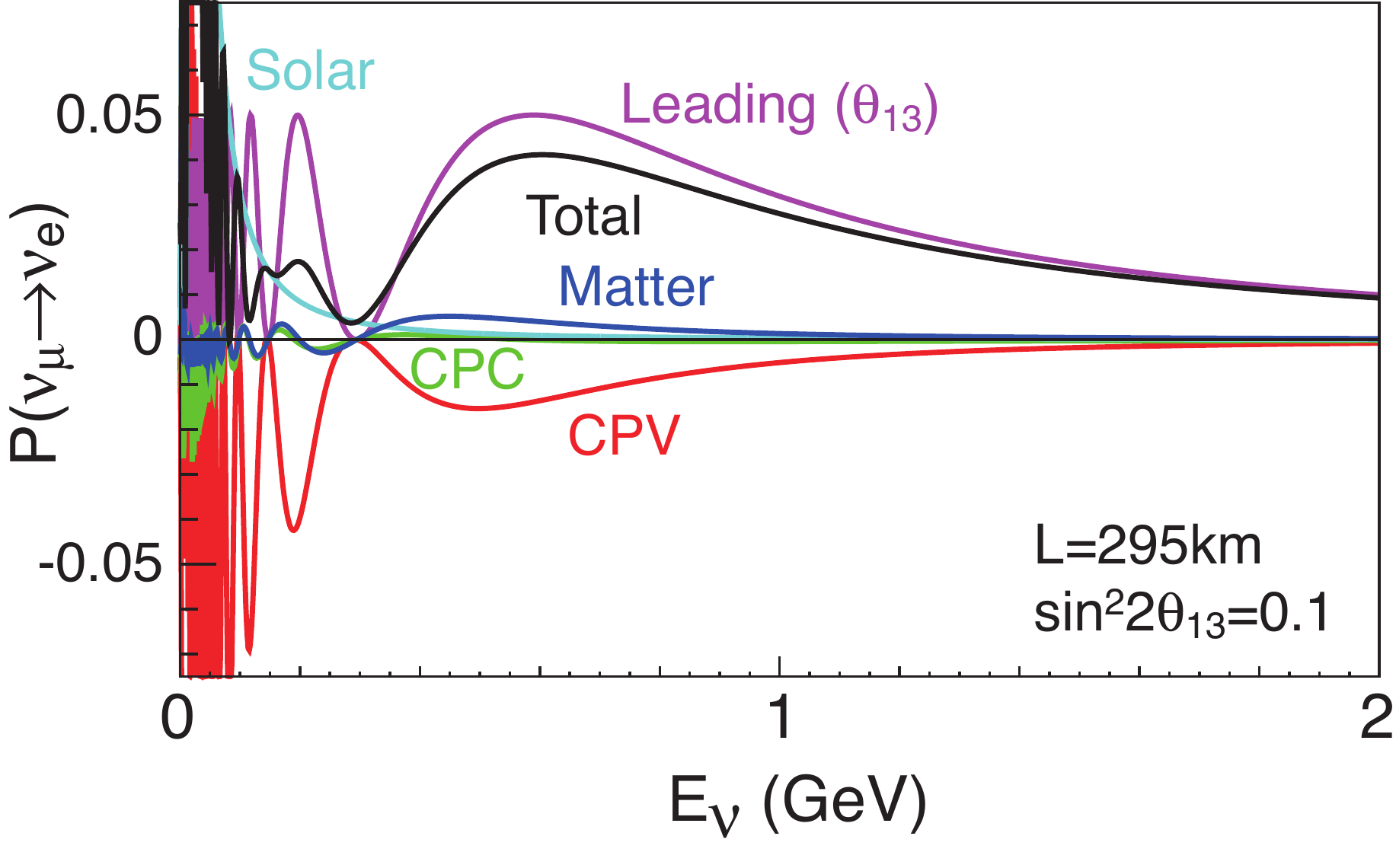}
\caption{Oscillation probability of $\numu \to \nue$  as a function of the neutrino energy with a baseline of 295~km.  $\sin^22\theta_{13}=0.1$
$\deltacp = \frac{1}{2}\pi$ and normal hierarchy is assumed.
Contribution from each term of the oscillation probability formula is shown separately.
\label{fig:cp-oscpob-bd}}
\end{figure}

Because there are a number of experiments planned to determine mass hierarchy
in the near future as shown in Table~\ref{tab:intro:2020},
it is expected that the mass hierarchy will be determined by the time Hyper-K
starts to take data.
If not, Hyper-K itself has a sensitivity to the mass hierarchy by the atmospheric neutrino measurements as shown in Table~\ref{tab:intro:phys}.
Furthermore, a combined analysis of the accelerator and atmospheric neutrino data in Hyper-K will enhance the sensitivity as shown in Sec.~\ref{sec:lbl-atm}.
Thus, the mass hierarchy is assumed to be known in this analysis, unless otherwise stated.

Figure~\ref{fig:cp-oscpob-bd} shows the contribution from each term of the $\numu \to \nue$ oscillation probability formula, Eq.(\ref{Eq:cpv-oscprob}).
For $\sin^22\theta_{13}=0.1$, $\sin^22\theta_{23}=1.0$, and $\deltacp = \pi/2$ with normal mass hierarchy, the contribution from the leading term, the $CP$ violating ($\sin\deltacp$) term, and the matter term to the $\numu \to \nue$ oscillation probability at 0.6\,GeV neutrino energy are 0.05, $-0.014$, and 0.004, respectively.
The effect of $CP$ violating term can be as large as 27\% of the leading term.
Due to the relatively short baseline and thus lower neutrino energy at the oscillation maximum, 
the contribution of the matter effect is smaller for the J-PARC to Hyper-Kamiokande experiment
compared to other proposed experiments like LBNE in the United States~\cite{Adams:2013qkq}.

In the previous study performed in 2011~\cite{Abe:2011ts}, the sensitivity was evaluated for a range of $\theta_{13}$ values
because the exact value of $\theta_{13}$ was not known at that time, 
although T2K collaboration had already reported an indication of electron neutrino appearance~\cite{Abe:2011sj}.
Now that the value of $\theta_{13}$ is known more precisely thanks to the reactor experiments~\cite{An:2012eh,An:2013zwz,Abe:2011fz,Abe:2014lus,Ahn:2012nd}, the sensitivity has been revised with the latest knowledge of the oscillation parameters.
In addition, the analysis method has been updated using a framework developed for the sensitivity study by T2K reported in~\cite{T2KPACreport}.
A binned likelihood analysis based on the reconstructed neutrino energy distribution is performed using both \nue\ (\nuebar) appearance and \numu\ (\numubar) disappearance samples simultaneously.
In addition to $\sin^22\theta_{13}$ and $\deltacp$, other parameters that were fixed in the previous study, $\sin^2\theta_{23}$ and $\Delta m^2_{32}$,  are also included in the fit.
Table~\ref{Tab:oscparam} shows the nominal oscillation parameters used in the study presented in this document, and the treatment during the fitting.
Systematic uncertainties are estimated based on the experience and prospects of the T2K experiment, and implemented as a covariance matrix which takes into account the correlation of uncertainties.

An integrated beam power of 7.5~MW$\times$10$^7$~sec is assumed in this study. It corresponds to $1.56\times10^{22}$ protons on target with 30\,GeV J-PARC beam.
The ratio of neutrino and anti-neutrino running time is assumed to be 1:3 so that the expected number of events are approximately the same for neutrino and anti-neutrino modes.

\begin{table}[htdp]
\caption{Oscillation parameters used for the sensitivity analysis and treatment in the fitting. The \textit{nominal} values are used for figures and numbers in this section, unless otherwise stated.}
\begin{center}
\begin{tabular}{cccccccc} \hline \hline
Parameter & $\sin^22\theta_{13}$ & $\deltacp$ & $\sin^2\theta_{23}$ & $\Delta m^2_{32}$ & mass hierarchy & $\sin^22\theta_{12}$ & $\Delta m^2_{12}$ \\ \hline 
Nominal & 0.10 & 0 & 0.50 & $2.4\times10^{-3}~\mathrm{eV}^2$ & Normal or Inverted & $0.8704$ & $7.6\times10^{-5}~\mathrm{eV}^2$ \\ 
Treatment & Fitted & Fitted & Fitted & Fitted & Fixed & Fixed & Fixed \\ \hline \hline
\end{tabular}
\end{center}
\label{Tab:oscparam}
\end{table}%

\subsection{Expected observables at Hyper-K}\label{sec:sens-events}
The neutrino flux presented in Sec.~\ref{sec:nuflux} is used as an input to the simulation.
Interactions of neutrinos in the Hyper-K detector are simulated with the NEUT program library~\cite{hayato:neut,Mitsuka:2007zz,Mitsuka:2008zz}, which is used in both Super-K and T2K.
The response of the detector is simulated using the Super-K full Monte Carlo simulation 
based on the GEANT3 package~\cite{Brun:1994zzo}.
The simulation is based on the SK-IV configuration with the upgraded electronics and DAQ system.
Events are reconstructed with the Super-K reconstruction software.
As described in Sec.~\ref{sec:det_performance}, the performance of Hyper-K detector for neutrinos with J-PARC beam energy is expected to be similar to that of Super-K.
Thus, the Super-K full simulation gives a realistic estimate of the Hyper-K performance.

The criteria to select \nue\ and \numu\ candidate events are based on those developed for and established with the Super-K and T2K experiments.
Fully contained (FC) events with a reconstructed vertex inside the fiducial volume (FV) and visible energy ($E_\mathrm{vis}$) greater than 30\,MeV are selected as FCFV neutrino event candidates.
In order to enhance charged current quasielastic (CCQE, $\nu_l + n \rightarrow l^- + p$ or $\overline{\nu}_l + p \rightarrow l^+ + n$) interaction, a single Cherenkov ring is required.

Assuming a CCQE interaction, the neutrino energy  ($E_\nu ^{\rm rec}$) is reconstructed from the energy of the final state charged lepton ($E_\ell$) and the angle between the neutrino beam and the charged lepton directions ($\theta_\ell$) as
\begin{eqnarray}
E_\nu ^{\rm rec}=\frac {2(m_n-V) E_\ell +m_p^2 - (m_n-V)^2 - m_\ell^2} {2(m_n-V-E_\ell+p_\ell\cos\theta_\ell)},
\label{eq:Enurec}
\end{eqnarray}
where $m_n, m_p, m_\ell$ are the mass of neutron, proton, and charged lepton, respectively, $p_\ell$ is the charged lepton momentum, and $V$ is the nuclear potential energy (27\,MeV).

\begin{figure}[tbp]%
\includegraphics[width=0.45\textwidth]{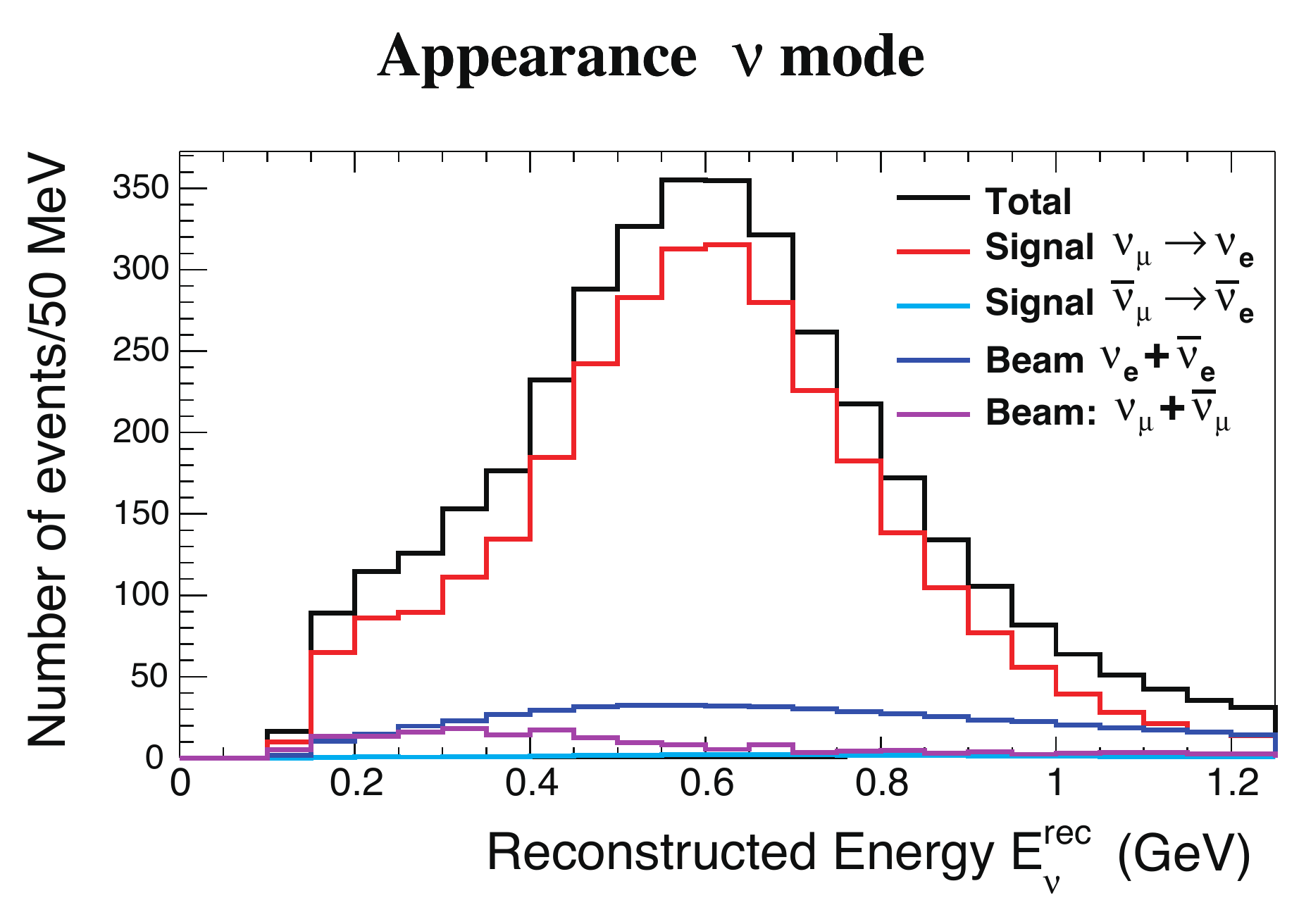}
\includegraphics[width=0.45\textwidth]{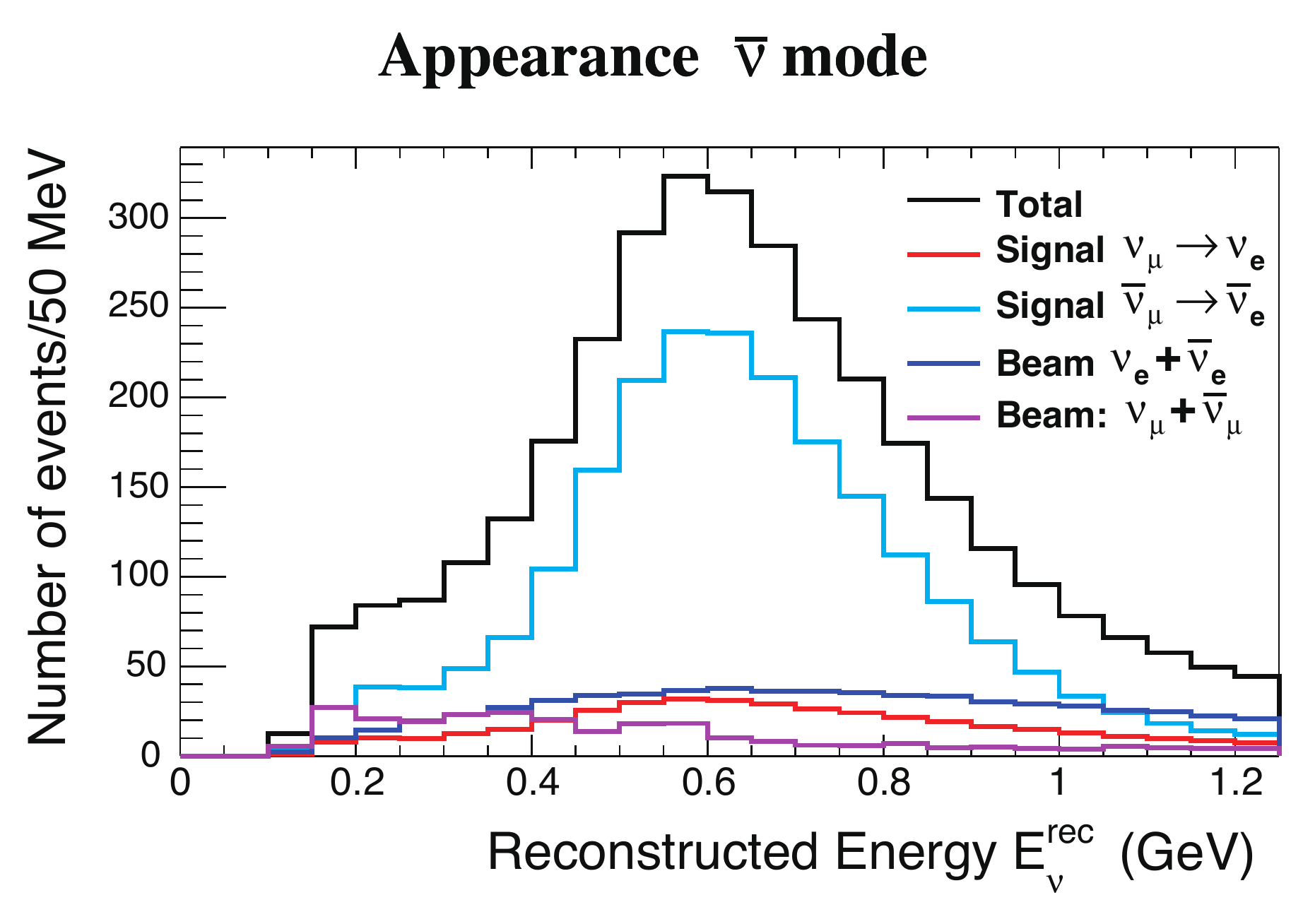}\\
\caption{
Reconstructed neutrino energy distribution of the $\nue$ candidate events.
\label{Fig:sens-enurec-nue}
}
\end{figure}

Then, to select \nue/\nuebar\ candidate events the following criteria are applied;
the reconstructed ring is identified as electron-like ($e$-like),
$E_\mathrm{vis}$ is greater than 100\,MeV,
there is no decay electron associated to the event, and
$E_\nu^\mathrm{rec}$ is less than 1.25\,GeV.
Finally, in order to reduce the background from mis-reconstructed $\pi^0$ events, additional criteria using a reconstruction algorithm recently developed for T2K (fiTQun, see Sec.~\ref{sec:det_performance}) is applied.
With a selection based on the reconstructed $\pi^0$ mass and the ratio of the best-fit likelihoods of the $\pi^0$ and electron fits as used in T2K~\cite{Abe:2013hdq}, the remaining $\pi^0$ background is reduced to about 30\% compared to the previous study~\cite{Abe:2011ts}.


\begin{table}[tbp]%
\caption{\label{Tab:sens-selection-nue}%
The expected number of \nue\ candidate events. $\sin^22\theta_{13}=0.1$ and $\deltacp=0$ are assumed.
Background is categorized by the flavor before oscillation. }
\begin{center}%
\begin{tabular}{c|cc|ccccc|c} \hline \hline
				& \multicolumn{2}{c|}{signal} & \multicolumn{5}{c|}{BG} & \multirow{2}{*}{~total~} \\ 
				&~$\numu \to \nue$~	& ~$\numubar \to \nuebar$~ 	&~$\numu$ CC~	&~$\numubar$ CC~	&~$\nue$  CC~& ~$\nuebar$ CC~ & ~~NC~~	&  \\ \hline  
~$\nu$ mode~~		& 3016				&	28						& 11		& 0				& 503	& 20		& 172		& 3750 \\ 
~$\bar{\nu}$ mode~~	& 396				&	2110					& 4			& 5				& 222	& 396		& 265		&3397 \\ \hline \hline
\end{tabular}%
\end{center}
\end{table}%

Figure~\ref{Fig:sens-enurec-nue} shows the reconstructed neutrino energy distributions of $\nue$ events after all the selections.
The expected number of \nue\ candidate events is shown in Table~\ref{Tab:sens-selection-nue} for each signal and background component.
In the neutrino mode, the dominant background component is intrinsic $\nue$ contamination in the beam.
The mis-identified neutral current $\pi^0$ production events, which was one of
the dominant background components in the previous study,
are suppressed thanks to the improved $\pi^0$ rejection. 
In the anti-neutrino mode, in addition to $\nuebar$ and $\numubar$, $\nue$ and $\numu$ components have non-negligible contributions due to larger fluxes and cross-sections compared to their counterparts in the neutrino mode.

\begin{figure}[tbp]%
\includegraphics[width=0.45\textwidth]{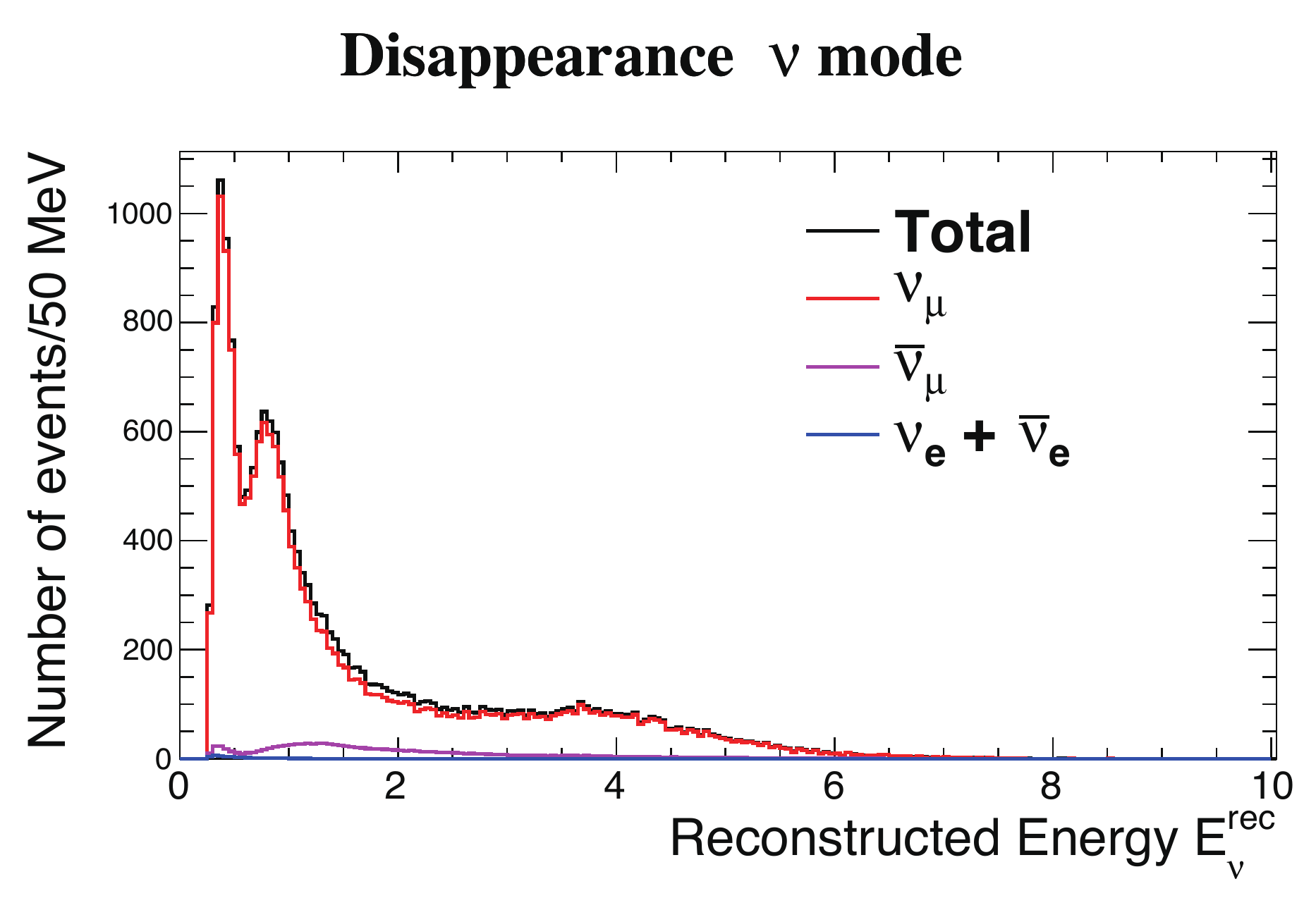}
\includegraphics[width=0.45\textwidth]{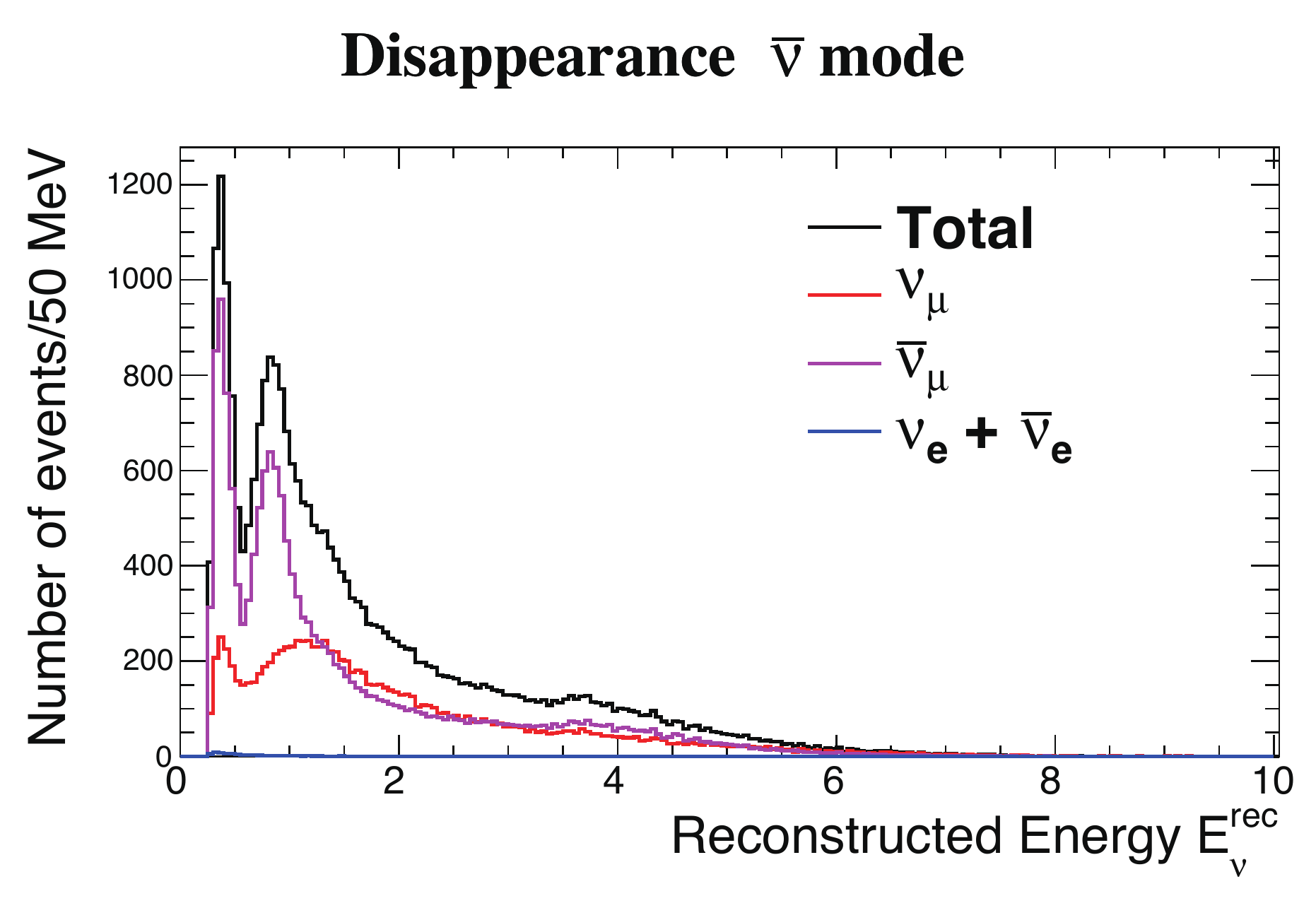}
\caption{%
Reconstructed neutrino energy distribution of the $\numu$ candidate events.
\label{Fig:sens-enurec-numu}
}
\end{figure}

\begin{table}[tbp]%
\caption{\label{Tab:sens-selection-numu}%
The expected number of $\numu$ candidate events.}
\begin{center}%
\begin{tabular}{cccccccc} \hline \hline
				&~$\numu$ CC~	& ~$\numubar$ CC~	&~$\nue$ CC~ & ~$\nuebar$ CC~ 	&~~NC~~ 	& ~$\numu \to \nue$~		& ~~total~~ 		\\ \hline 
$\nu$ mode~~		& 17225		&	1088			& 11			& 1				& 999 		& 49			& 19372		 \\ 
$\bar{\nu}$ mode~~	& 10066		&	15597			& 7				& 7				& 1281		& 6  			& 26964		 \\ \hline \hline
\end{tabular}%
\end{center}
\end{table}%

For the \numu/\numubar\ candidate events the following criteria are applied;
the reconstructed ring is identified as muon-like ($\mu$-like),
the reconstructed muon momentum is greater than 200\,MeV/$c$, and
the number of decay electron associated to the event is 0 or 1.


Figure~\ref{Fig:sens-enurec-numu} shows the reconstructed neutrino energy distributions of the selected $\numu$/$\numubar$ events.
Table~\ref{Tab:sens-selection-numu} shows the number of $\numu$ candidate events for each signal and background component.
For the neutrino mode, most of the events are due to $\numu$, 
while in the anti-neutrino mode the contribution from wrong-sign $\numu$
components is significant.

\begin{figure}[tbp]
\includegraphics[width=0.45\textwidth]{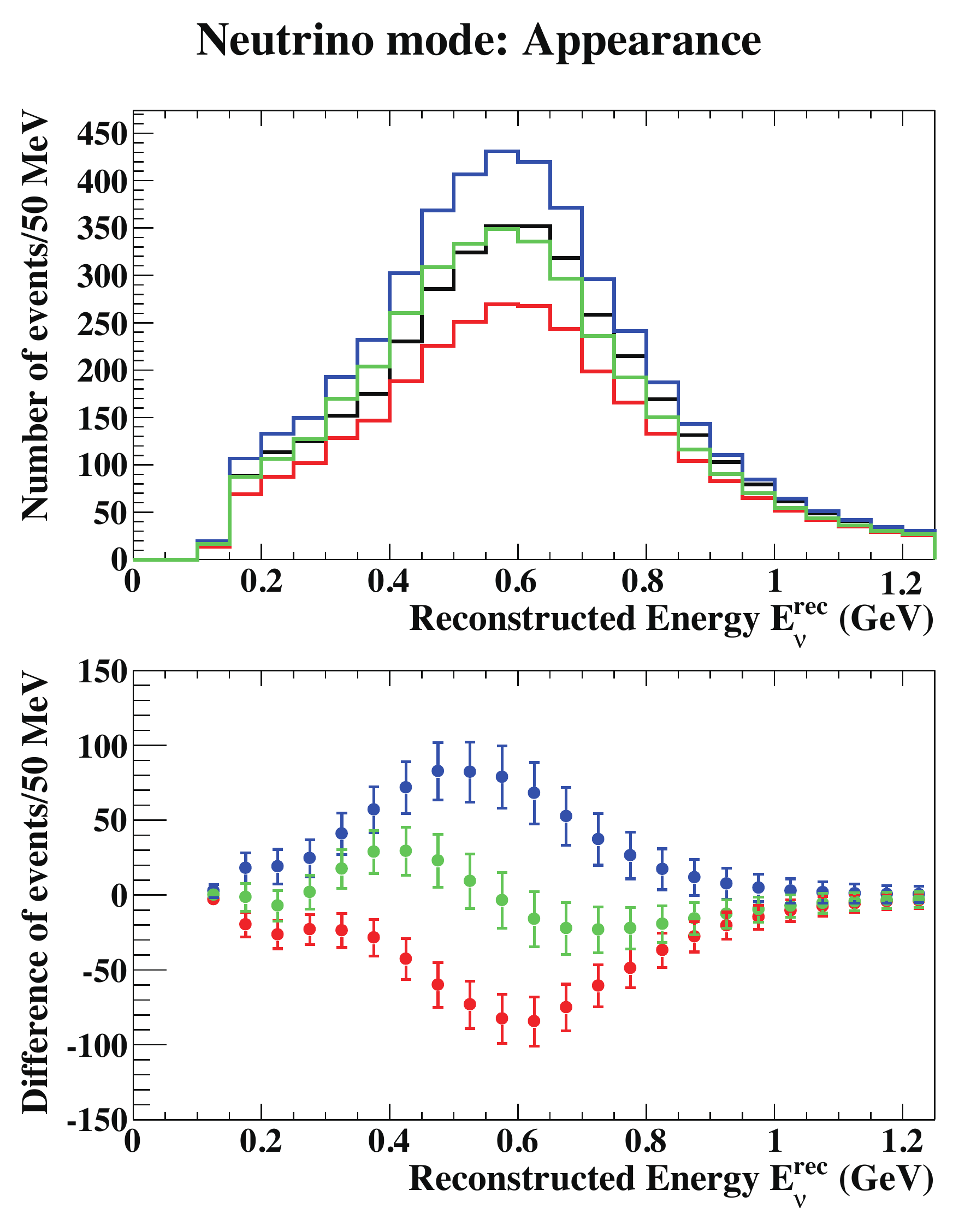}
\includegraphics[width=0.45\textwidth]{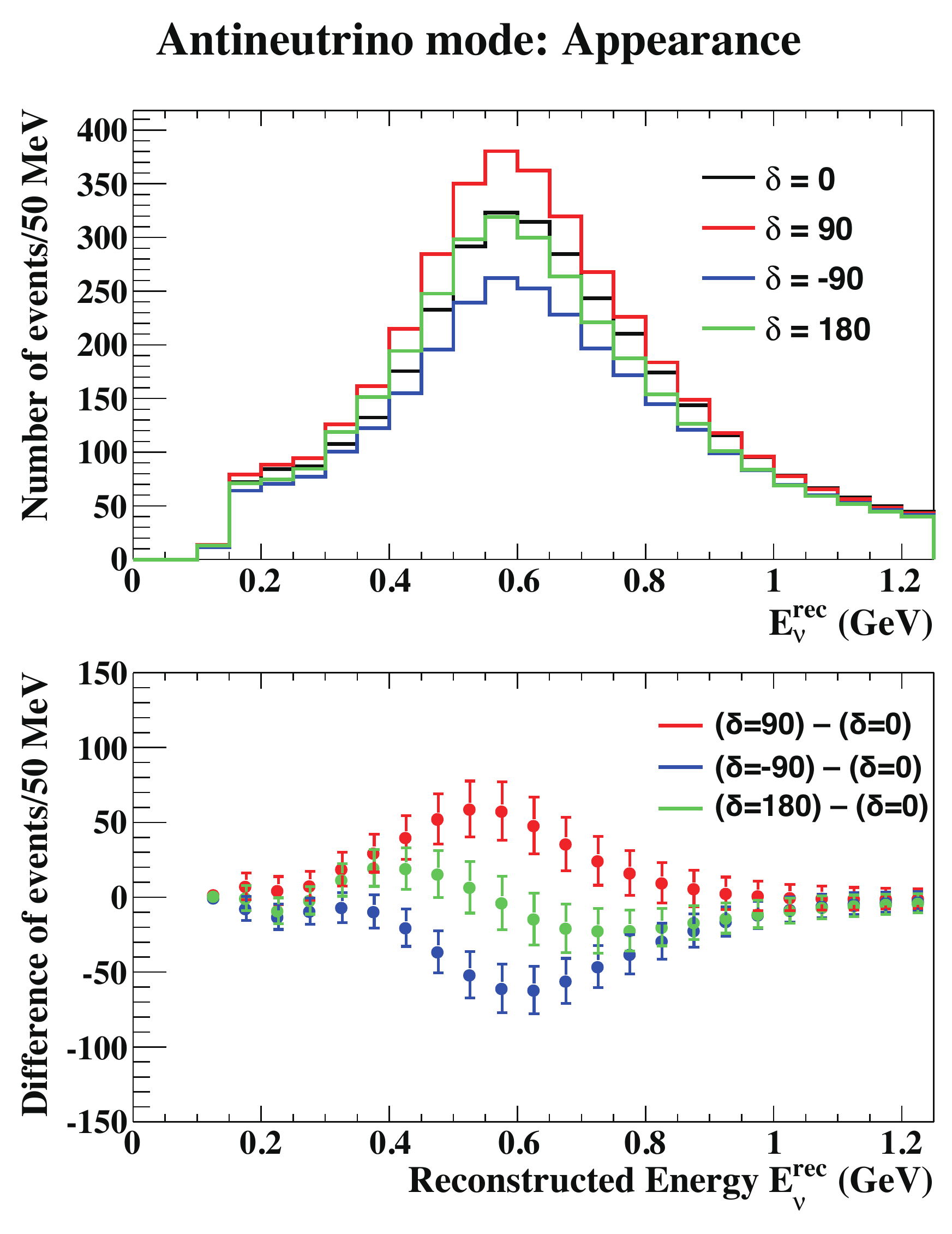}
\caption{
Top: Reconstructed neutrino energy distribution for several values of $\deltacp$.  
$\sin^22\theta_{13}=0.1$ and normal hierarchy is assumed. 
Bottom: Difference of the reconstructed neutrino energy distribution from the case with $\deltacp=0^\circ$.
The error bars represent the statistical uncertainties of each bin.
}
\label{enurecdiff-nue}
\end{figure}

\begin{figure}[tbp]
\includegraphics[width=0.45\textwidth]{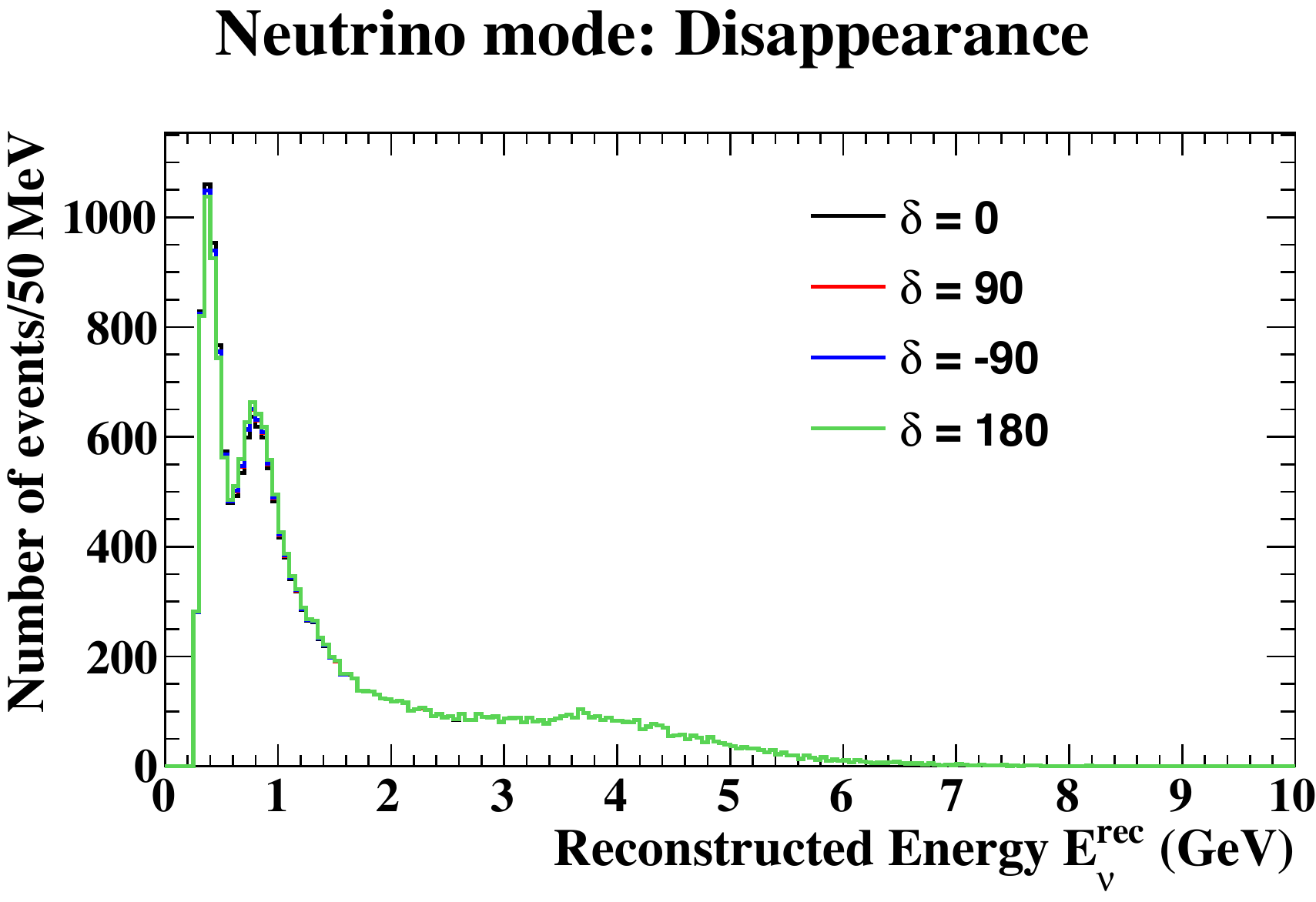}
\includegraphics[width=0.45\textwidth]{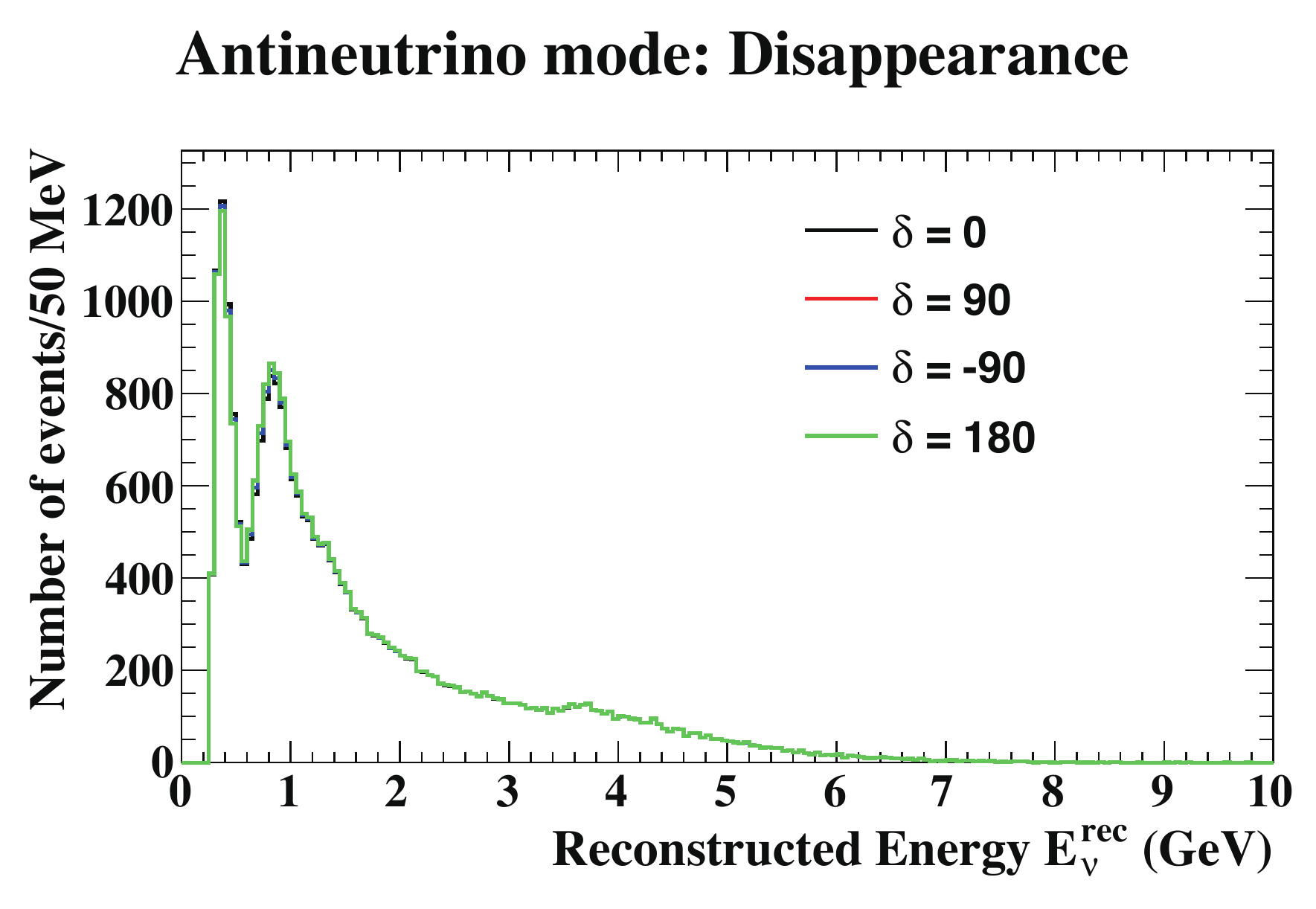}
\caption{
Reconstructed neutrino energy distribution of $\numu$ candidates for several values of $\deltacp$.
}
\label{enurecdiff-numu}
\end{figure}

The reconstructed neutrino energy distributions of $\nue$ events for several values of $\deltacp$
are shown in the top plots of Fig.~\ref{enurecdiff-nue}.
The effect of $\deltacp$ is clearly seen using the reconstructed neutrino energy.
The bottom plots show the difference of reconstructed energy spectrum from $\deltacp=0^\circ$ for the cases $\delta = 90^\circ, -90^\circ$ and $180^\circ$.
The error bars correspond to the statistical uncertainty.
By using not only the total number of events but also the reconstructed energy distribution,
the sensitivity to $\deltacp$ can be improved, 
and one can discriminate all the values of $\deltacp$, including the difference between $\deltacp = 0$ and $\pi$.
Figure~\ref{enurecdiff-numu} shows the reconstructed neutrino energy
distributions of the $\numu$ sample for several values of $\deltacp$.
As expected the difference is very small for $\numu$ events.


\subsection{Analysis method}
The sensitivity of a long baseline experiment using Hyper-K and J-PARC neutrino beam is studied using a binned likelihood analysis based on the reconstructed neutrino energy distribution.
Both \nue\ appearance and \numu\ disappearance samples, in both neutrino and
antineutrino runs, are simultaneously fitted.

The $\chi^2$ used in this study is defined as 
\begin{equation} \label{eq:sens:chi2}
\chi^2 =  -2 \ln \mathcal{L}  + P,
\end{equation}
where $\ln \mathcal{L}$ is the log likelihood for a Poisson distribution,
\begin{equation}
-2\ln \mathcal{L} = \sum_k \left\{ -{N_k^\mathrm{test}(1+f_i)} + N_k^\mathrm{true} \ln \left[ N_k^\mathrm{test}(1+f_i) \right] \right\}.
\end{equation}
Here, $N_k^\mathrm{true}$ ($N_k^\mathrm{test}$) is the number of events in $k$-th reconstructed energy bin for the true (test) oscillation parameters.
The index $k$ runs over all reconstructed energy bins for
muon and electron neutrino samples and for neutrino and anti-neutrino mode running.
The binning of the systematic parameter $f_i$ is coarser than the reconstructed energy bins,
which are grouped based on the behavior against the systematics uncertainty, with variable widths.
For anti-neutrino mode samples, an additional overall normalization parameter
with 6\% prior uncertainty is introduced to account for a possible uncertainty in the anti-neutrino interaction, which is less known experimentally in this energy region.
A normalization weight $(1+f^{\overline{\nu}}_\mathrm{norm})$ is multiplied to $N_k^\mathrm{test}$ in the anti-neutrino mode samples.

The penalty term $P$ in Eq.~\ref{eq:sens:chi2} constrains the systematic parameters $f_i$ with the normalized covariance matrix $C$,
\begin{equation}
P = \sum_{i,j} f_i (C^{-1})_{i,j} f_j.
\end{equation}

The size of systematic uncertainty is evaluated based on the experience and prospects of the T2K experiment, as it provides the most realistic estimate as the baseline.
We estimate the systematic uncertainties assuming the T2K neutrino beamline and
near detectors, taking into account improvements expected with future T2K
running and analysis improvements.
For Hyper-K a further reduction of systematic uncertainties will be possible
with upgrade of beamline and near detectors, improvements in detector calibration and analysis techniques, and improved understanding of neutrino interaction with more measurements.
In particular, as described in Sec.~\ref{sec:ND}, studies of near detectors are ongoing with a goal of further reducing systematic uncertainties.
The sensitivity update is expected in the near future as the near detector design studies advance.

There are three main categories of systematic uncertainties. We assume improvement from the current T2K uncertainties for each category as follows.
\begin{description}
\item[i) Flux and cross section uncertainties constrained by the fit to current near detector data] 
These arise from systematics of the near detectors. 
The understanding of the detector will improve in the future, but this category of
uncertainties is conservatively assumed to stay at the same level as currently estimated.
\item[ii) Cross section uncertainties that are not constrained by the fit to current near detector data]
Errors in this category will be reduced as more categories of samples are added to the near detector data fit, which constrains the cross section models.
We assume the uncertainties arising from different target nucleus between the near and the far detectors will become negligible by including the measurement with the water target in the near detector.
\item[iii) Uncertainties on the far detector efficiency and reconstruction modeling]
Because most of them are estimated by using atmospheric neutrinos as a control sample, errors in this category are expected to decrease with more than an order of magnitude larger statistics available with Hyper-K than currently used for T2K.
Uncertainties arising from the energy scale 
is kept the same because 
it is not estimated by the atmospheric neutrino sample.
\end{description}
The flux and cross section uncertainties are assumed to be uncorrelated between the neutrino and anti-neutrino running, except for the uncertainty of \nue/\numu\ cross section ratio which is treated to be anti-correlated considering the theoretical uncertainties studied in~\cite{Day:2012gb}.
Because some of the uncertainties, such as those from the cross section
modeling or near detector systematics, are expected to be correlated and give
more of a constraint, this is a conservative assumption.
The far detector uncertainty is treated to be fully correlated between the neutrino and anti-neutrino running.

\begin{figure}[tbp]
\includegraphics[width=0.45\textwidth]{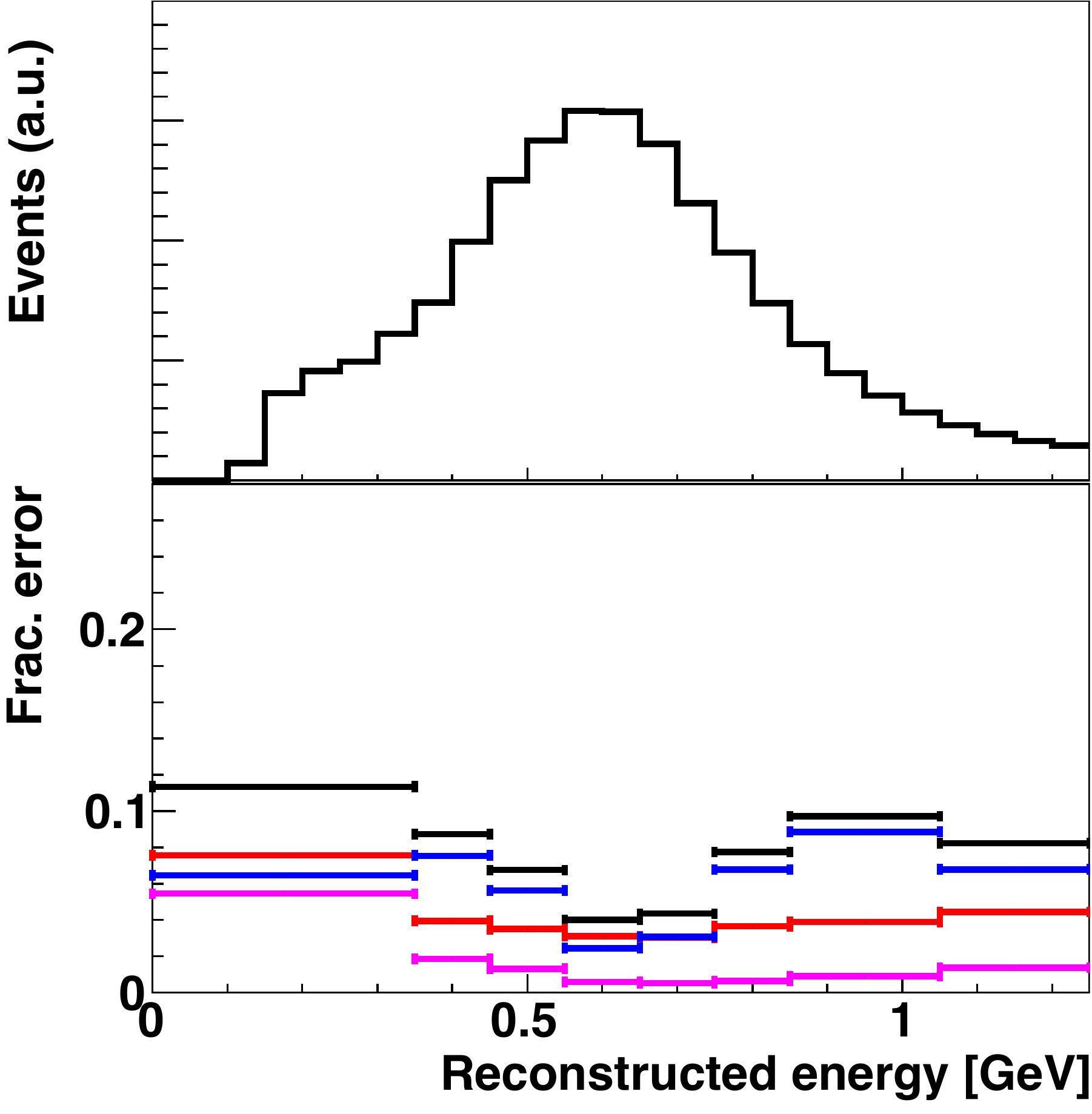}
\includegraphics[width=0.45\textwidth]{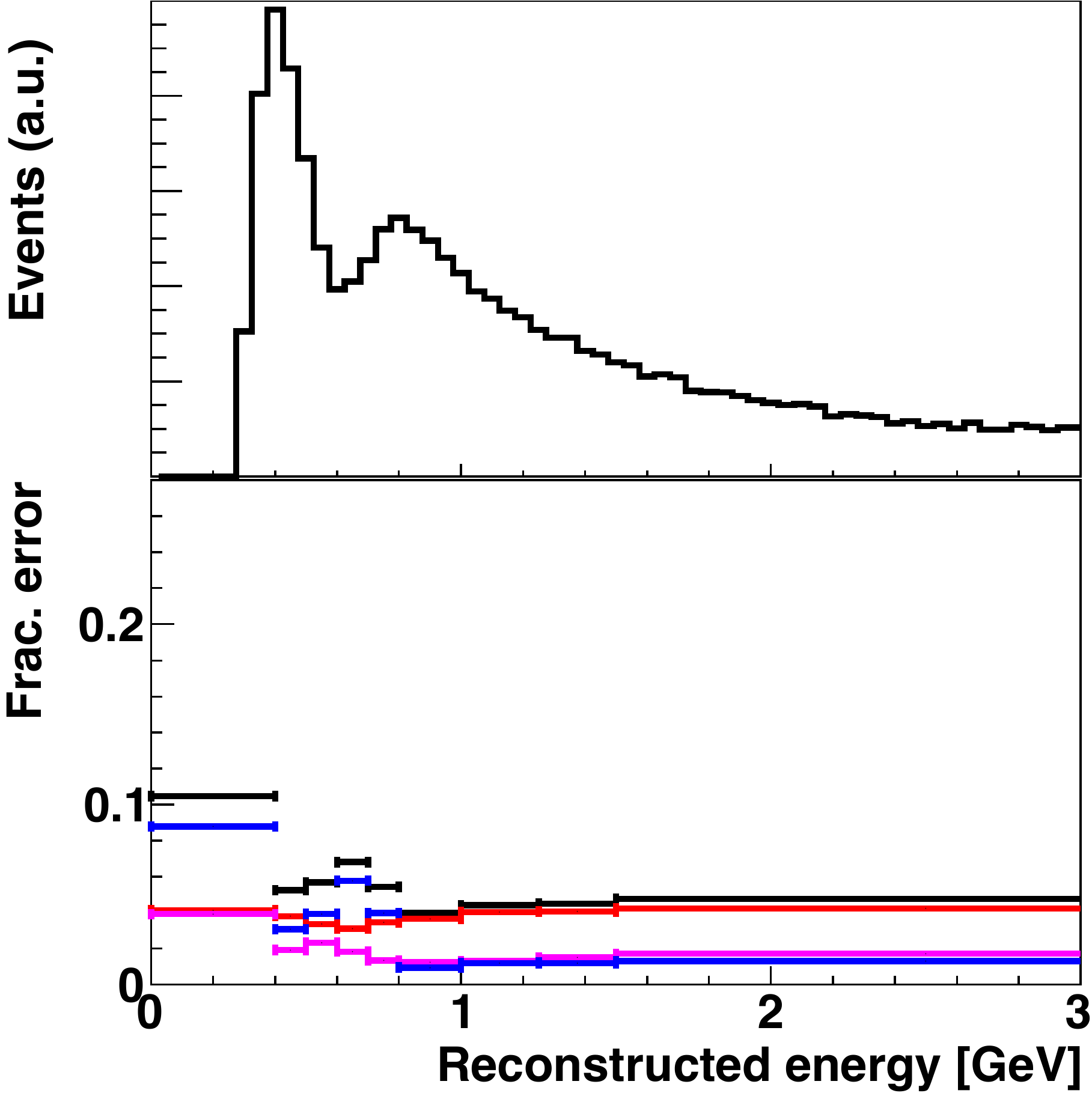}
\caption{
Fractional error size for the appearance (left) and the disappearance (right) reconstructed energy spectra (bottom plots) in the neutrino mode.
Black: total uncertainty, red: the flux and cross-section constrained by the near detector, 
magenta: the near detector non-constrained cross section,
blue: the far detector error.
\label{Fig:systerror}
}
\end{figure}

\begin{figure}[tbp]
\includegraphics[width=0.45\textwidth]{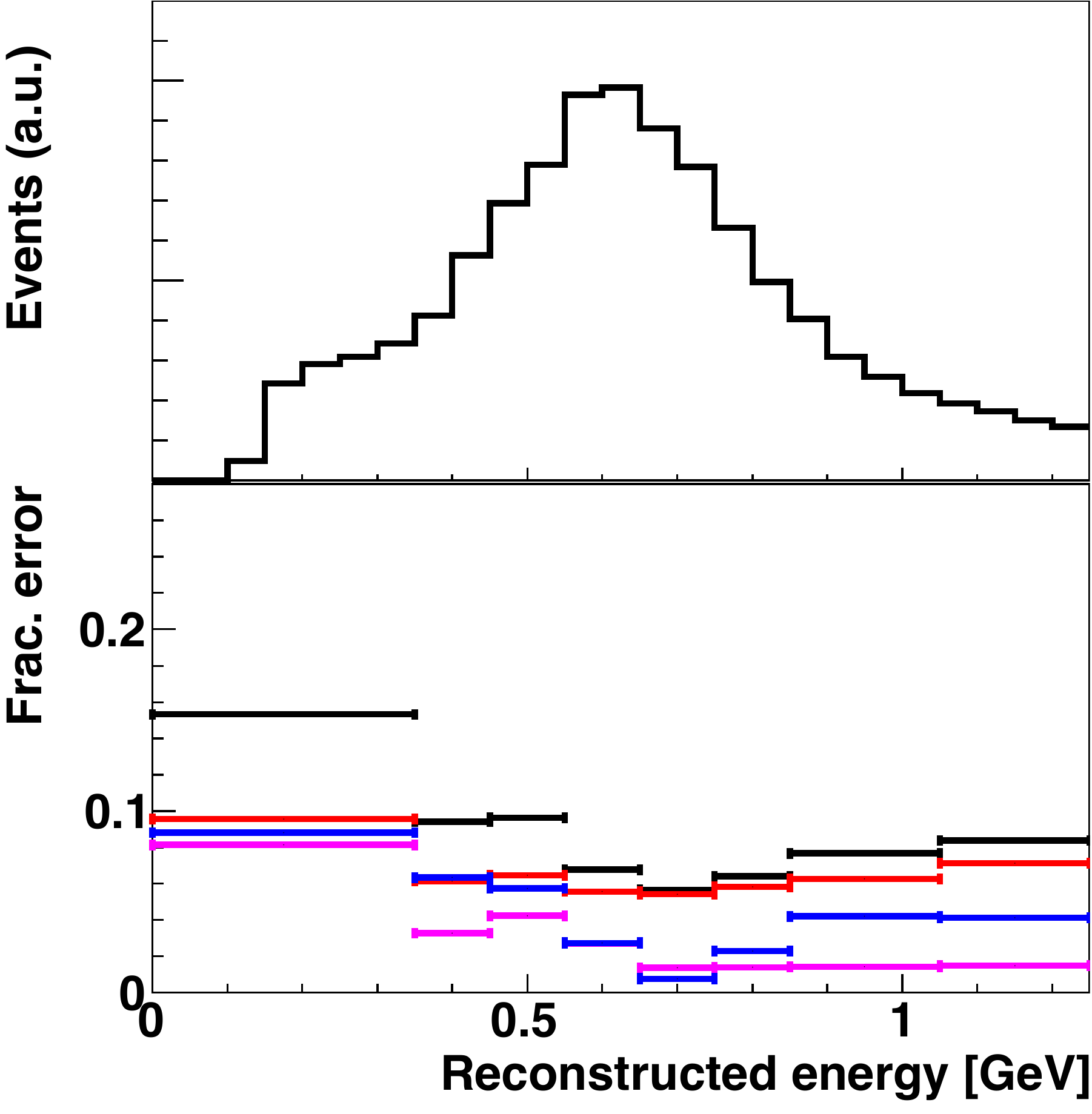}
\includegraphics[width=0.45\textwidth]{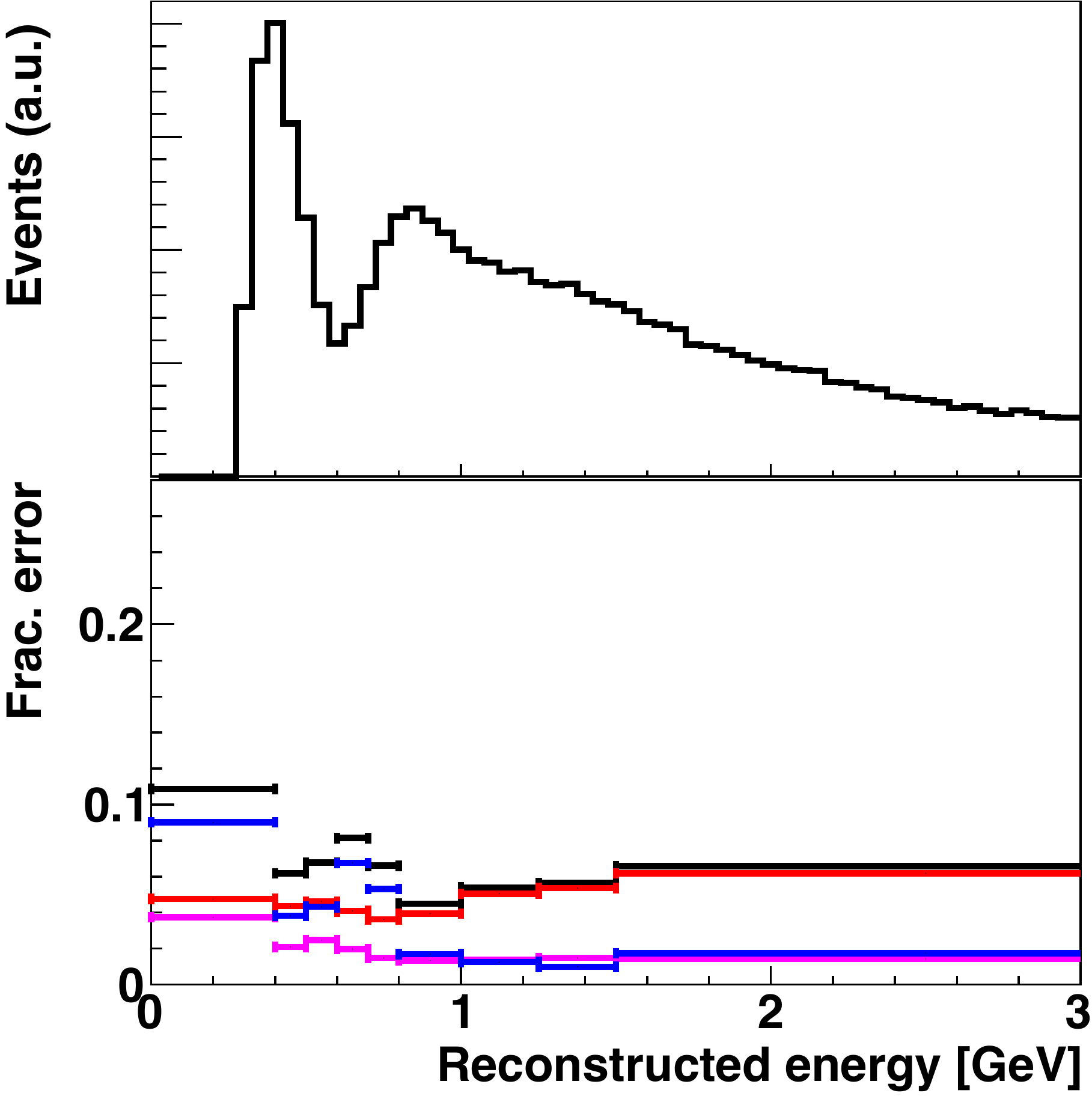}
\caption{
Fractional error size for the appearance (left) and the disappearance (right) reconstructed energy spectra (bottom plots) in the anti-neutrino mode.
Black: total uncertainty, red: the flux and cross-section constrained by the near detector, 
magenta: the near detector non-constrained cross section,
blue: the far detector error.
\label{Fig:systerror-anti}
}
\end{figure}

\begin{figure}[tbp]
\includegraphics[width=0.7\textwidth]{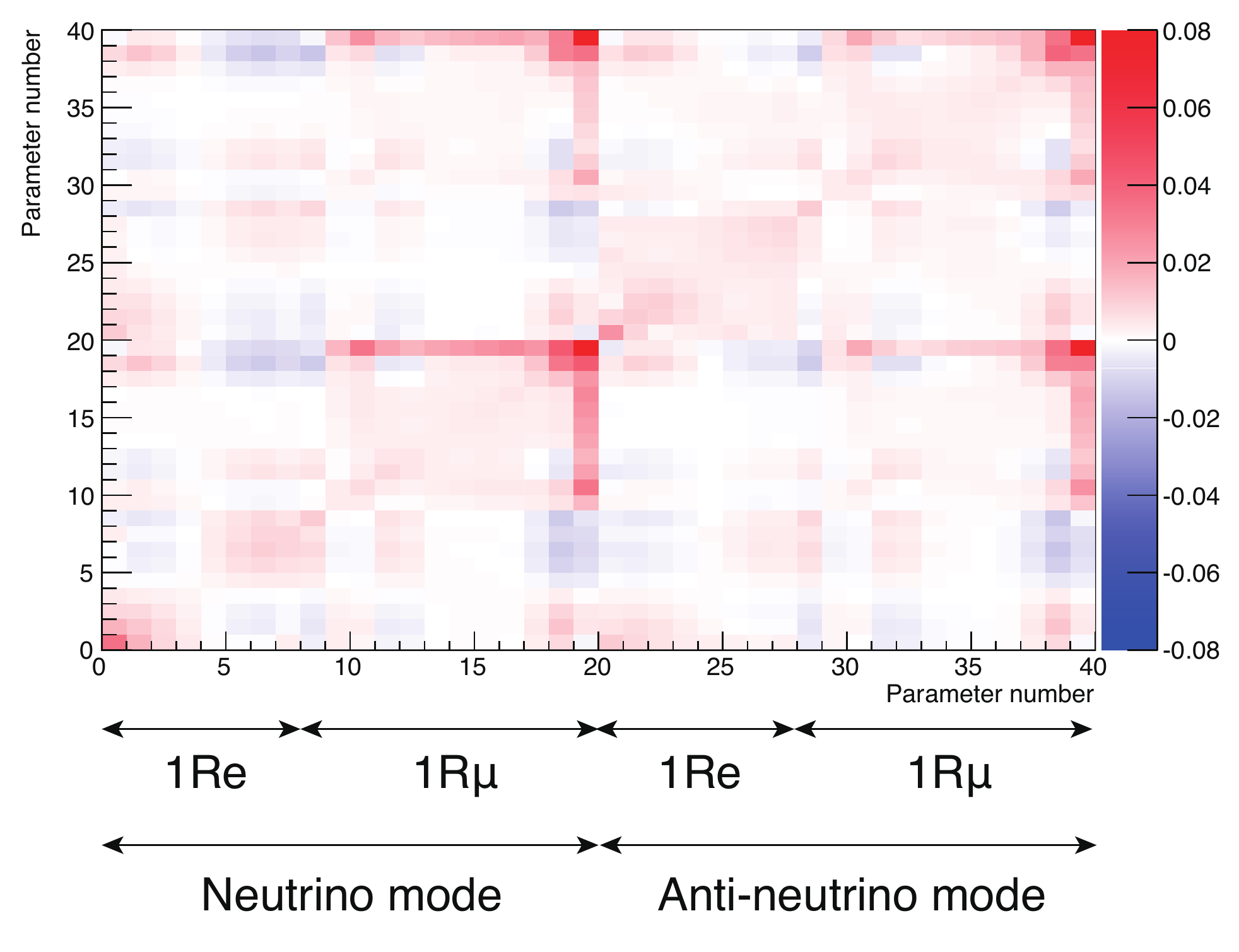}
\caption{
Covariance matrix between reconstructed energy bins of the four samples due to the systematic uncertainties.
Bins 1--8, 9--20, 21--28, and 29--40 correspond to 
the neutrino mode single ring $e$-like,
the neutrino mode single ring $\mu$-like,
the anti-neutrino mode single ring $e$-like, and
the anti-neutrino mode single ring $\mu$-like samples, respectively.
\label{Fig:errormatrix}
}
\end{figure}

Figures~\ref{Fig:systerror} and \ref{Fig:systerror-anti} show the fractional systematic uncertainties for the appearance and disappearance reconstructed energy spectra in neutrino and anti-neutrino mode, respectively.
Black lines represent the prior uncertainties and bin widths of the systematic parameters $f_i$,
while colored lines show the contribution from each uncertainty source.
Figure~\ref{Fig:errormatrix} shows the covariance matrix of the systematic uncertainties between the reconstructed neutrino energy bins of the four samples.
The systematic uncertainties (in \%) of the number of expected events at the far detector are summarized in Table~\ref{tab:sens:systsummary}.

\begin{table}[htdp]
\caption{Uncertainties (in \%) for the expected number of events at Hyper-K from the systematic uncertainties assumed in this study.}
\begin{center}
\begin{tabular}{cccccc}  \hline \hline
\multirow{2}{*}{Source} & \multicolumn{2}{c}{$\nu$ mode} & \multicolumn{2}{c}{$\overline{\nu}$ mode}  \\
 & Appearance & Disappearance & Appearance & Disappearance\\ \hline
Flux \& ND-constrained cross section & 3.0 & 2.8 & 5.6 & 4.2 \\
ND-independent cross section& 1.2 & 1.5 & 2.0 & 1.4 \\
Far detector & 0.7 & 1.0 & 1.7 & 1.1 \\ \hline
Total & 3.3 & 3.3 & 6.2 & 4.5 \\
\hline \hline
\end{tabular}
\end{center}
\label{tab:sens:systsummary}
\end{table}%

\subsection{Expected sensitivity to CP violation}

\begin{figure}[tbp]
\centering
\includegraphics[width=0.65\textwidth]{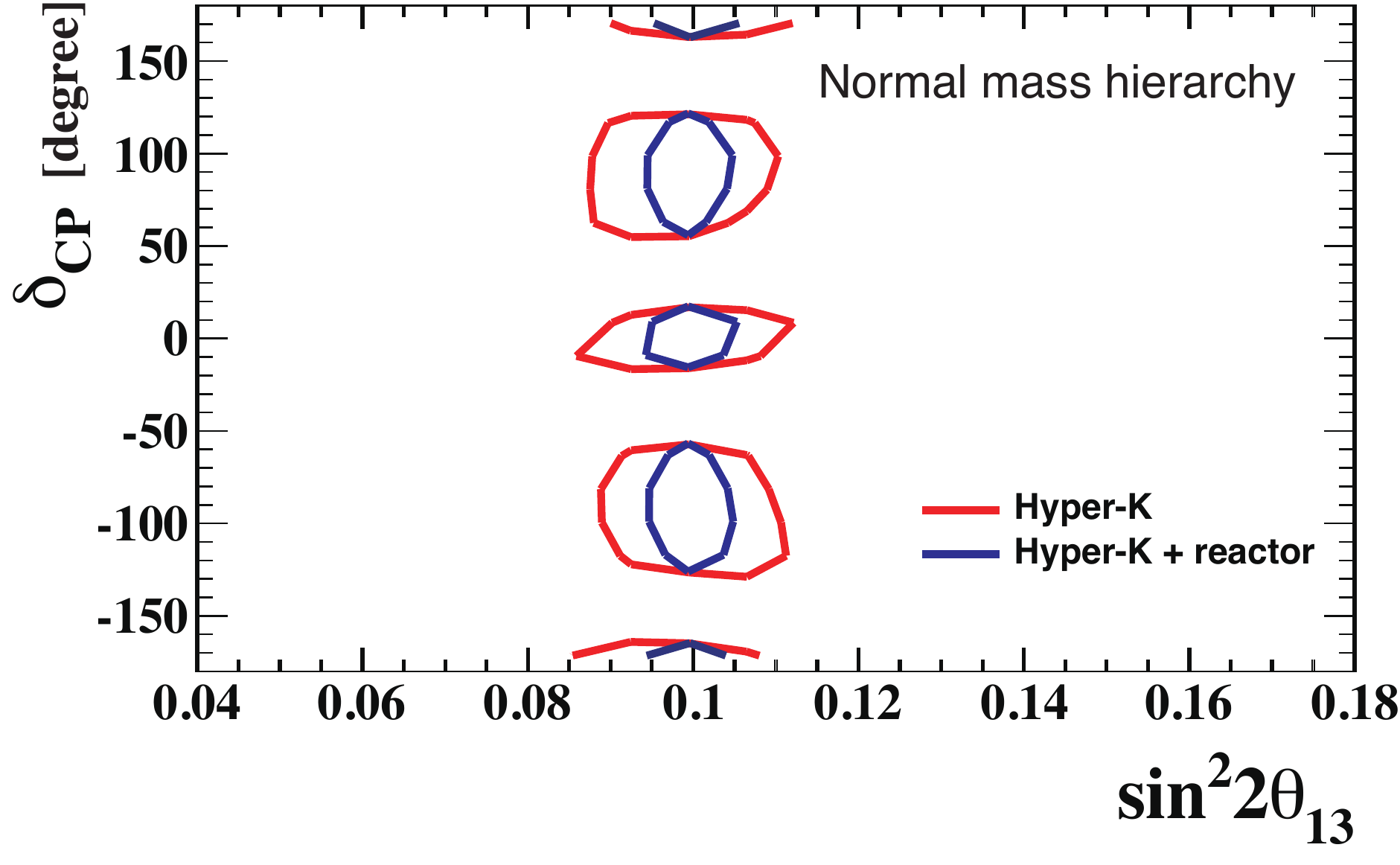}
\includegraphics[width=0.65\textwidth]{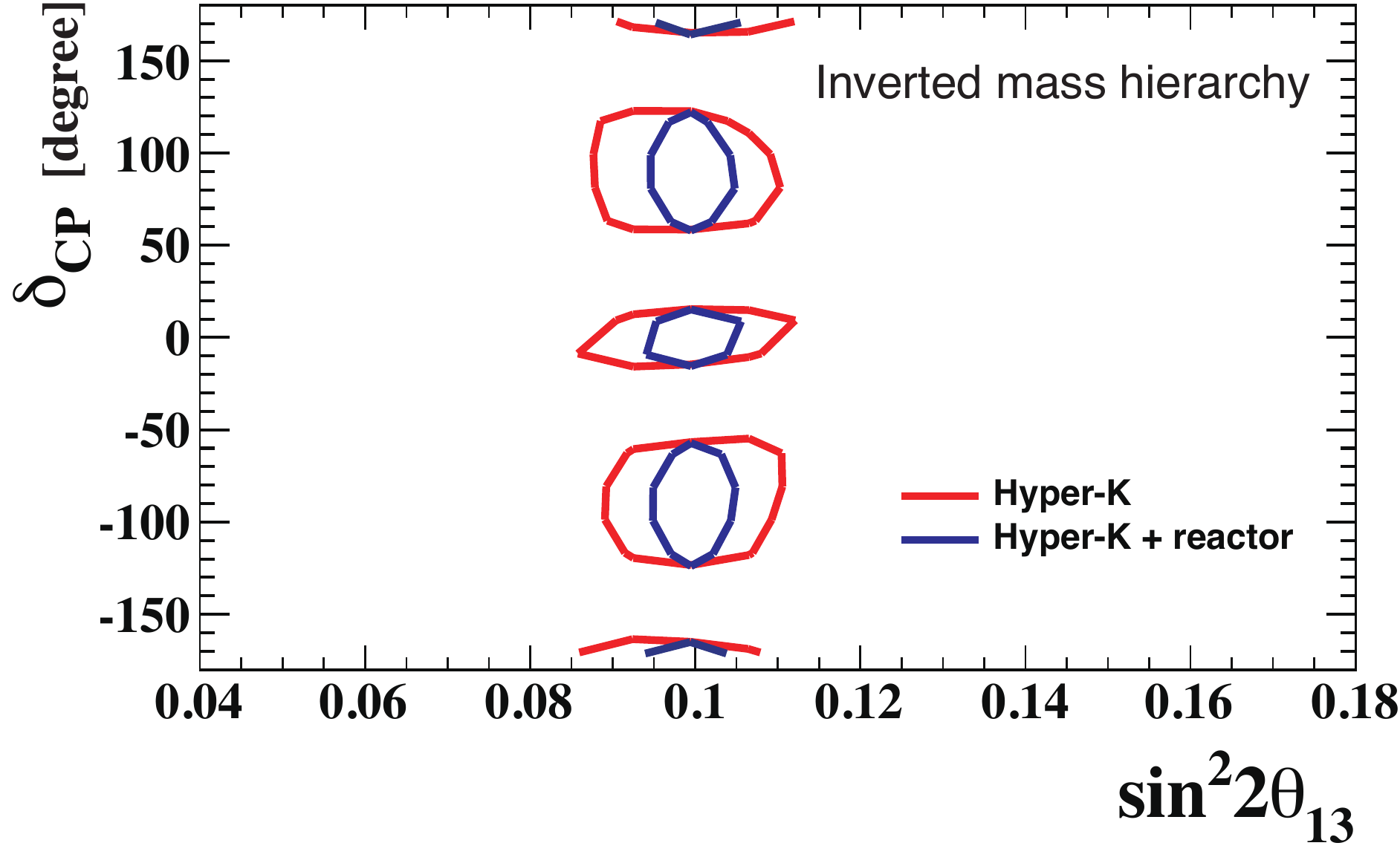}
\caption{The 90\% CL allowed regions in the $\sin^22\theta_{13}$-$\deltacp$ plane.
The results for the true values of $\deltacp = (-90^\circ, 0, 90^\circ, 180^\circ)$ are overlaid.
Top: normal hierarchy case. Bottom: inverted hierarchy case.
Red (blue) lines show the result with Hyper-K only (with $\sin^22\theta_{13}$ constraint from reactor experiments).
\label{fig:CP-contour}}
\end{figure}

\begin{figure}[tbp]
\includegraphics[width=0.65\textwidth]{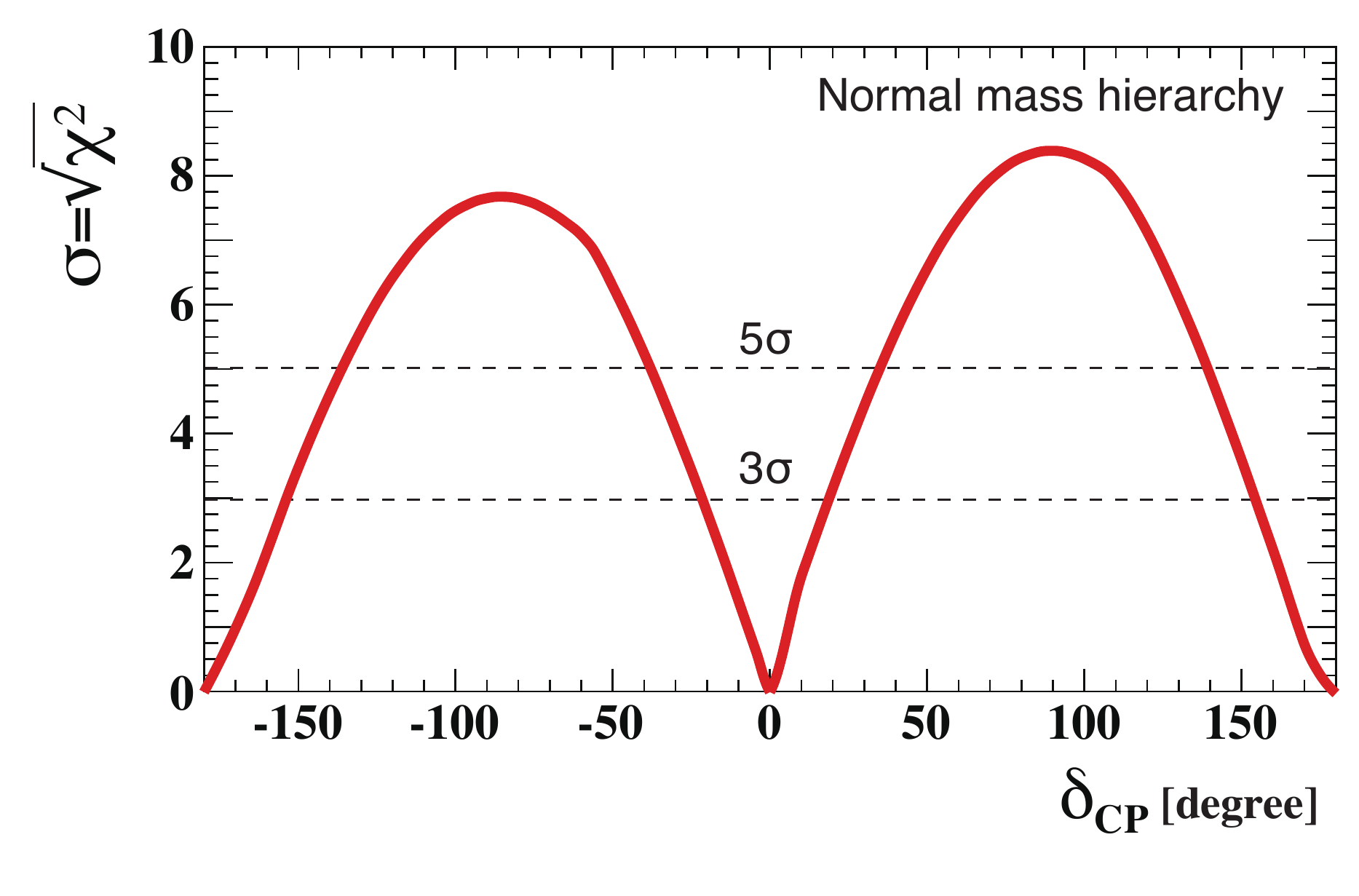}
\includegraphics[width=0.65\textwidth]{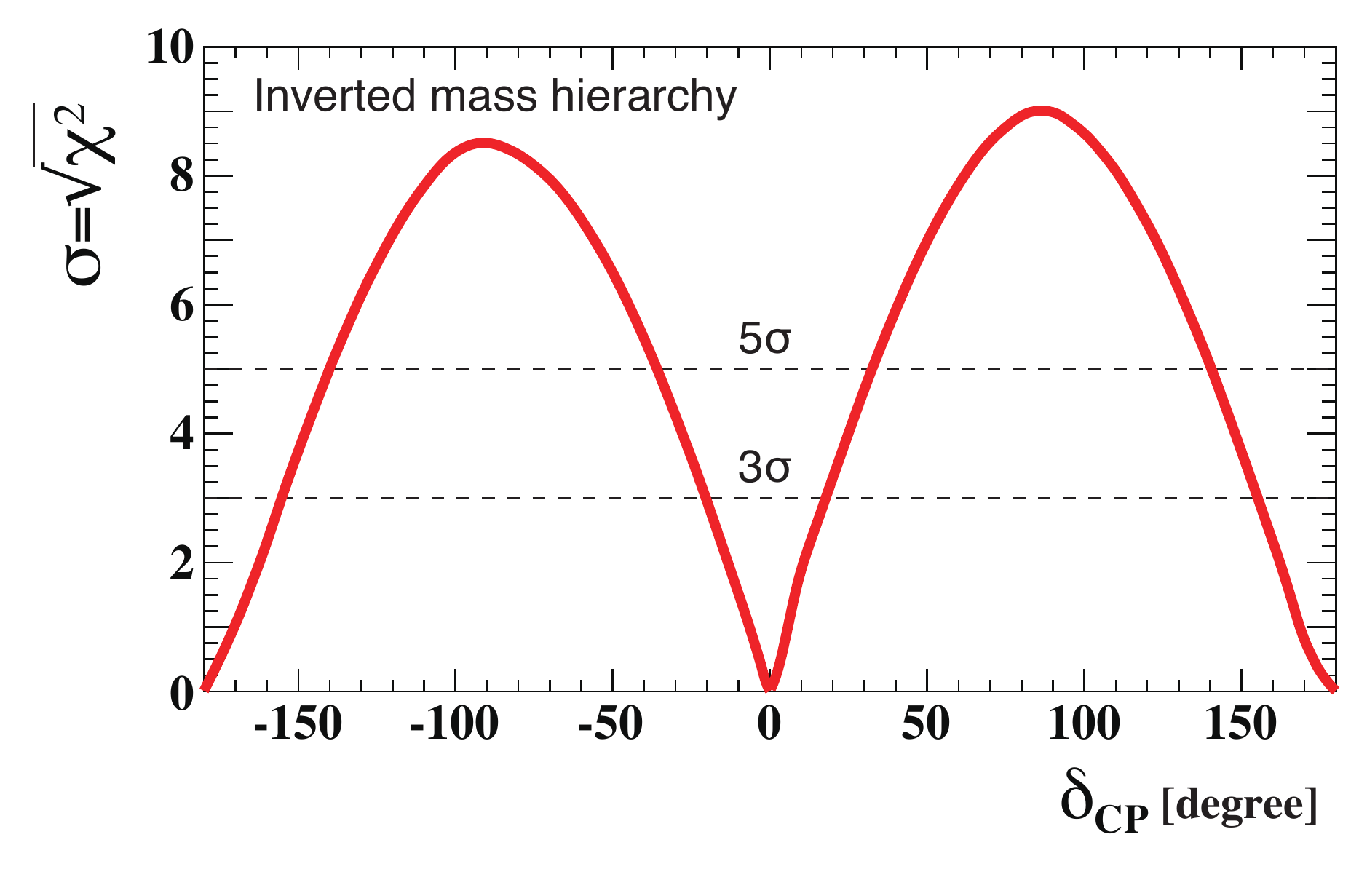}
\caption{Expected significance to exclude $\sin\deltacp = 0$.
Top: normal hierarchy case. Bottom: inverted hierarchy case.
\label{fig:CP-chi2}}
\end{figure}

\begin{figure}[tbp]
\centering
\includegraphics[width=0.65\textwidth]{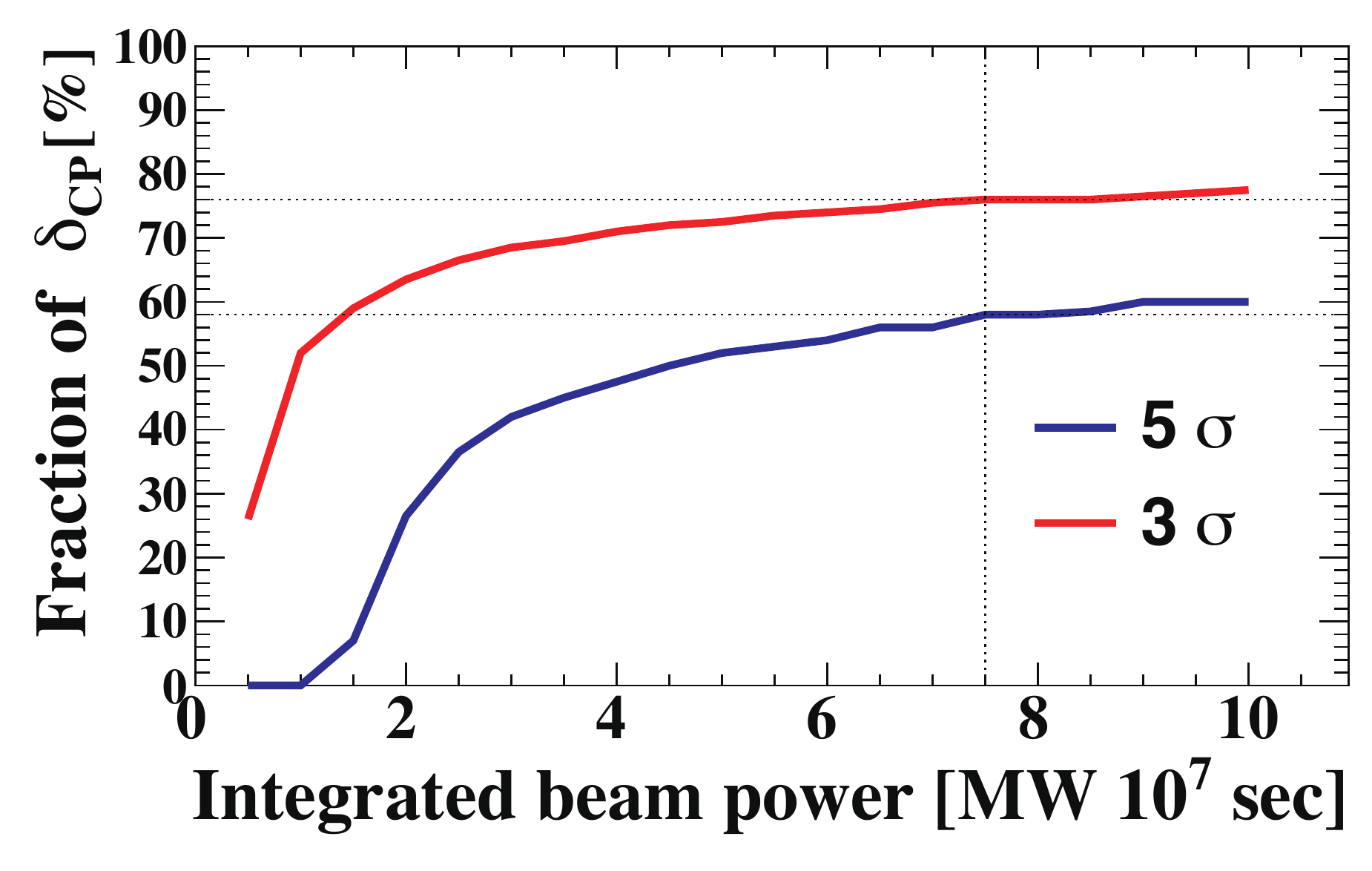}
\caption{Fraction of $\deltacp$ for which $\sin\deltacp= 0$ can be excluded with 3\,$\sigma$ (red) and 5\,$\sigma$ (blue) significance as a function of the integrated beam power. For the normal hierarchy case.
The ratio of neutrino and anti-neutrino mode is fixed to 1:3. 
\label{fig:delta-sens-time}}
\end{figure}

Figure~\ref{fig:CP-contour} shows the 90\% CL allowed regions on the $\sin^22\theta_{13}$-$\deltacp$ plane.
The results for the true values of $\deltacp = (-90^\circ, 0, 90^\circ, 180^\circ)$ are overlaid.
The top (bottom) plot shows the case for the normal (inverted) mass hierarchy.
The value of \deltacp\ will be determined well.
Also shown are the allowed regions when we include a constraint on $\sin^22\theta_{13}$ from the reactor experiments.
The $\sin^22\theta_{13}$ uncertainty of  0.005 is assumed.
With reactor constraints, although the contour becomes narrower in the direction of $\sin^22\theta_{13}$, the sensitivity to $\deltacp$ does not significantly change.

Figure~\ref{fig:CP-chi2} shows the expected significance to exclude $\sin\deltacp = 0$ (the $CP$ conserved case).
The significance is calculated as $\sqrt{\Delta \chi^2}$, where $\Delta \chi^2$ is the difference of $\chi^2$ for the \textit{trial} value of \deltacp\ and for $\deltacp = 0^\circ$ or 180$^\circ$ (the smaller value of difference is taken).
We have also studied the case with a reactor constraint, but the result changes only slightly.
Figure~\ref{fig:delta-sens-time} shows the fraction of $\deltacp$ for which $\sin\deltacp= 0$ is excluded with more than 3\,$\sigma$ and 5\,$\sigma$ of significance as a function of the integrated beam power.
The ratio of integrated beam power for the neutrino and anti-neutrino mode is fixed to 1:3.
The normal mass hierarchy is assumed.
The results for the inverted hierarchy is almost the same.
$CP$ violation in the lepton sector can be observed with more than 3(5)\,$\sigma$ significance for 76(58)\% of the possible values of $\deltacp$.

\begin{figure}[tbp]
\centering
\includegraphics[width=0.65\textwidth]{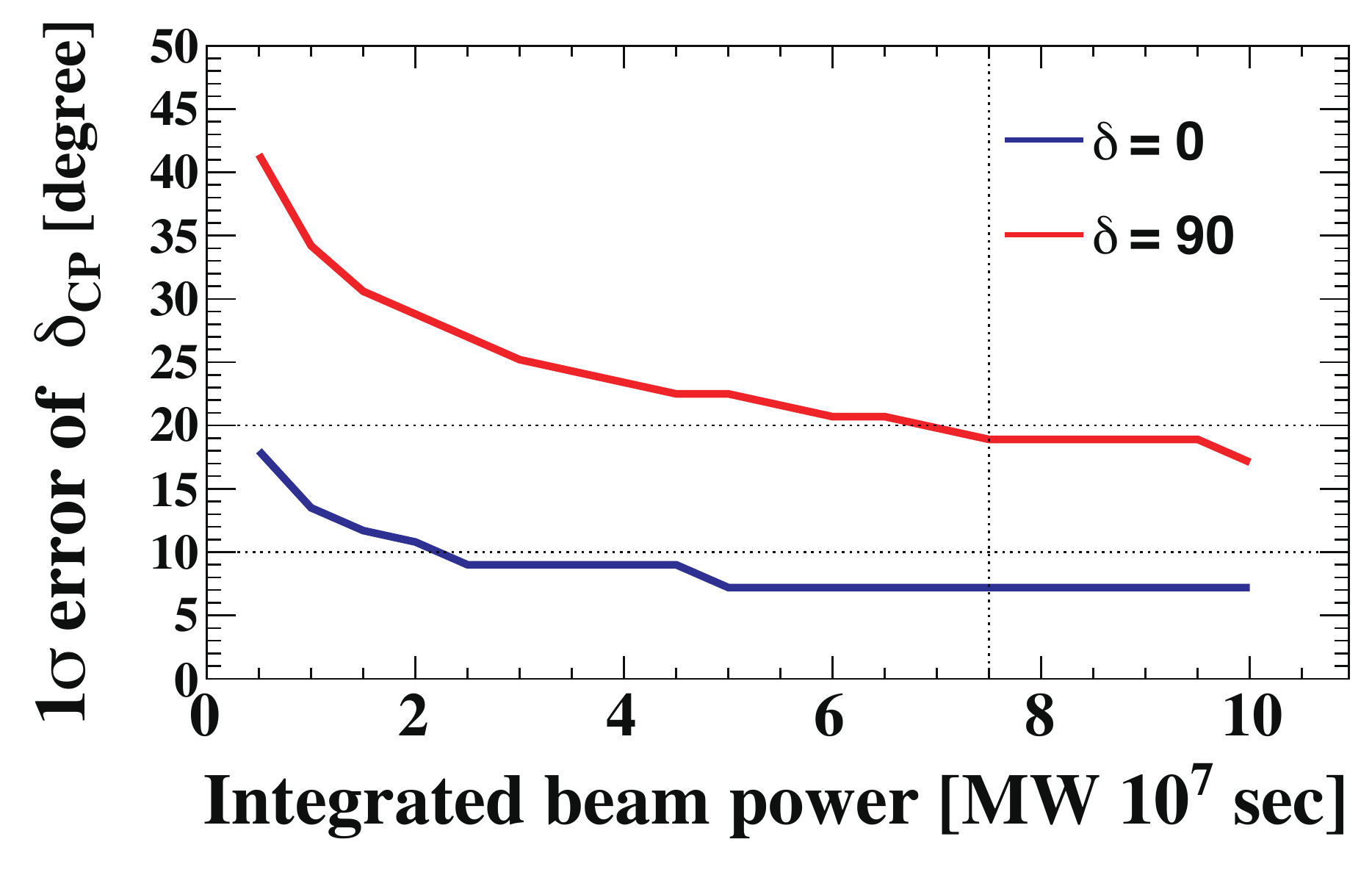}
\caption{Expected 1$\sigma$ uncertainty of $\deltacp$ as a function of integrated beam power. 
\label{fig:delta-error-time}}
\end{figure}

Figure~\ref{fig:delta-error-time} shows the 1$\sigma$ uncertainty of $\deltacp$ as a function of the integrated beam power.
With 7.5~MW$\times$10$^7$sec of exposure (1.56$\times$10$^{22}$ protons on target), 
the value of $\deltacp$ can be determined to better than 19$^\circ$ for all values of $\deltacp$.

\subsection{Sensitivity to $\Delta m^2_{32}$ and $\sin^2\theta_{23}$}
\begin{figure}[tbp]
\centering
\includegraphics[width=0.65\textwidth]{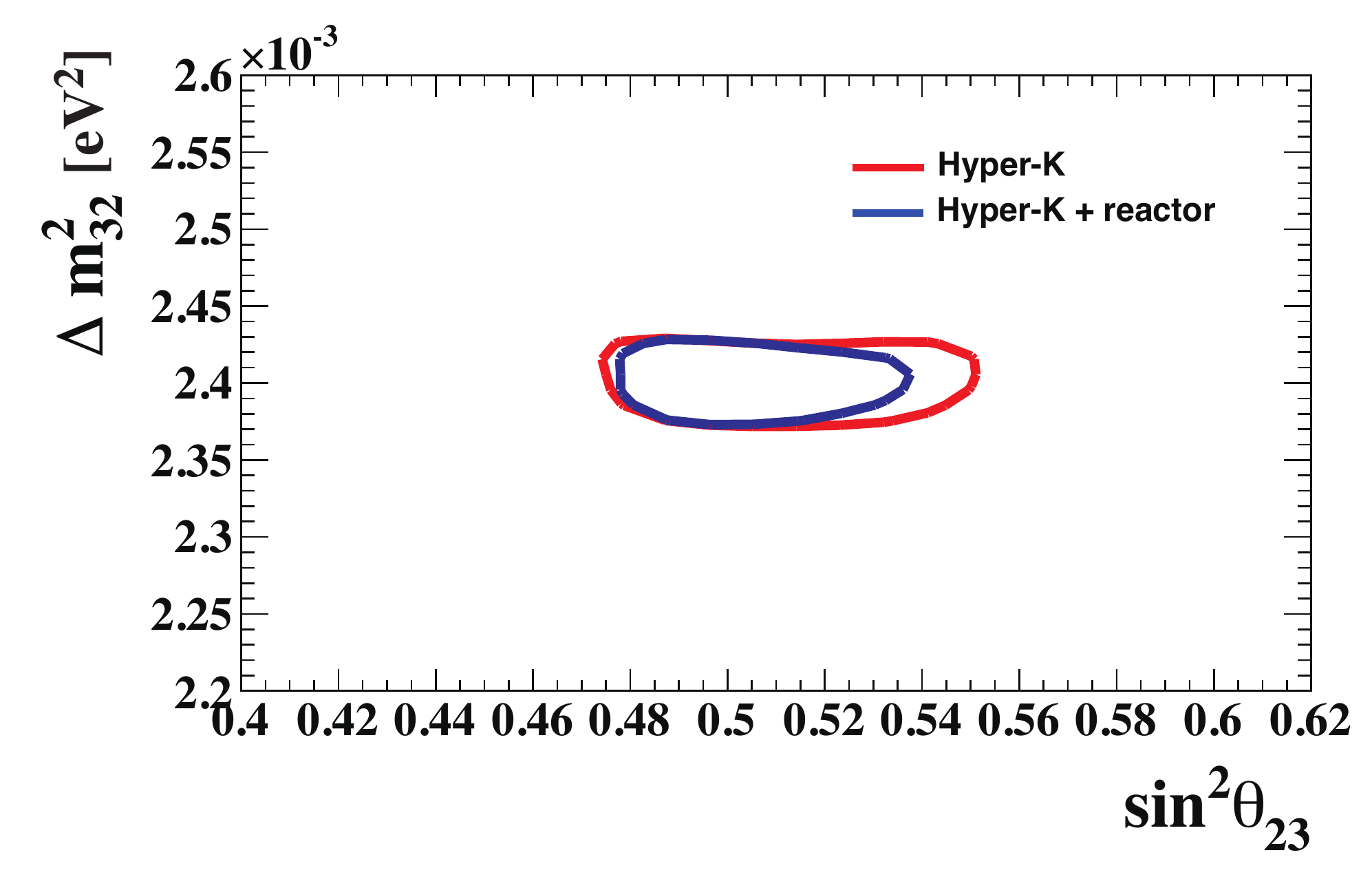}
\caption{ The 90\% CL allowed regions in the $\sin^2\theta_{23}$--$\Delta m^2_{32}$ plane.
The true values are $\sin^2\theta_{23}=0.5$ and $\Delta m^2_{32} = 2.4 \times 10^{-3}$~eV$^2$.
Effect of systematic uncertainties is included. The red (blue) line corresponds to the result with Hyper-K alone (with reactor constraints on $\sin^22\theta_{13}$).
\label{fig:theta23-0.50}}
\end{figure}

\begin{figure}[tbp]
\centering
\includegraphics[width=0.65\textwidth]{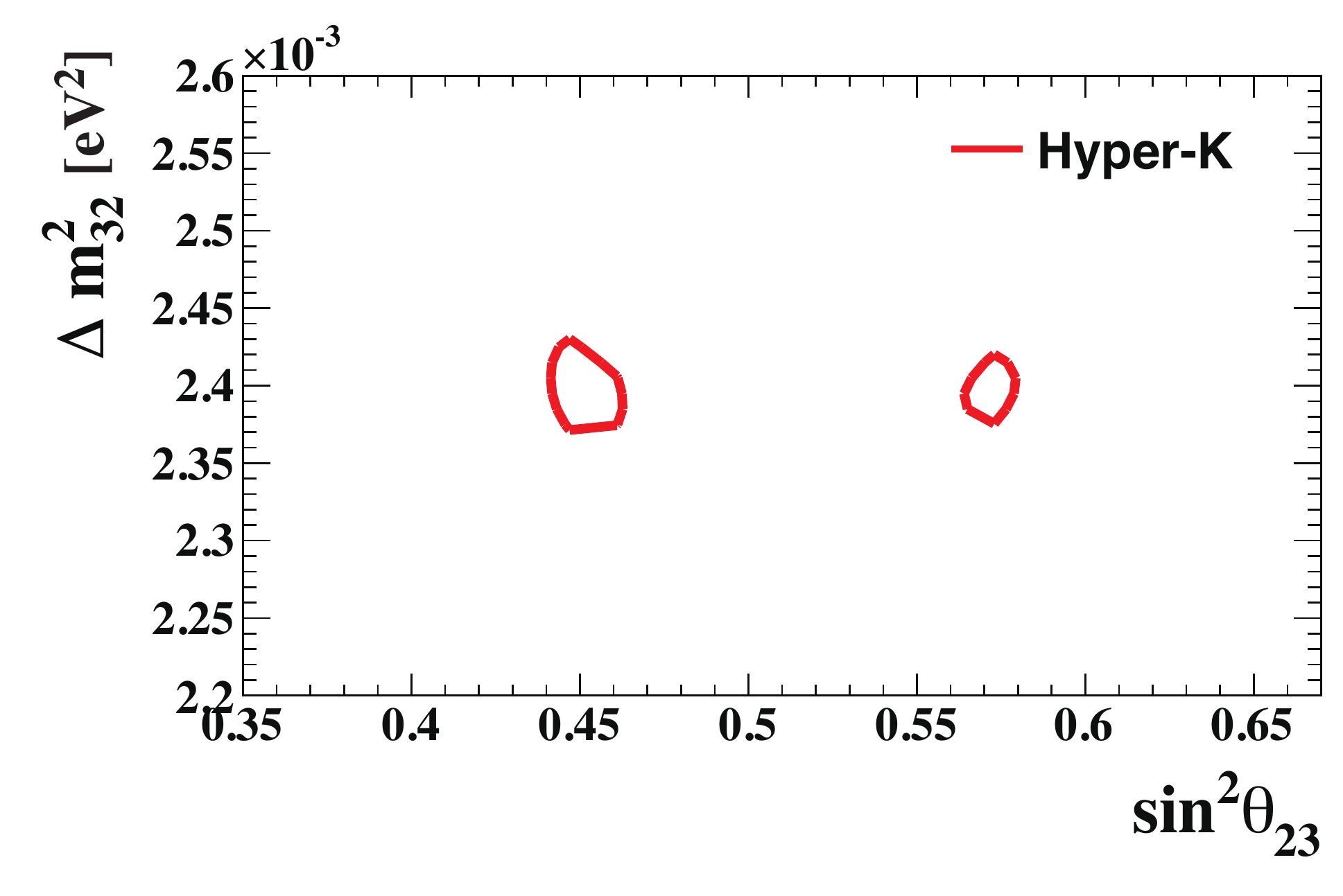}
\includegraphics[width=0.65\textwidth]{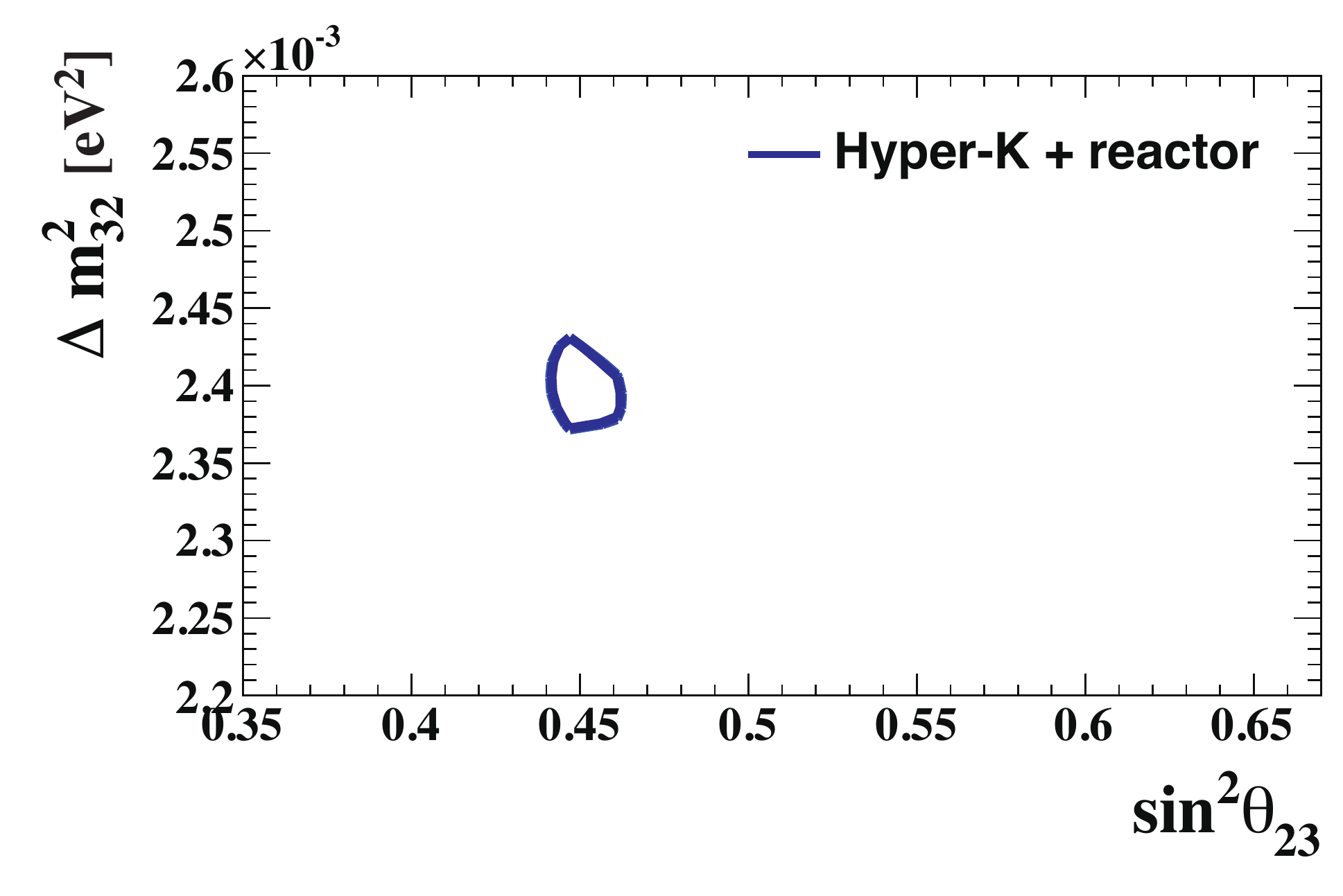}
\caption{ 90\% CL allowed regions in the $\sin^2\theta_{23}$--$\Delta m^2_{32}$ plane.
The true values are $\sin^2\theta_{23}=0.45$ and $\Delta m^2_{32} = 2.4 \times 10^{-3}$~eV$^2$.
Effect of systematic uncertainties is included.
Top: Hyper-K only. Bottom: With reactor constraint.
\label{fig:theta23-0.45}}
\end{figure}

\begin{table}[tdp]
\caption{Expected 1$\sigma$ uncertainty of $\Delta m^2_{32}$ and $\sin^2\theta_{23}$  for true $\sin^2\theta_{23}=0.45, 0.50, 0.55$. 
Reactor constraint on $\sin^22\theta_{13}=0.1\pm 0.005$ is imposed.}
\begin{center}
\begin{tabular}{ccccccc} \hline \hline
True $\sin^2\theta_{23}$	& \multicolumn{2}{c}{$0.45$}  		& \multicolumn{2}{c}{$0.50$} 		& \multicolumn{2}{c}{$0.55$}\\ 
Parameter  				& $\Delta m^2_{32}$ 	& $\sin^2\theta_{23}$  & $\Delta m^2_{32}$		& $\sin^2\theta_{23}$ 	& $\Delta m^2_{32}$	& $\sin^2\theta_{23}$\\ \hline
 Normal hierarchy	& $1.4\times10^{-5}$~eV$^2$		& 0.006 			& $1.4\times10^{-5}$~eV$^2$		& 0.015				& $1.5\times10^{-5}$~eV$^2$			& 0.009\\
 Inverted hierarchy	& $1.5\times10^{-5}$~eV$^2$		& 0.006 			& $1.4\times10^{-5}$~eV$^2$		& 0.015				& $1.5\times10^{-5}$~eV$^2$			& 0.009\\
\hline \hline
\end{tabular}
\end{center}
\label{tab:23sensitivity}
\end{table}%

The result shown above is obtained with $\sin^2\theta_{23}$ and $\Delta m^2_{32}$ as free parameters as well as $\sin^22\theta_{13}$ and $\deltacp$, with a nominal parameters shown in Table~\ref{Tab:oscparam}.
The use of the $\numu$ sample in addition to $\nue$ enables us to also measure $\sin^2\theta_{23}$ and $\Delta m^2_{32}$.
Figure~\ref{fig:theta23-0.50} shows the 90\% CL allowed regions for the true value of $\sin^2\theta_{23}=0.5$. 
Hyper-K will be able to provide a precise measurement of $\sin^2\theta_{23}$ and $\Delta m^2_{32}$.
Figure~\ref{fig:theta23-0.45} shows the 90\% CL allowed regions on the $\sin^2\theta_{23}$-$\Delta m^2_{32}$ plane, for the true values of $\sin^2\theta_{23}=0.45$ and $\Delta m^2_{32} = 2.4 \times 10^{-3}$~eV$^2$. 
With a constraint on $\sin^22\theta_{13}$ from the reactor experiments, the octant degeneracy is resolved and $\theta_{23}$ can be precisely measured.

The expected precision of $\Delta m^2_{32}$ and $\sin^2\theta_{23}$ for true $\sin^2\theta_{23}=0.45, 0.50, 0.55$ with reactor constraint on $\sin^22\theta_{13}$ is summarized in Table~\ref{tab:23sensitivity}.

\subsection{Combination with atmospheric neutrino data \label{sec:lbl-atm}}
\begin{figure}[tbp]
  \begin{center}
  \includegraphics[width=0.45\textwidth]{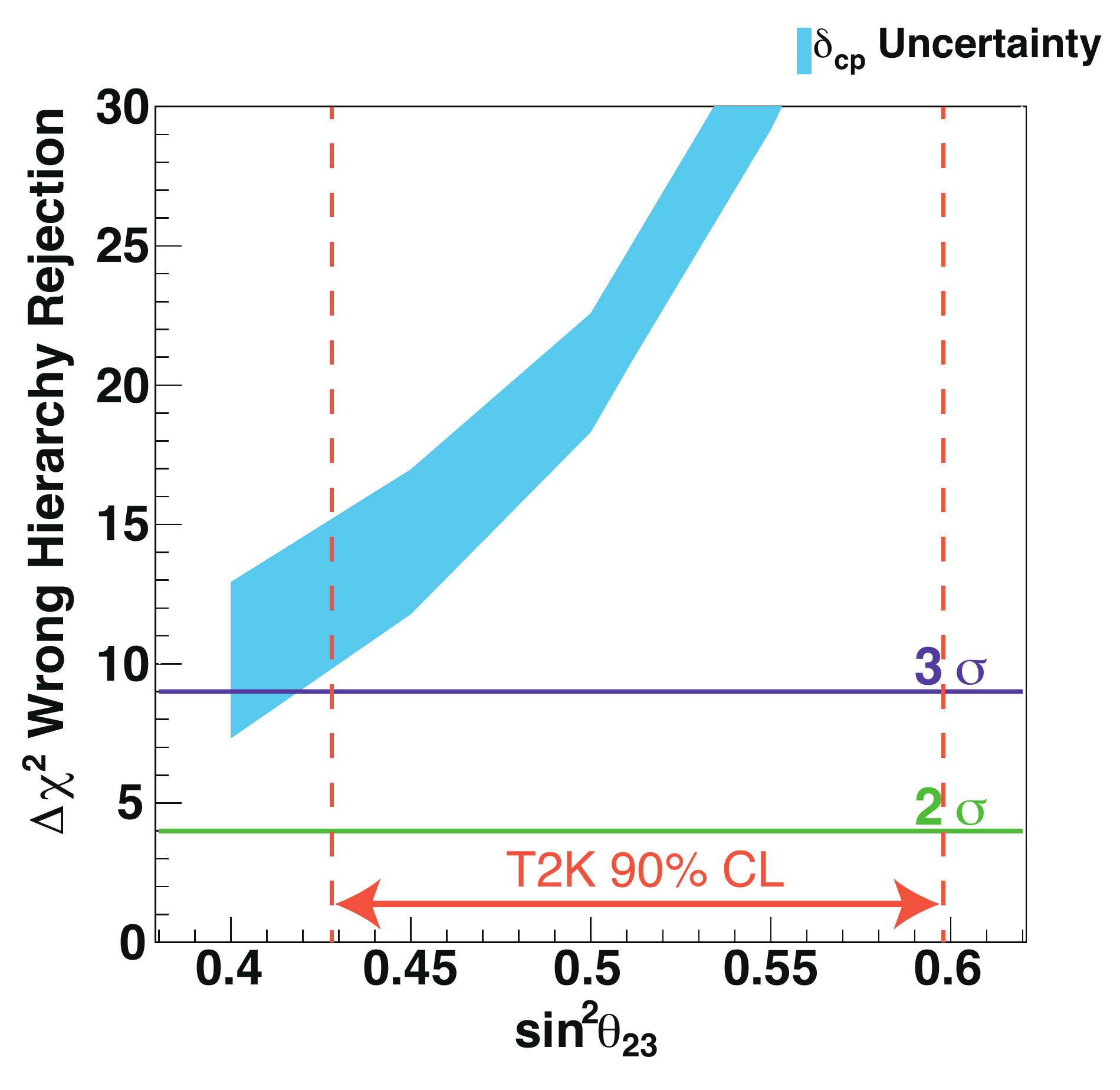}
  \includegraphics[width=0.45\textwidth]{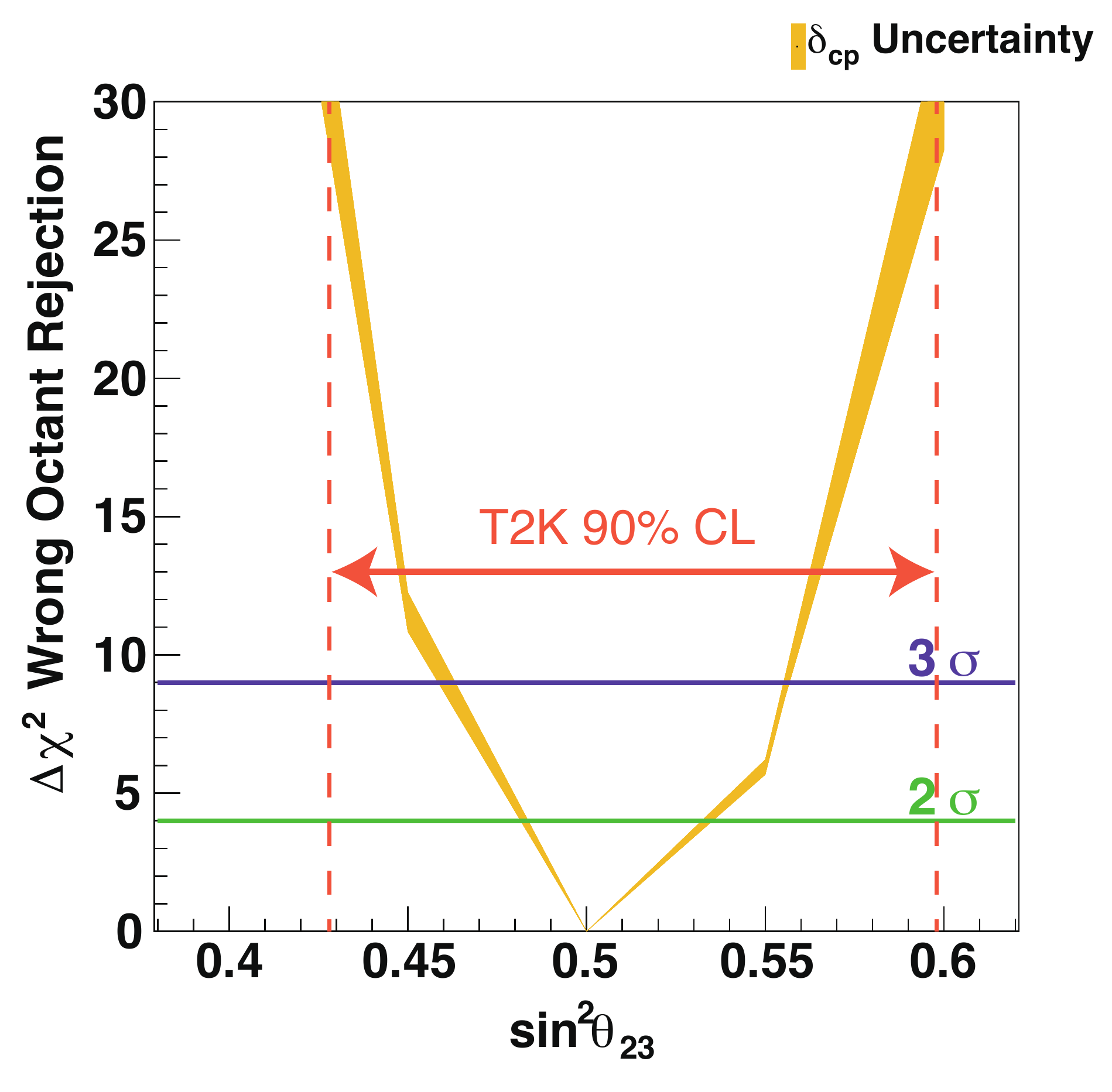}
     \caption{Atmospheric neutrino sensitivities for a ten year exposure of Hyper-K 
              assuming the mass hierarchy is normal. 
              Left: the $\Delta \chi^{2}$ discrimination of the wrong 
              hierarchy hypothesis as a function of the assumed true value of 
              $\sin^{2} \theta_{23}$.   
              Right: the discrimination between the wrong octant for each value of 
               $\sin^{2} \theta_{23}$. 
               The uncertainty from $\deltacp$ is 
              represented by the thickness of the band.
              Vertical dashed lines indicate 90\% confidence intervals of $\sin^2\theta_{23}$ from the recent T2K measurement~\cite{Abe:2014ugx}.
              }
     \label{fig:atmsens}
  \end{center}
\end{figure}

Atmospheric neutrinos can provide an independent and complementary information to the accelerator beam program on the study of neutrino oscillation.
For example, through the matter effect inside the Earth, a large statistics sample of atmospheric neutrinos by Hyper-K will have a good sensitivity to the mass hierarchy and $\theta_{23}$ octant.
Because Hyper-K will observe both accelerator and atmospheric neutrinos with the same detector, the physics capability of the project can be enhanced by combining two complementary measurements.

Assuming a 10 year exposure, Hyper-K's sensitivity to the mass hierarchy and
the octant of $\theta_{23}$ by atmospheric neutrino data are shown in Fig.~\ref{fig:atmsens}.
Depending on the true value of $\theta_{23}$ the sensitivity changes considerably,
but for all currently allowed values of this parameter the mass hierarchy 
sensitivity exceeds $3\,\sigma$ independent of the assumed hierarchy. 
If $\theta_{23}$ is non-maximal, the atmospheric neutrino data can be used to discriminate 
the octant at $3\,\sigma$ if $ \sin^{2}\theta_{23} < 0.46 $ or $ \sin^{2}\theta_{23} > 0.56 $. 

In the previous sections, the mass hierarchy is assumed to be known prior to the Hyper-K measurements.
This is a reasonable assumption considering the increased opportunities, thanks to a large value of $\theta_{13}$, of ongoing and proposed projects for mass hierarchy determination.
However, even if the mass hierarchy is unknown before the start of experiment, Hyper-K itself will be able to determine it with the atmospheric neutrino measurements.

\begin{figure}[tbp]
\includegraphics[width=0.45\textwidth]{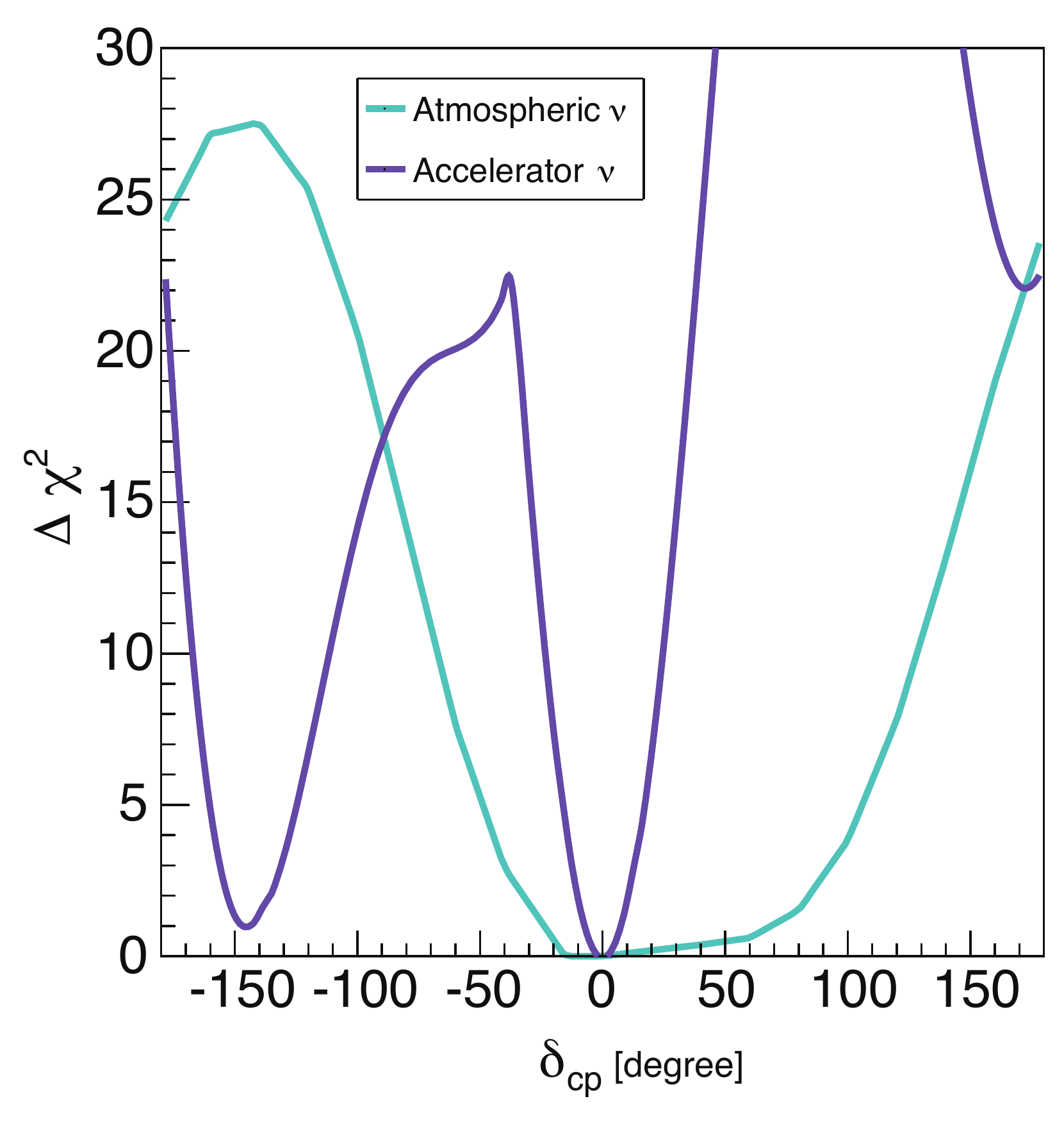}
\includegraphics[width=0.45\textwidth]{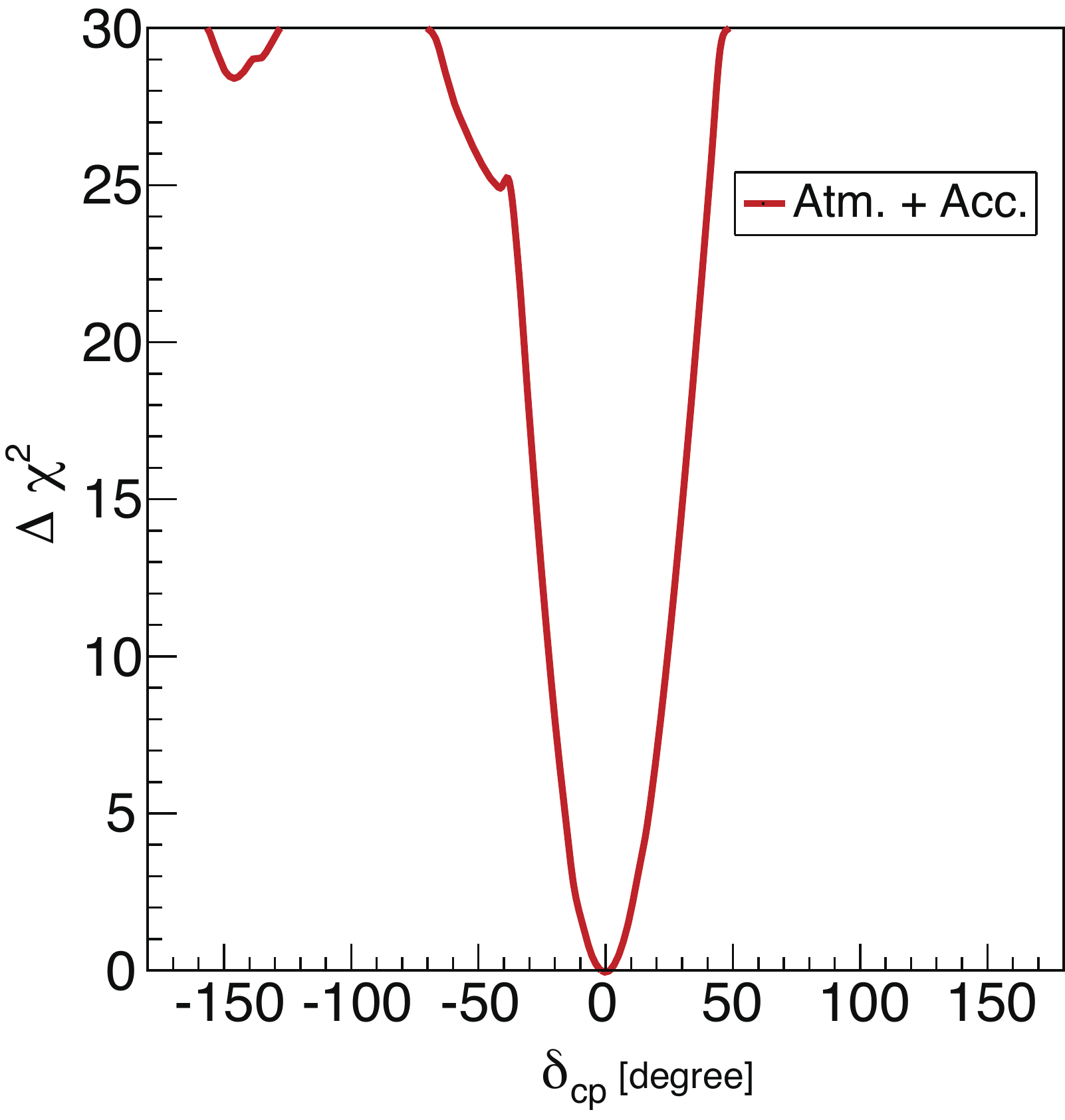}\\
\caption{Combination of the accelerator and atmospheric data. 
Left: Expected $\Delta \chi^2$ values for accelerator and atmospheric neutrino measurements assuming that the mass hierarchy is unknown. 
The true mass hierarchy is normal hierarchy and the true value of $\deltacp=0$.
Right: By combining the two measurements, the sensitivity can be enhanced. In this example study, the $\Delta \chi^2$ is simply added.
\label{fig:lbl-atm}}
\end{figure}

Because Hyper-K will observe both accelerator and atmospheric neutrinos with the same detector, the physics capability of the project can be enhanced by combining two complementary measurements.
As a demonstration of such a capability, a study has been done by simply adding
$\Delta \chi^2$ from two measurements, although in a real experiment a more sophisticated analysis is expected.
Assuming the true mass hierarchy of normal hierarchy and the true value of $\deltacp=0$,
the values of expected $\Delta \chi^2$ as a function of $\deltacp$ for each of
the accelerator and atmospheric neutrino measurements,
\textit{without} assumption of the prior mass hierarchy knowledge,
are shown in the left plot of Fig.~\ref{fig:lbl-atm}.
For the accelerator neutrino measurement, there is a second minimum near $\deltacp=150^\circ$ because of a degeneracy with mass hierarchy assumptions.
On the other hand, the atmospheric neutrino measurement can discriminate the
mass hierarchy, but the sensitivity to the $CP$ violating phase $\deltacp$ is worse than the accelerator measurement.
By adding the information from both measurements, as shown in the right plot of Fig.~\ref{fig:lbl-atm}, the fake solution can be eliminated and a precise measurement of $\deltacp$ will be possible.

\section{Conclusion}
The sensitivity to leptonic $CP$ asymmetry of a long baseline experiment using 
a neutrino beam directed from J-PARC to the Hyper-Kamiokande detector has been studied
based on a full simulation of beamline and detector.
With an integrated beam power of 7.5~MW$\times$10$^7$~sec,
the value of $\deltacp$ can be determined to better than 19$^\circ$ for all values of $\deltacp$
and
$CP$ violation in the lepton sector can be observed with more than 3~$\sigma$ (5~$\sigma$) significance for 76\% (58\%) of the possible values of $\deltacp$.

Using both $\nu_e$ appearance and $\nu_\mu$ disappearance data, a precise measurement of $\sin^2\theta_{23}$ will be possible.
The expected 1$\sigma$ uncertainty is 0.015(0.006) for $\sin^2\theta_{23}=0.5(0.45)$.

\bibliography{hyperk-loi-2014}

\end{document}